\documentclass[journal]{IEEEtran}

\usepackage{graphicx}

\usepackage{fancyhdr}
\usepackage{subcaption}

\usepackage{tabularx}
\usepackage{array} 
\usepackage{booktabs} 
\usepackage{amssymb}
\usepackage{comment}
\usepackage[linesnumbered,ruled,vlined]{algorithm2e}
\usepackage{svg}
\usepackage{amsmath}
\usepackage[acronym, nopostdot, nogroupskip]{glossaries}
\usepackage[style=ieee,
            maxbibnames=1,
            minbibnames=1,
            doi=false,
            url=false
            ]{biblatex}
\AtEveryBibitem{%
  \clearfield{note}%
  \clearfield{issn}%
  \clearfield{isbn}%
  \clearfield{pages}%
  \clearfield{volume}%
  \clearfield{number}%
  \clearfield{keywords}%
  \clearlist{language}
  \clearlist{eprint}
  \clearlist{eprinttype}
  \clearlist{eprintclass}
}
\renewbibmacro*{title}{%
  \iffieldundef{shorttitle}
    {\printfield[title]{title}}
    {\printfield[title]{shorttitle}}%
}

\newacronym{eo}{EO}{Earth observation}
\newacronym{rs}{RS}{remote sensing}
\newacronym{esa}{ESA}{European Space Agency}
\newacronym{hypso}{HYPSO}{HYPerspectral Small satellite for Ocean Observation}
\newacronym{ntnu}{NTNU}{Norwegian University of Science and Technology}
\newacronym{cubesat}{CubeSat}{Cube Satellite}

\newacronym{calval}{cal/val}{calibration and validation}
\newacronym{pics}{PICS}{pseudo-invariant calibration sites}
\newacronym{radcalnet}{RadCalNet}{Radiometric Calibration Network}
\newacronym{rolo}{ROLO}{Robotic Lunar Observatory}
\newacronym{lime}{LIME}{Lunar Irradiance Model}
\newacronym{dcc}{DCC}{deep convective clouds}
\newacronym{obcd}{OBCD}{onboard calibration device}

\newacronym{leo}{LEO}{low Earth orbit}
\newacronym{cots}{COTS}{commercial-off-the-shelf}

\newacronym{brdf}{BRDF}{bi-directional reflectance distribution function}
\newacronym{srf}{SRF}{spectral response function}
\newacronym{toa}{ToA}{top-of-atmosphere}
\newacronym{lvf}{LVF}{linear variable filter}
\newacronym{etf}{ETF}{electronically tunable filter}

\newacronym{gsd}{GSD}{ground sampling distance}
\newacronym{grd}{GRD}{ground resolvable distance}
\newacronym{mtf}{MTF}{modulation transfer function}
\newacronym{lsf}{LSF}{line spread function}
\newacronym{psf}{PSF}{point spread function}
\newacronym{snr}{SNR}{signal-to-noise ratio}
\newacronym{fwhm}{FWHM}{full width at half maximum}

\begin{document}

\title{In-orbit Intra-calibration of a Hyperspectral Satellite Constellation}

\author{Aria Alinejad,
    Cameron L. Penne,
    Dennis D. Langer,
    Simen Berg,
    Marie B. Henriksen,
    P. S. Vishnu,
    Roger Birkeland,
    Tor Arne Johansen,
    Joseph L. Garrett 
\thanks{This work has been submitted to the IEEE for possible publication. Copyright may be transferred without notice, after which this version may no longer be accessible.

Aria Alinejad, Joseph L. Garrett, Simen Berg, P. S. Vishnu, Marie B. Henriksen,  and Tor Arne Johansen are with the Department of Engineering Cybernetics, Norwegian University of Science and Technology (NTNU), Trondheim, Norway.

Dennis D. Langer is with the Department of Marine Technology, Norwegian University of Science and Technology (NTNU), Trondheim, Norway.

Roger Birkeland and Cameron L. Penne are with the Department of Electronic Systems, Norwegian University of Science and Technology (NTNU), Trondheim, Norway.

}}

\maketitle

\begin{abstract}
Small satellites offer frequent revisits, fine spatial resolution, and short development cycles, but lack the continuity that has enabled larger Earth observation missions to make fundamental contributions to science. Calibration and validation (cal/val) is therefore critical for their data to be used alongside that of larger missions, and for constellations, harmonized cal/val is essential for data from multiple satellites to be used interchangeably.

This article presents a cal/val procedure for hyperspectral small satellite constellations that requires no onboard calibration hardware, and tests it on the two CubeSats of the HYPSO constellation. Lunar captures, analyzed against the ESA Lunar Irradiance Model (LIME), are used for spectral and radiometric calibration, spectral resolution estimation, and assessment of post-launch misalignment, while smile is characterized in-orbit by edge detection of absorption features in Libya-4 captures. These procedures rely on resolving narrow absorption features, such as the solar Fraunhofer lines, which is enabled by the hyperspectral instrumentation. Libya-4 captures are also used, together with NASA's PACE satellite, to harmonize the radiometric magnitude of the two satellites, while RadCalNet sites and PACE are used for independent validation.

Two sets of calibration coefficients are derived, correcting post-launch misalignment of laboratory coefficients and harmonizing the radiometric response across the constellation. After calibration, Fraunhofer lines are within 0.14 nm of their expected positions on average, the remaining smile is below 0.8 nm, and the difference in spectral shape is 1.3\% between the satellites and about 3\% relative to PACE. Over Libya-4, the HYPSO spectra differ from PACE by 4.0\% on average after correcting for differences in viewing geometry, compared to 7.1\% and 17.3\% with the laboratory coefficients, and normalized difference indices over RadCalNet sites are consistent with PACE. These results demonstrate that hyperspectral small satellite constellations can provide data products suitable for joint analysis, within the constellation and alongside larger missions, provided the analysis is robust to the remaining uncertainty in radiometric magnitude. Spectral resolution, SNR, spatial resolution, and wavelength-dependent blur are also compared.
\end{abstract}

\IEEEpeerreviewmaketitle

\newpage 
\section{Introduction}
\IEEEPARstart{H}{yperspectral} small satellite constellations offer a cost-effective and flexible platform for \gls{eo}, with applications across environmental monitoring, resource management, and scientific research. Compared to multispectral instruments, which are designed around a fixed set of spectral bands selected for anticipated downstream tasks, hyperspectral sensors capture a continuous spectral range, with at least 30 contiguous bands \cite{skauli2022specification}. This allows the same dataset to be used for a wide variety of applications, including ones not foreseen at the time of instrument design, and enables analysis techniques that make use of the full spectrum, such as separation of spectra into different identifiable materials through unmixing \cite{rasti_image_2024}.

Satellite constellations provide shorter revisit times compared to single satellites, and can effectively extend the lifetime of each satellite if joint calibration is performed: as individual satellites de-orbit, their successors can continue to gather ground-truth matchups and utilize previously acquired data for algorithm development. When satellites are launched incrementally, this also creates a window to learn from earlier instruments and introduce improvements between launches. Despite these benefits, a constellation also places further demands on calibration, as it not only pertains to the accuracy of each satellite's measurement relative to the scene, but also to its consistency with measurements from the other satellites in the constellation.

As the new space era unfolds, independent actors are increasingly entering the \gls{eo} field, and the CubeSat design standard has made space missions accessible to projects with significantly lower budgets \cite{welle_overview_2020}. Small satellites are typically operating in \gls{leo}, giving fine \gls{gsd} and revisit times, while reduced costs allow for rapid deployment cycles. These factors are connected to the shift from large satellites to constellations of small spacecraft \cite{marcuccio_smaller_2019}. Moreover, many small satellite operators buy hyperspectral instrumentation commercially \cite{rijlaarsdam2024next, angulo2026kaustsat}, which creates constellations through common instrumentation, if not in operation. However, while this development pace is well suited to iterative scientific improvements and commercially competitive operations, it also places new demands on \gls{calval}. Smaller budgets and tighter mass constraints mean that small satellites are less likely to carry \glspl{obcd}, making terrestrial and extraterrestrial targets the primary sources for in-orbit calibration. The short in-orbit lifetime of CubeSats, typically three to seven years, further motivates efficient use of imaging, ideally covering several cal/val procedures per capture.

Cal/val is essential to ensure that satellite measurements are physically interpretable, comparable with data from other missions, and suitable for scientific use. It is also necessary to detect, characterize, and correct for post-launch changes, artifacts, and inter-sensor deviations that would otherwise undermine the interoperability of constellation data and the transferability of analysis algorithms across satellites. It will enable data from small-satellite constellations to be used alongside that of high-budget, large satellites, which in turn allows ground-truth matchups from these missions to be used for transfer-learning approaches, and facilitates systems-of-systems approaches that combine satellite data with other measurement modalities \cite{dallolio_satelliteusv_2021}.

\begin{table*}[htbp]
    \centering
    \caption{Overview of radiometric intra-calibration methods.}
    \label{tab:intra-calibration_overview}
    \begin{tabular}{||>{\raggedright\arraybackslash}p{0.3\linewidth}|>{\raggedright\arraybackslash}p{0.22\linewidth}|>{\raggedright\arraybackslash}p{0.3\linewidth}||}
    \hline
    \textbf{Target} & \textbf{Satellite} & \textbf{Instrument} \\
    \hline \hline

    RadCalNet & Sentinel-2A/2B; Zhuhai-1; HYPSO& Multispectral; Hyperspectral; Hyperspectral\\\hline
    Cross-calibration & PlanetScope & Multispectral \\\hline
    Tandem flight & Sentinel-3A/3B; Landsat-8/9& Multispectral; Multispectral \\\hline
    Rayleigh scattering & Sentinel-2A/2B; Sentinel-2A/2C& Multispectral; Multispectral \\\hline
    DCC& Sentinel-2A/2B; Sentinel-2A/2C& Multispectral; Multispectral \\\hline
    Lunar calibration & RapidEye; HYPSO & Multispectral; Hyperspectral \\
    \hline

    \end{tabular}
\end{table*}

\subsection{Related Work and Research Gap}
 Table \ref{tab:intra-calibration_overview} gives an overview of established radiometric intra-calibration methods for satellite constellations that do not rely on \glspl{obcd}. Firstly, one can use a common terrestrial reference with ground radiometry, such as a \gls{radcalnet} site; this has been done by both the Zhuhai-1 hyperspectral small-satellite constellation \cite{zhang_onorbit_2022} and the Sentinel-2A/2B constellation \cite{revel_sentinel2a_2019}. A second method is to use another well-calibrated satellite as the reference, as has been done for the PlanetScope small-satellite constellation using Landsat 8 OLI \cite{houborg_cubesat_2018, leach_normalization_2019}. It is also possible to calibrate using deep ocean captures, since the signal will be dominated by Rayleigh scattering and can therefore be simulated and used as a reference \cite{revel_sentinel2a_2019}. \Gls{dcc} are also used as calibration references, since they have little atmosphere above them and exhibit a near-Lambertian reflectance, making the signal easy to simulate \cite{revel_sentinel2a_2019}.

All of the methods mentioned above benefit greatly from temporal proximity and similar viewing geometry. An approach that has therefore emerged is near-simultaneous inter-satellite comparison, in which two platforms observe the same target within a short time window through precise synchronization of their orbits, as part of their early operation campaigns. Examples include the Sentinel-3 A/B tandem phase, where the two satellites flew 30 seconds apart along the same ground track \cite{lamquin_olci_2020}, and the Landsat-8/9 underfly \cite{gross_initial_2022}. Another approach is to use the Moon as a common calibration target, which avoids the need for temporal proximity due to the Moon's lack of atmosphere and temporal stability. A recent comparison of Sentinel-2C against Sentinel-2A utilized all of the methods mentioned above \cite{clerc_copernicus_2026}; Moon captures were also taken with Sentinel-2C, but could not be used for intra-comparison since Sentinel-2A had not taken any. The only peer-reviewed work we found to use Moon captures for radiometric intra-comparison of a satellite constellation is on the RapidEye constellation \cite{steyn_rapideye_2009}; where no Lunar irradiance model, such as \gls{rolo} or \gls{lime}, is utilized. Furthermore, of the above-mentioned constellations, Zhuhai-1 is the only one equipped with hyperspectral instruments. In this work we therefore investigate the utility of the Moon for radiometric intra-calibration of a hyperspectral constellation, utilizing the \gls{esa} \gls{lime}.

Regarding spectral calibration there are important differences between multispectral and hyperspectral instruments. For multispectral instruments, Spectral Band Adjustment Factors (SBAF) is a common approach, as used by the Landsat and Sentinel-2 constellations \cite{gross_initial_2022, clerc_copernicus_2026}. For hyperspectral sensors, the higher spectral resolution enables utilization of narrow atmospheric absorption features, such as the oxygen A-band, over terrestrial targets; which is the established approach across most hyperspectral missions \cite{gao_vicarious_2012, benhadj_hyperscout1_2024, cogliati_prisma_2021, yamamoto_initial_2022, raj_postlaunch_2025, brell_assessment_2026, sanders_orbital_2024, jr_initial_2024}. The Moon offers an alternative spectral reference, with Fraunhofer lines well preserved in the absence of a Lunar atmosphere. PACE is the only hyperspectral mission found to have used the Moon for spectral calibration; validating the positions of individual Fraunhofer absorption dips against expected values \cite{meister_onorbit_2024}. In this work we perform a full-spectrum comparison against the LIME-modeled Lunar irradiance, which can characterize higher-order distortions of the spectral axis rather than a simple positional shift. Several of the procedures presented here rely on resolving narrow absorption features, which is only possible with a hyperspectral instrument: spectral calibration and estimation of spectral resolution against the Fraunhofer lines of the LIME spectrum, in-orbit characterization of smile using absorption features across the slit, and validation of the spectral axis using Fraunhofer lines and the oxygen A-band in terrestrial captures.

\subsection{Mission Description and Paper Contributions}
The HYPerspectral Small satellite for ocean Observation (HYPSO) mission consists of two CubeSats equipped with identically designed hyperspectral imaging payloads, developed by the Small Satellite Lab at the Norwegian University of Science and Technology (NTNU) \cite{grotte_ocean_2022, bakken_hypso1_2023}. In this paper we present a cal/val procedure for hyperspectral small satellite constellations that requires no \glspl{obcd}, only Lunar captures and a small selection of terrestrial targets, and demonstrate it on the HYPSO constellation. The contributions are threefold. First, we show how Moon captures, analyzed against \gls{lime}, can be used for spectral calibration, radiometric intra-calibration, estimation of spectral resolution, assessment of post-launch misalignment of laboratory coefficients, and artifact analysis, while terrestrial targets are used for smile correction, harmonization of the radiometric magnitude, and assessment of \gls{grd} and \gls{snr}. Second, we quantify which parts of the calibration are more and less reliable, by validating the spectral and radiometric coefficients against references that are independent of the Moon: Fraunhofer lines and the oxygen A-band in terrestrial captures, and PACE and \gls{radcalnet}. Since PACE is used only as a transfer reference to harmonize the magnitude of HYPSO-2 to that of HYPSO-1, while HYPSO-1 is not adjusted to it, this shows how closely a small satellite with \gls{cots} instrumentation and no \glspl{obcd} can match a well-calibrated reference mission. Third, we characterize two identically designed \gls{cots} hyperspectral instruments in orbit, showing how similar their responses can be made, and where the limitations are. Two sets of calibration coefficients are derived: one correcting for post-launch misalignment of laboratory coefficients, and one harmonizing the radiometric response across the constellation to enable joint algorithms that utilize data from both satellites interchangeably.

\section{Background}
\begin{figure*}[tb]
    \centering
    \includegraphics[width=\linewidth]{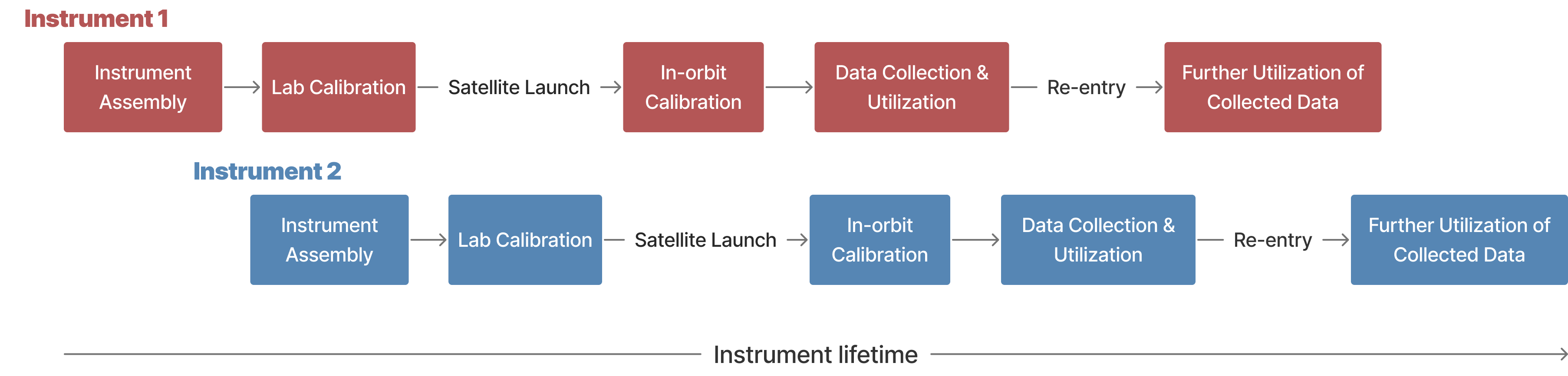}
    \caption{Constellation lifetime stages for a two satellite hyperspectral constellation. }
    \label{fig:instrument_life}
\end{figure*}

In hyperspectral constellations, even if all instruments have identical design, differences in sensor response are inevitable, undermining the interchangeable and complementary use of multi‑satellite data. Each satellite undergoes separate processes of assembly, lab calibration, launch, and deployment. These are all processes that may introduce distinct deviations. Asynchronous production timelines across sensors further amplify these effects, and different launch times result in unequal time in orbit, resulting in different degradation. Figure \ref{fig:instrument_life} shows the instrument lifetime stages for a two-satellite hyperspectral constellation. The subsequent paragraphs first discuss how calibration uncertainties affect downstream data products, and then outline relevant laboratory calibration procedures and how in‑orbit calibration can build on them. This is followed by a broad overview of different types of hyperspectral instruments, and finally, a description of the HYPSO constellation.

\subsection{Data Products from Spectral Imagers} \label{sec:data_products}
One critical aspect of a calibration is how the uncertainties it uncovers affect downstream data products. The simplest data product, and the one on which most validation efforts concentrate, is the measured radiance or reflectance at each wavelength, often at the \gls{toa}. Other common data products are normalized difference indices (NDIs) \cite{philpot2021nd}, typified by the normalized difference vegetation index, and extended band-ratio algorithms, such as those used for ocean color \cite{o2019chlorophyll}. Since these are formed from ratios between bands, they are invariant to an overall multiplicative radiometric error that does not alter the spectral shape of the radiometric response. Band ratios and normalized difference indices can also be used as inputs to more recent machine learning approaches \cite{o2023hyperspectral, peng2025novel}. Some algorithms bring remote sensing even closer to spectroscopy by taking the wavelength at which a particular feature occurs to be the object of interest \cite{ruddick2001optical}. Spectral features can also be emphasized by continuum removal, where the spectrum is divided by its convex hull \cite{clark1984reflectance}, or by derivative analysis \cite{dierssen2015space}. For data products of these types, the shape of the radiometric response is therefore more important than its magnitude.

\subsection{Laboratory Calibration}
Two of the most important lab calibration types are radiometric and spectral calibration \cite{skauli2022specification}. These are used, respectively, to convert instrument counts into SI‑traceable radiance values with unit $[\text{Wm}^{-2}\text{sr}^{-1}\text{nm}^{-1}]$ and to map the instrument response onto a spectral axis. Radiometric calibration is performed in the lab using a target with well‑known irradiance output and a uniform, or well‑characterized, spectral response. By using the expected radiometric response for each pixel, a set of radiometric calibration coefficients can be calculated. For spectral calibration, gas lamps that emit radiation at specific wavelengths can be used to construct a mapping between measured peaks and known wavelength values. This mapping can, in turn, be used to extrapolate a set of spectral coefficients. Spectral and radiometric lab calibration are highlighted here as they are useful in contextualizing the subsequent sections; however, there are many other important lab calibration types and procedures. A more complete description of hyperspectral instrument calibration can be found in \cite[Chapter~10]{eismann_hyperspectral_2012}.

\subsection{In-orbit Calibration}
Despite thorough laboratory calibration, launch, deployment, and the space environment place the instrument under substantial mechanical and thermal stress, as well as radiation, which can lead to changes in the optical system. In‑orbit cal/val therefore aims to detect, characterize, and correct post‑launch changes relative to the pre‑launch baseline.

\subsection{References and Targets for Vicarious Calibration}
Vicarious calibration has emerged as an important approach, in which well‑characterized, stable, and uniform targets are used for calibration \cite{turpie_calibration_2023, eismann_hyperspectral_2012}.

\subsubsection{Terrestrial Targets}
In practice, some terrestrial sites have been designated as \gls{pics} and have been found to be radiometrically constant over time, with uncertainties of 1\% to 3\% after accounting for atmospheric effects \cite{yoon_guidelines_2015}. These sites are typically spatially homogeneous, have a low probability of cloud cover, and exhibit a relatively constant \gls{brdf} over time. In their most basic use, PICS provide a stable target imaged by the sensor itself, from which radiometric stability can be tracked or a reference reflectance can be modeled; they do not, on their own, provide an independent ground-truth measurement.

Some sites are additionally instrumented with automated in-situ radiometers that supply an external reference. \gls{radcalnet} is one such network: it provides quality-controlled, SI-traceable \gls{toa} reflectances and their associated uncertainties at a set of sites, sampled at regular intervals  \cite{bouvet_radcalnet_2019}. This makes RadCalNet suitable for validation against an independent reference, rather than only for internal consistency checks, and it has been widely used for vicarious calibration of hyperspectral missions, including Zhuhai-1, EnMAP, and HyperScout-1 \cite{zhang_onorbit_2022, storch_enmap_2023, benhadj_hyperscout1_2024}. HYPERNETS is a more recent network offering higher spectral resolution \cite{ruddick_hypernets_2024}.

\subsubsection{Lunar Target}\label{sec:lunar_target}
Apart from terrestrial targets, the Moon can also be used for several important in-orbit cal/val procedures, and has several important benefits. It is uniquely stable over time and has virtually no atmosphere, resulting in less distortion and uncertainty. Furthermore, it is a relatively bright target against a dark background and does not exhibit significant spectral absorption features, largely preserving the features of the solar spectrum. However, to obtain reliable irradiance estimates, an analytic model is needed to account for variations due to phase, libration, and the Moon’s non-Lambertian reflectance behavior \cite{stone_moon_2009}. One such model is \gls{lime}, which builds on \gls{rolo} \cite{toledano_lime_2024}. LIME accounts for Earth–Sun–Moon geometry and Lunar libration and outputs the integrated disk irradiance of the Moon at a given time and satellite viewing geometry, with an expanded radiometric uncertainty of approximately 2\%. The modeled irradiance can then be compared to satellite observations for cal/val.

At present, the Moon has already served as a stable reference source for cal/val across multiple satellite missions, including SeaWiFS, MODIS, VIIRS, GaoFen-4, COMS, Jilin-1 GP02, MethaneSat, Sentinel-3B, Hyperion, and PACE \cite{barnes_changes_1999,  xiong_intercomparison_2008, wang_viirs_2014, cai_using_2024, oh_coms_2018, shu_absolute_2023, nasr_methanesat_2025, neneman_use_2020, kieffer_initial_2002, lee_pace_2024, meister_onorbit_2024}, of which Sentinel-3B and MethaneSat used LIME, while Hyperion and PACE are the only satellites that host hyperspectral instruments.

\subsection{Instrument Harmonization}
When the goal is to harmonize several instruments, intra‑calibration and cross‑calibration are useful paradigms. These terms are often used interchangeably \cite{turpie_calibration_2023}; however, in this paper we use intra‑calibration to denote the use of a common reference to calibrate two instruments in the same constellation to agree with each other, and cross‑calibration and -validation to denote the use of a well‑characterized instrument to calibrate or validate another. Both approaches are demonstrated for the HYPSO constellation: Lunar captures are used to derive intra‑calibration radiometric gain factors that harmonize the shape of the radiometric response, and NASA's PACE satellite \cite{gorman_nasa_2019} is used as a common reference over Libya-4 to harmonize its magnitude, whereas PACE and RadCalNet are used for cross‑validation over PICS targets.

\subsection{In-orbit Performance Characterization} \label{sec:inorbit_char}
Several of the in-orbit characterization procedures depend on each other's results. Table~\ref{tab:cal_dependencies} summarizes how smile correction, spectral calibration, correction of misalignment, estimation of spectral resolution, and radiometric calibration each rely on the outputs of the others. Since no order removes all dependencies, we propose the order in which they are presented in this section, where each procedure uses the in-orbit coefficients of the preceding procedures and laboratory coefficients otherwise. The order is chosen so that the remaining dependencies on laboratory coefficients are weak. Smile is placed first, as it does not require accurate spectral or radiometric calibration and mostly depends on the relative position of features along the slit. Spectral calibration follows, since the misalignment is derived from the spectral shift, and the spectral resolution is estimated against a reference at the calibrated wavelengths. Radiometric calibration is placed last, as it depends on all of the above, whereas the preceding procedures rely mainly on the shape and position of spectral features rather than the absolute radiance.

\begin{table}[htbp]
\centering
\caption{Dependencies between the in-orbit characterization procedures, listed in the order they are performed. \textbullet: uses the in-orbit result of a preceding procedure. $\circ$: uses the laboratory value, as the procedure is performed later. --: not required.}
\label{tab:cal_dependencies}
\begin{tabular}{@{}lccccc@{}}
\toprule
& \multicolumn{5}{c}{Requires} \\
\cmidrule(l){2-6}
Procedure & Smile & Spectral & Misalign. & Resol. & Radiom. \\
\midrule
Smile        & --          & $\circ$     & $\circ$     & --          & $\circ$ \\
Spectral     & \textbullet & --          & $\circ$     & $\circ$     & $\circ$ \\
Misalignment & --          & \textbullet & --          & --          & -- \\
Resolution   & \textbullet & \textbullet & \textbullet & --          & $\circ$ \\
Radiometric  & \textbullet & \textbullet & \textbullet & \textbullet & -- \\
\bottomrule
\end{tabular}
\end{table}

\subsubsection{Smile}
An important distortion in hyperspectral pushbroom instruments is the smile effect, a spectral misregistration where the wavelengths recorded at the edges of the swath are shifted compared to the center. In-orbit, smile is commonly characterized from the position of atmospheric absorption features across the swath, as done for EnMAP, PRISMA, HISUI, and GHOSt \cite{brell_assessment_2026, cogliati_prisma_2021, yamamoto_initial_2022, sanders_orbital_2024}. In this work we also use the position of absorption features, but instead of estimating the spectral shift of each across-track pixel by modeling a few strong atmospheric absorption bands, we apply a Sobel edge detection filter \cite{gonzalez2008digital} to bright captures of a uniform target. This extracts solar and atmospheric absorption lines as continuous curves along the slit, at several locations across the spectral axis, similar to the emission lines used for smile calibration in the lab. The smile is then given directly by the curvature of these lines, without radiative transfer modeling, and is measured across the spectral range rather than at a few absorption bands.

\subsubsection{Spectral Calibration}
Spectral calibration is an important step prior to radiometric calibration, since a misaligned spectral axis leads to radiometric coefficients being applied at the wrong wavelengths. Accurate spectral alignment also matters downstream: atmospheric correction is central to many hyperspectral EO applications and remains a challenge for tasks such as the estimation of phytoplankton groups \cite{bracher_obtaining_2017}. A misaligned spectral axis propagates into the atmospheric model and introduces errors in the retrieved surface products.

Some hyperspectral satellites, such as HISUI, EnMAP, and PACE, carry \glspl{obcd} for in-orbit spectral calibration \cite{urai_initial_2021, storch_enmap_2023, jr_initial_2024}. However, such hardware is not always available, and even when it is, ground-based references remain valuable, as shown for HISUI \cite{yamamoto_initial_2022}. The established ground-based approach is to compare the measured position of atmospheric absorption features, such as the oxygen A-band, against their expected wavelengths. A limitation of this approach is that such features do not span the full visible range, so the correction is constrained by only a small number of reference points and cannot fully characterize how the spectral axis is distorted.

If the spectral resolution is sufficient, the Fraunhofer lines caused by absorption in the solar atmosphere can be used instead. These are preserved in the Lunar spectrum (Section \ref{sec:lunar_target}), and comparing the full measured spectrum against a modeled reference such as LIME characterizes the entire wavelength axis rather than a few isolated points, allowing both a uniform shift and higher-order distortions to be characterized.

Once the spectral coefficients are calibrated, the Fraunhofer lines detected by the Sobel filter can also be used for validation. Here, their measured positions in the bright uniform targets are compared against those in the TSIS-1 Hybrid Solar Reference Spectrum \cite{coddington_version_2023} convolved with the instrument response, providing a check independent of the Lunar calibration target.

\subsubsection{Misalignment}
The mechanical shock of launch can cause movement of optical components, leading to misalignment between detector pixels and their assigned spectral and radiometric coefficients. Such misalignment should, if possible, be corrected in-orbit.

\subsubsection{Spectral Resolution}
If the reference spectrum used for spectral calibration has a higher spectral resolution than the instrument under assessment, it can also serve as a reference for estimating the instrument's spectral resolution. This is done by convolving \glspl{srf} of varying widths to the reference spectrum and identifying which convolved version most closely agrees with the instrument's measured response. Bakken et al. \cite{bakken_hypso1_2023} demonstrate this approach on HYPSO-1 using the solar spectrum directly, building on the method of Ostrikov et al. \cite{ostrikov_estimation_2019}. Here we follow the same approach but build on it by using LIME as the reference, because it is better suited (Section \ref{sec:lunar_target}).

\subsubsection{Radiometric Calibration}
In-orbit assessment of the radiometric response follows an approach analogous to laboratory calibration, but the choice of reference source becomes central. Established methods for vicarious radiometric assessment include PICS-based absolute models, RadCalNet, cross-calibration against a well-characterized satellite, traditional reflectance-based vicarious calibration (TRBVC), \gls{dcc}, Rayleigh scattering over deep ocean, and Lunar calibration \cite{kabir_vicarious_2020}. TRBVC relies on dedicated in-situ measurements of surface reflectance and atmospheric state during overpass; it can be applied over a wide range of target types but requires field campaigns and is correspondingly costly. The RadCalNet, PICS, and cross-calibration approaches require only imagery of a suitable target by the sensor under assessment, and are therefore cheaper and more commonly used; across hyperspectral missions they are the dominant choice, as used by HICO, PRISMA, HISUI, EnMAP, Zhuhai-1, HyperScout-1, GHOSt, and Pixxel \cite{gao_vicarious_2012, meygret_prisma_2024, mizuochi_first_2024, storch_enmap_2023, zhang_onorbit_2022, benhadj_hyperscout1_2024, sanders_orbital_2024, raj_postlaunch_2025}.

A common difficulty with these approaches is that the compared measurements should be acquired close in time, and the off-nadir viewing angle influences the result. Using the Moon avoids both issues (Section \ref{sec:lunar_target}); among hyperspectral missions, only Hyperion and PACE have previously used it for radiometric calibration \cite{kieffer_initial_2002, lee_pace_2024}. Kieffer et al. \cite{kieffer_initial_2002} further show that any two spacecraft that have observed the Moon can be cross-calibrated regardless of the time between observations, and here we build on this for a hyperspectral constellation.

When using the Moon as a radiometric reference source it requires separation of the Lunar signal from the scene background. A straightforward strategy is to classify pixels into Moon and background and apply a mask, but this can be unreliable if the Moon's spatial extent varies with wavelength, or if cross-talk (smearing) causes signal to bleed between Moon and background pixels (see Section \ref{sec:artifacts}). In such cases, it is preferable to estimate and subtract the average background signal from all pixels to extract the Lunar signal. The method used in this paper is detailed in Section~\ref{ax:Extraction of Moon Signal} of the Supplementary Material. As the magnitude of the extracted Lunar signal is also affected by these factors, we use the Moon to calibrate the shape of the radiometric response, and harmonize the magnitude by cross-calibration against a well-characterized satellite over a PICS. Since the satellites typically pass over the site at different times of day, the difference in viewing and illumination geometry must then be accounted for, as the reflectance of the site depends on its \gls{brdf}.

If coefficients are adjusted, they should subsequently be validated against a reference that is as independent as possible from the one used for calibration. Here, cross-validation with PACE over RadCalNet sites serves this purpose for both the spectral and radiometric coefficients, since the oxygen A-band position is defined by atmospheric physics rather than by the Lunar model.

\subsubsection{Geometric Response}
When performing cal/val of satellite constellations it is also useful to investigate and compare the spatial resolution of the instruments. For this the \gls{gsd} is an important metric

\begin{equation}
    GSD_{across-track} = \frac{dp}{f} \; \text{[m / pixel]},
    \label{eq:gsd}
\end{equation}
where $d$ [m] is the distance to the target, $p$ [m / pixel] is the pixel size, and $f$ [m] is the focal length \cite{eismann_hyperspectral_2012}. The GSD quantifies how much ground area each pixel represents, and for a constellation of identically designed instruments it will be nearly constant across satellites, with altitude and off-nadir angle being the primary sources of variation. A more diagnostic metric for inter-satellite comparison is therefore the \gls{grd}, which estimates the smallest distance between two objects that the instrument can still distinguish. Unlike the GSD, the GRD depends on the \gls{mtf} of the optical system, which is more susceptible to variation introduced by instrument assembly, launch, and deployment. Given the \gls{fwhm} (in pixels) of the \gls{lsf}, the GRD can be estimated using the following equation

\begin{equation}
    GRD_{across-track} = GSD_{across-track} \cdot FWHM \; \text{[m]}.
\end{equation}

Methods for estimating the GRD or MTF fall into three categories based on the target used: edge, pulse/line, and impulse methods \cite{kabir_vicarious_2020}. Among these, we adopt the line method, deriving the LSF from a thin, bright bridge target. To compare the spatial resolution of the HYPSO satellites, we restrict our investigation to the across‑track GRD. For a pushbroom instrument, Equation~(\ref{eq:gsd}) is valid only for across‑track calculations. Consequently, for the remainder of this paper, we simplify the notation by using $GSD$ and $GRD$ for $GSD_{across-track}$ and $GRD_{across-track}$. Atmospheric scattering also influences the GRD. We report the GRD prior to atmospheric correction, consistent with the data product currently provided to HYPSO users.

\subsubsection{Artifacts and SNR}
Noise and artifacts may also account for unexpected differences between instruments in a constellation, and are therefore important to characterize. As a generalization, noise refers to unwanted contributions to the signal due to random processes, such as thermal noise, and can be quantified by the \gls{snr}. Artifacts, on the other hand, are structured unwanted signal contributions, such as ghosting from internal reflections. Because of the high contrast between the Moon and the background, Lunar captures are particularly useful for investigating both cross-talk and ghosting effects. Other artifacts such as striping, keystone, and dead pixels are not addressed in this paper but should in general be included for complete instrument characterization.

There are two common ways of Earth-imagery-based SNR estimation: homogeneous area method (HA), and the local means and local standard deviations method (LMLSD) \cite{kabir_vicarious_2020}. Here we use the HA method, with the same implementation as was previously used for HYPSO-1 and described in \cite{bakken_hypso1_2023}. Let $Y$ be a $B \times N$ dimensional matrix of $B$ bands and $N$ pixels. $y_i$ is then the vector containing the value of the $i$'th band for each pixel. Assuming $y_i$ can be explained as a linear combination of the $(B-1)$ other bands, we can write

\[
y_i=Y_{\delta_i} \beta_i + \xi_i
\]

where $Y_{\delta_i}$ is a matrix containing all rows of $Y$ except for $y_i$, $\beta_i$ is the regression vector of size $(B-1)\times 1$, and $\xi_i$ is the modeling error. Using the least squares estimator for regression this can be used to estimate the spectral noise

\[
\hat\beta_i = (Y_{\delta_i}^TY_{\delta_i})^{-1}Y_{\delta_i}^Ty_i,
\]
\[
\xi_i = y_i - Y_{\delta_i}\hat\beta_i,
\]
where $\xi_i$ estimates the spectral noise in band $i$ \cite{bioucas-dias_hyperspectral_2008}. While the spatial noise can be estimated as
\[
\xi_n = (x_n-x_{n+1}) ~~\forall n \in \{1, 2,\dots, N-1\},
\]
where $x_n$ is the vector containing all the bands of the $n$'th pixel, and $\xi_n$ estimates the spatial noise in pixel $n$ \cite{green_transformation_1988}.

It is important to note that the SNR estimates are dependent on lighting conditions. As highlighted in \cite{ren_noise_2014}, SNR estimation via the HA method relies on uniform surface and atmospheric conditions; deviations from this ideal inevitably introduce inaccuracies. In this work, we primarily use SNR estimates to compare HYPSO‑1 and HYPSO‑2. For comparisons between HYPSO‑1 and other hyperspectral satellites, we refer the reader to \cite{bakken_hypso1_2023}.

\subsection{Hyperspectral Instruments}
Hyperspectral instruments are central to advancements in spaceborne EO, and more than a dozen hyperspectral satellites have already been launched for EO purposes. In \cite{qian_hyperspectral_2021}, the authors provide a good overview of current and future hyperspectral satellite missions between 1996 and 2029, as of June 2021.

There are several approaches for hyperspectral instrument design, of which the three main ones are dispersive element-based, spectral filter-based, and Fourier transform-based. The dispersive element-based approach is the most common \cite{qian_hyperspectral_2021} and the one used by the HYPSO satellites. Other satellites using dispersive element-based instruments include HICO, PRISMA, HISUI, EnMAP, PACE OCI, and the planned CHIME \cite{lucke_hyperspectral_2011, cogliati_prisma_2021, matsunaga_hisui_2019, guanter_enmap_2015, gorman_nasa_2019, nieke_copernicus_2023}.

Dispersive element-based instruments use prisms or diffraction gratings (or a combination) to disperse light onto a detector, and can be implemented using either whiskbroom or pushbroom scanning. The whiskbroom method measures a single point at a time and requires moving mechanical parts, whereas pushbroom scanning measures a full spatial line at a time and can utilize the movement of the satellite itself for scanning.

Spectral filter-based designs encompass a wide range of implementations, with two prominent approaches being \gls{lvf} and \gls{etf}-based instruments. LVF systems typically operate in a pushbroom configuration, but unlike dispersive-based pushbroom imagers, they capture the entire 2D scene at each time instance, with each line recorded at a different wavelength. The satellite must therefore re-visit each line multiple times to reconstruct the full spectral cube. Examples of satellites using the LVF design include the Zhuhai-1 constellation, HyperScout-1, and IMS-1 \cite{zhang_onorbit_2022, esposito_inorbit_2019, kumar_hyperspectral_2015}. ETF-based designs, in contrast, acquire the entire scene at a single wavelength per exposure and rely on sequential tuning to build up the spectral dimension, giving a trade-off between SNR and spectral resolution. An example of a constellation using this concept is the Hyperfield constellation \cite{tikka_hyperfield_2023}.

Fourier transform-based instruments are not as common in EO and will not be detailed here. A more in-depth overview of hyperspectral imaging instruments is provided by \cite{qin_chapter_2010}.

\subsection{The HYPSO Constellation}
At present, the HYPSO constellation consists of two 6U CubeSats equipped with hyperspectral imaging payloads of identical optical design. The instruments were built using \gls{cots} components, and utilize transmission gratings for spectral dispersion, with a nominal operating spectral range of 400--800\,nm, limited by low signal below 400 nm and second order diffraction effects above 800 nm \cite{henriksen_prelaunch_2022}. Within this interval, the 1080 spectral pixels are binned by a factor of nine, yielding 120 effective spectral bands, and 1092 of the 1216 available spatial pixels are used for nominal captures. Unless otherwise stated, all captures in this paper are taken in nominal mode with $9\times$ binning applied. Further detail about the instruments and their pre-launch calibration can be found in \cite{grotte_ocean_2022, henriksen_prelaunch_2022}.

Despite the identical optical design, small differences in assembly and laboratory calibration, together with different orbits and time in orbit (HYPSO-1 launched in early 2022 and HYPSO-2 in mid 2024, both in sun-synchronous polar orbits with different local time of descending node), introduce inter-sensor variations. Additionally, incremental upgrades to the satellite platform lead to operational differences among the satellites. HYPSO-2 is equipped with improved data transfer capabilities, both onboard and from space to the ground. The satellite is equipped with deployable solar panels to accommodate the expected increase in payload duty cycle resulting from higher data throughput. Changes in the mechanical mounting of an attitude sensor also allow for more favorable viewing angles during imaging for HYPSO-2 compared to HYPSO-1. A joint calibration and harmonization effort is thus required to ensure the two satellites’ data remain interpretable and interchangeable.

\section{Cal/Val Experiments}
In this section, we present details of the different types of captures for in-orbit calibration. Each type is analyzed and discussed in the following.

\subsection{Lunar Captures} \label{sec:lunarcaptures}
\begin{figure}[htbp]
    \centering
    \includegraphics[width=\linewidth]{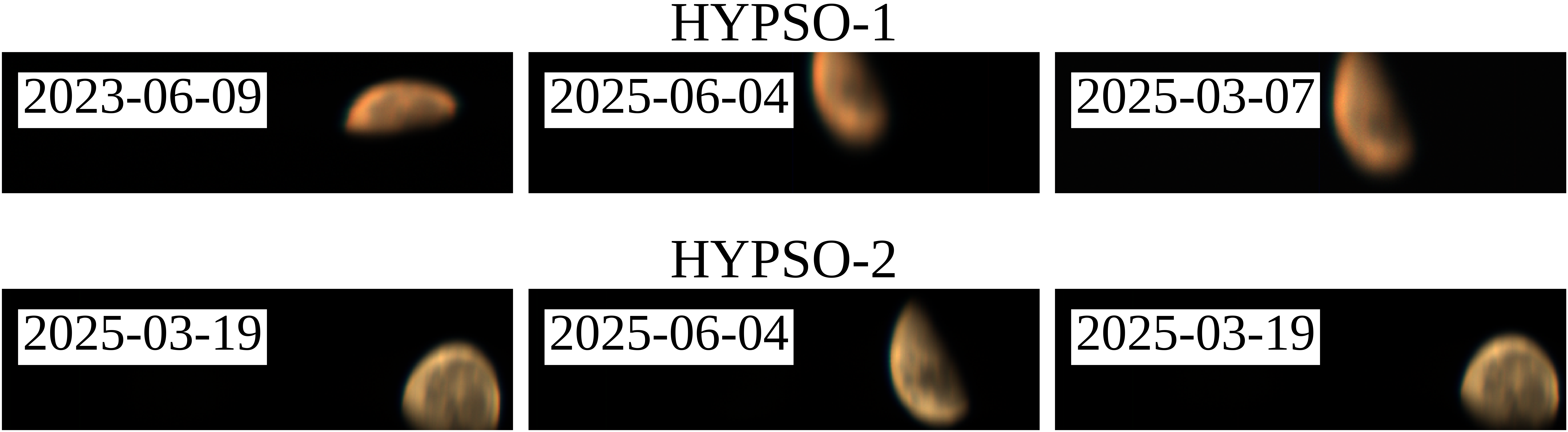}
    \caption{Moon captures by HYPSO-1 and HYPSO-2, used for spectral calibration, assessment of post-launch misalignment, estimation of spectral resolution, radiometric calibration, and artifact analysis.}
    \label{fig:moon}
\end{figure}

At time of writing HYPSO-1 has taken 145 Moon captures, while HYPSO-2 has taken 70; of these 6 are selected and analyzed in this paper, see Figure~\ref{fig:moon}. This selection is based on captures that have an absolute Moon phase angle between $2^\circ$ and $90^\circ$, and where the Moon is sufficiently within frame. The Moon phase angle is restricted since LIME only gives a reliable simulation within this range \cite{toledano_lime_2024}. Moreover, captures with similar phase angle are preferred as it reduces uncertainty~\cite{stone_moon_2009}.

These captures are used for spectral calibration, assessment of post-launch misalignment, estimation of spectral resolution, radiometric calibration, and investigation of artifacts. Furthermore, it is important to note that spectral binning is not applied to the captures.

\subsection{PICS and RadCalNet}
\begin{figure}[htbp]
    \centering
    \includegraphics[width=1\linewidth]{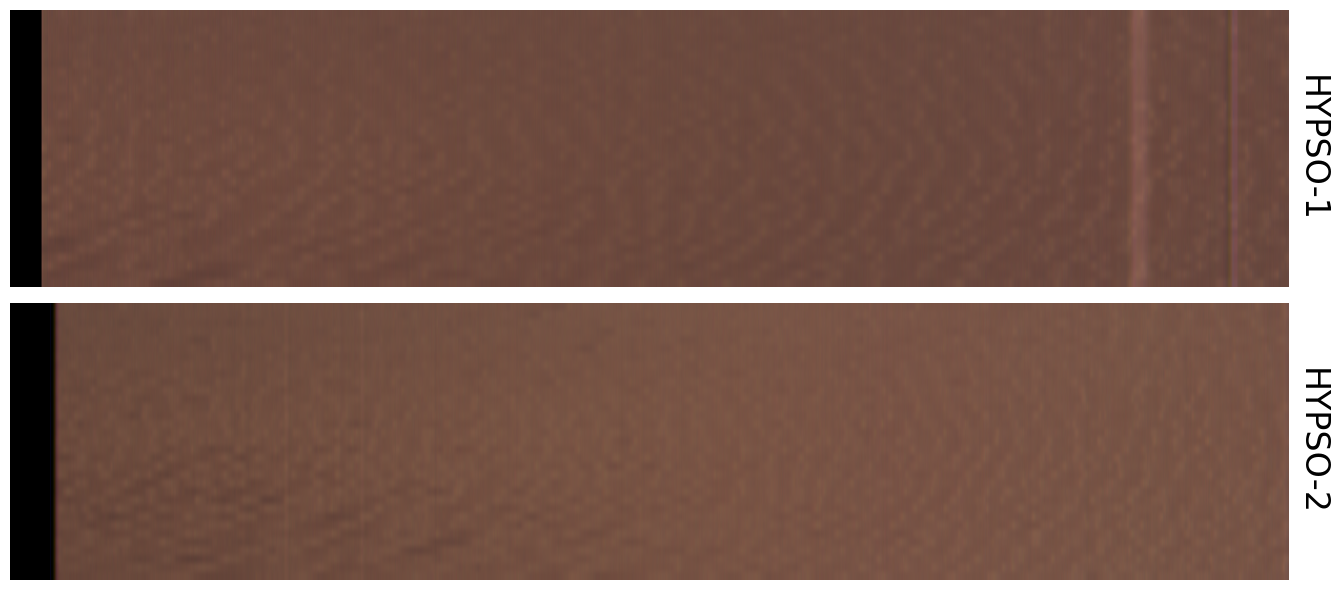}
    \caption{Libya-4 captures held out for validation of the smile correction, acquired by HYPSO-1 (2025-05-06, 08:42) and HYPSO-2 (2025-09-12, 09:10). The artifact visible in the HYPSO-1 capture falls outside the nominal capture area.}
    \label{fig:libya_fullframe}
\end{figure}

Smile is characterized using the Libya-4 PICS \cite{_libya_}. Libya-4 has been found to be the most stable of all PICS \cite{tuli_new_2019}, and it is well-suited to this task because it is bright and homogeneous, and, as opposed to the Moon, covers the whole frame. Four captures were acquired by each satellite using the full swath width of the instrument and no binning, see Figure~\ref{fig:libya_fullframe}. For each satellite, three captures are used to characterize the smile and the fourth is held out to validate the correction. As the solar Fraunhofer lines are clearly visible in these captures, all four are also used to validate the in-orbit calibrated spectral coefficients. In addition, full-frame Libya-4 captures acquired on the same day as a PACE overpass are used to harmonize the radiometric magnitude of the two satellites (Section~\ref{sec:magnitude}). Of the 8 such captures by HYPSO-1 and 10 by HYPSO-2, 7 and 5 remain after screening for viewing angle, homogeneity over the site, and saturation.

For validation of the spectral and radiometric coefficients after calibration, the Gobabeb \cite{_gobabeb_} and La Crau \cite{_crau_} RadCalNet sites are used. These are chosen because RadCalNet actively provides SI-traceable \gls{toa} reflectances for these sites, increasing the likelihood of good matchups between satellite and ground measurements. The \gls{toa} reflectance is defined as
\begin{equation}
    R = \pi \frac{\left(\frac{r}{r_{\mathrm{o}}}\right)^2 L}{F_{\mathrm{o}} \cos \theta_{\mathrm{s}} } \; \text{[a.u.]},
\end{equation}
where $L$ is the radiance, $F_{\mathrm{o}}$ is the solar irradiance at mean Earth--Sun distance, $r_\mathrm{o}$ is the mean Earth--Sun distance, $r$ is the current Earth--Sun distance, and $\theta_\mathrm{s}$ is the solar zenith angle \cite{OceanOpticsBook2022}. Data from HYPERNETS was also considered, but was not publicly available at the time of writing \cite{ruddick_hypernets_2024}.

\subsection{Bridge Target}
For GRD assessment, terrestrial targets are preferable since the GRD depends on the distance to the target. For the HYPSO satellites, captures of two parallel bridges over Lake Pontchartrain in Louisiana are used for this purpose. The target is well suited because the bridges are bright against the dark lake surface, and their combined width, including the gap between them, is approximately 50 m, as estimated from satellite imagery. Since this is smaller than the GSD of both HYPSO satellites, the bridges subtend less than one pixel and the response resembles that of a line target, which is the condition required for LSF estimation. The GSD of HYPSO-1 and HYPSO-2, calculated using Equation~(\ref{eq:gsd}) with $d$ taken as the viewing path to target $h/\cos\theta$ to account for the off-nadir angle $\theta$, are $GSD_\text{h1} = 54$ m/pixel and $GSD_\text{h2} = 70$ m/pixel, respectively.

\begin{figure}[htbp]
    \centering
    \includegraphics[width=1\linewidth]{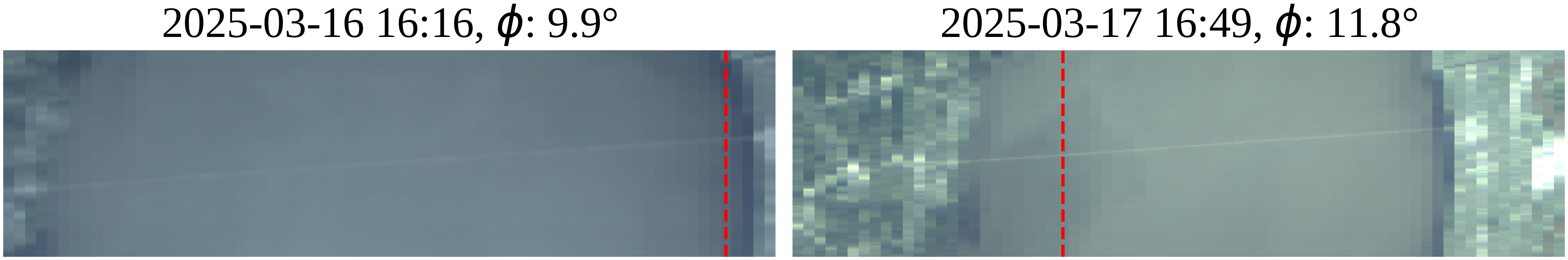}
    \caption{Cut-outs from Lake Pontchartrain bridge captures by HYPSO-1 (left) and HYPSO-2 (right). The red lines indicate the across-track frame in which we investigate the GRD of the satellites. The along-track direction is horizontal in these images.}
    \label{fig:grd_bridge}
\end{figure}

Figure \ref{fig:grd_bridge} shows the captures of the Lake Pontchartrain bridge used for across-track GRD estimation, with HYPSO-1 to the left and HYPSO-2 to the right. These captures were selected to make the acquisition conditions as comparable as possible between the two satellites: they have similar off-nadir viewing angles of $9.9^\circ$ and $11.8^\circ$, and were taken one day apart at a comparable time of day, so that lake and atmospheric conditions are as similar as achievable. The red lines indicate the across-track frame in which we investigate the GRD of the satellites.

\subsection{Uniform Ocean Targets}
\begin{figure}[htbp]
    \centering
    \includegraphics[width=1\linewidth]{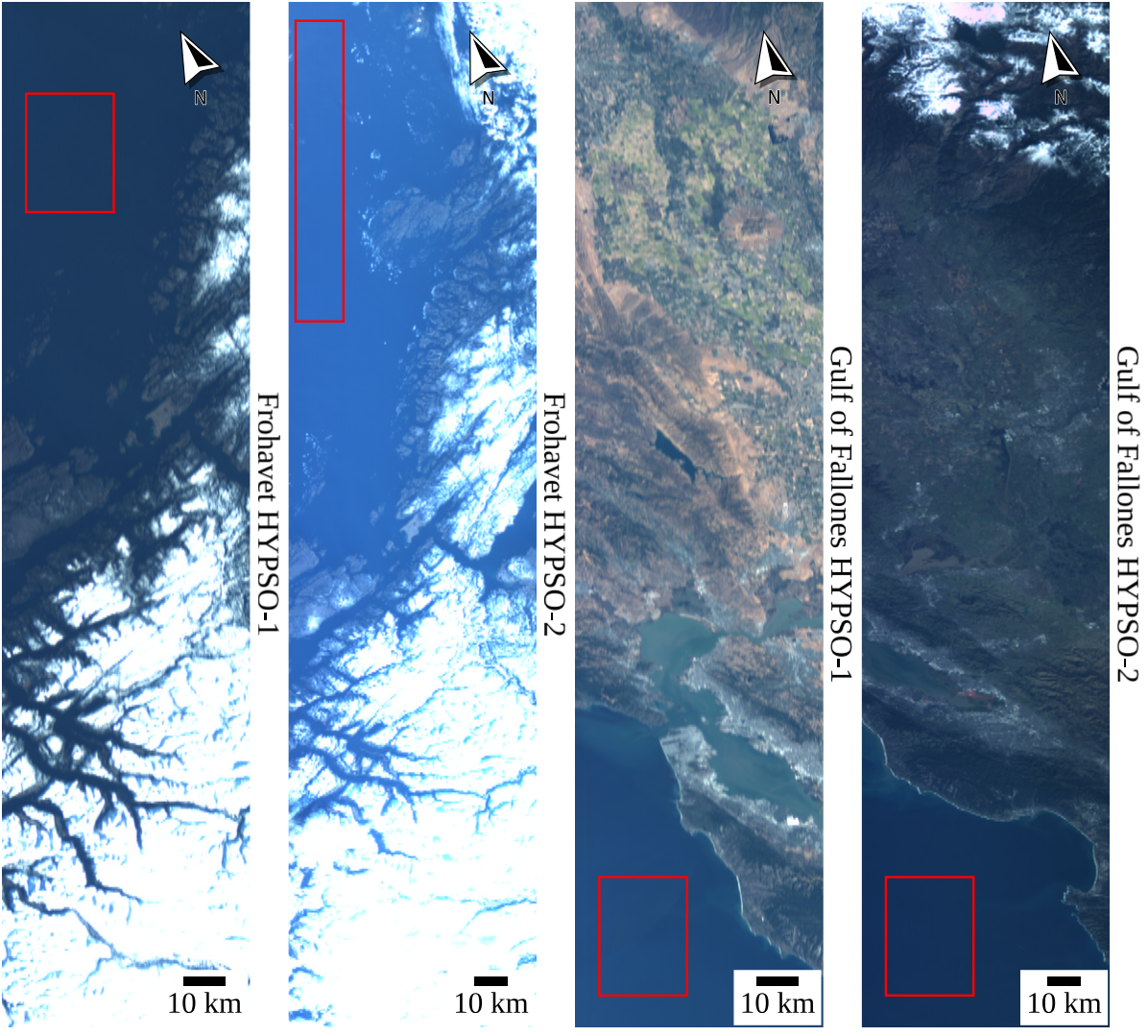}
    \caption{RGB renders of Frohavet and the Gulf of Farallones, captured by HYPSO-1 and HYPSO-2. The red boxes highlight the regions used for SNR analysis. The brightness of the renders has been increased for visualization purposes.}
    \label{fig:snr_targets}
\end{figure}

The main objective of the HYPSO constellation is ocean monitoring; we are therefore most interested in the SNR of ocean targets, and investigate the SNR using uniform dark ocean patches. The two targets used are Frohavet outside Trondheim and the Gulf of Farallones outside San Francisco. The Frohavet captures were acquired by HYPSO-1 (2024-03-05, 10:25) and HYPSO-2 (2025-02-25, 11:26) with off-nadir viewing angles of 17.2$^\circ$ and 25.4$^\circ$, respectively. The Gulf of Farallones captures were acquired by HYPSO-1 (2024-09-30, 18:25) and HYPSO-2 (2025-01-08, 19:13) with off-nadir viewing angles of 12.6$^\circ$ and 10.9$^\circ$, respectively.\footnote{These captures were chosen because of the lack of clouds over the target area.} Figure \ref{fig:snr_targets} shows RGB renders of these sites, where red boxes highlight the regions used for SNR analysis.

\section{Results}
The procedures are applied to HYPSO-1 and HYPSO-2 in the order given in Section~\ref{sec:inorbit_char} and Table~\ref{tab:cal_dependencies}.

\subsection{Smile Correction}
\begin{figure}[htbp]
    \centering
    \includegraphics[width=1\linewidth]{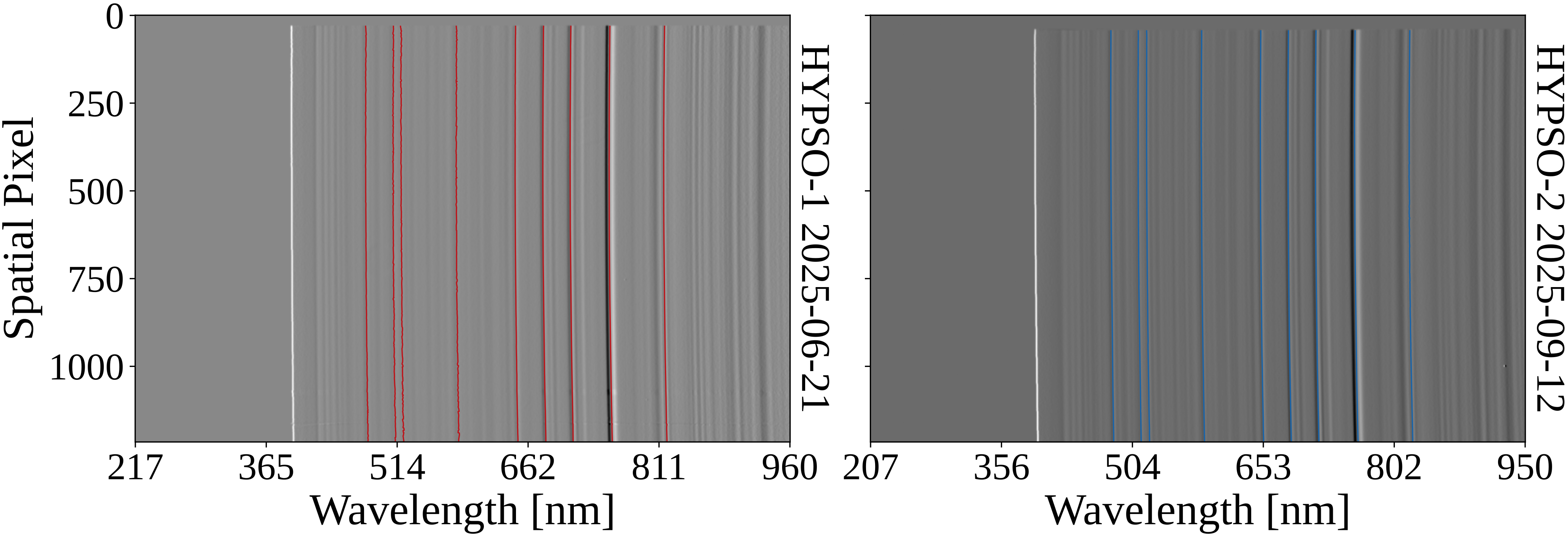}
    \caption{Sobel edges of a Libya-4 capture for HYPSO-1 (left) and HYPSO-2 (right), with the tracked center of each absorption feature along the slit overlaid in red and blue, respectively. The curvature of the tracked lines shows the smile.}
    \label{fig:smile_sobel_method}
\end{figure}

\begin{figure}[htbp]
    \centering
    \includegraphics[width=1\linewidth]{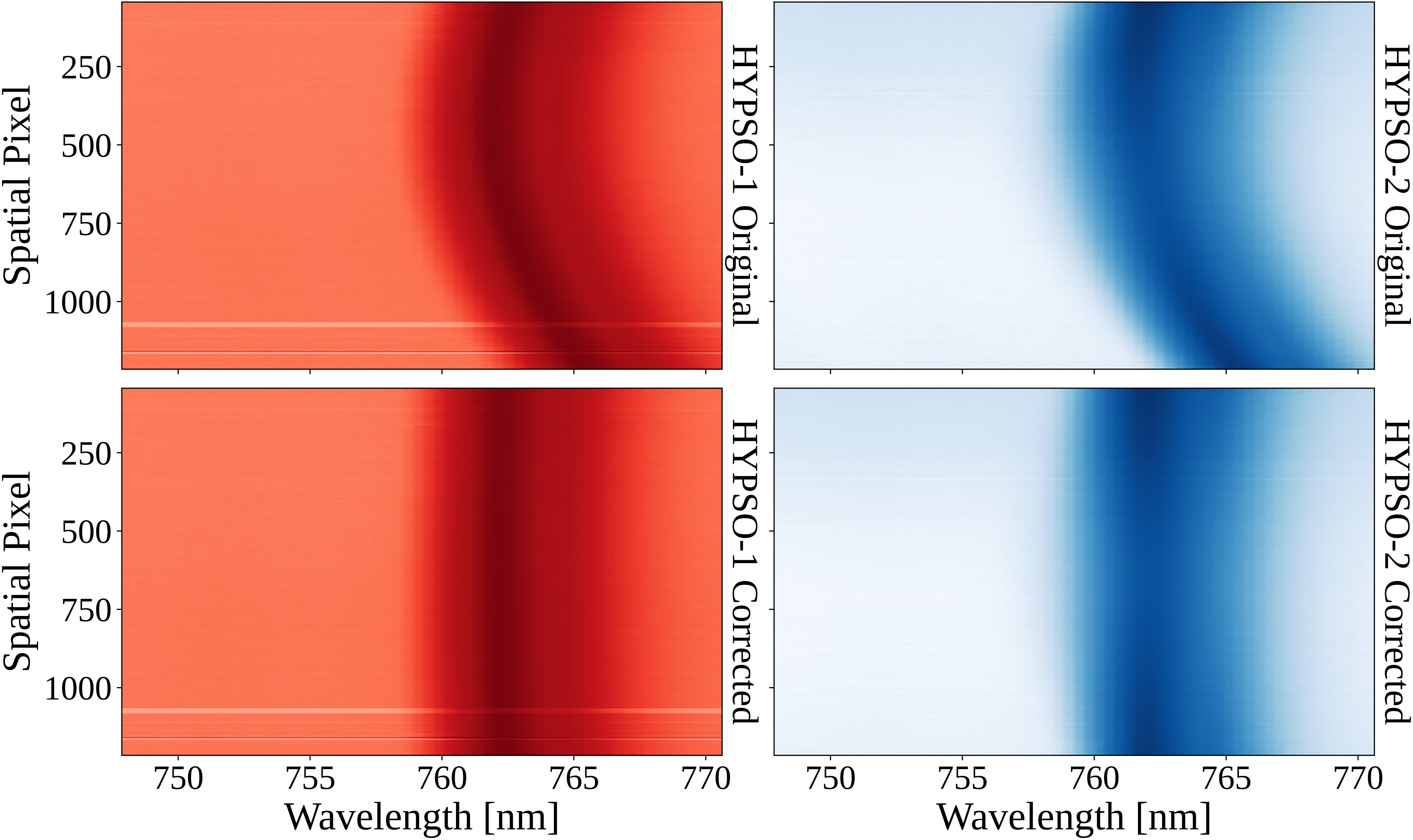}
    \caption{Zoom of the O$_2$ A-band in the validation capture of HYPSO-1 (left) and HYPSO-2 (right), before (top) and after (bottom) smile correction. The curved absorption feature becomes straight after correction.}
    \label{fig:smile_corrected}
\end{figure}

Using three Libya-4 captures for each satellite, calibrated with the laboratory radiometric coefficients, the average of all along-track pixels is computed, giving the average frame. A horizontal Sobel filter followed by a Gaussian blur is then applied to this frame, highlighting horizontal spectral features. The algorithm then looks at a window around where each absorption dip is expected at the center of the swath width, using the laboratory wavelengths only to place this window. The center of the dip is taken as the midpoint between the minimum and maximum of the Sobel response, corresponding to the two flanks of the dip, each refined to sub-pixel accuracy by fitting a second order polynomial to a three-pixel window. A window around this location is then used to look for the same absorption dip in the neighboring pixels, while disregarding saturated pixels. Following this procedure in both directions along the swath width then gives the location of the absorption dip across the swath width, see Figure~\ref{fig:smile_sobel_method}. From here we follow the exact same method as was used for lab calibration \cite{henriksen_prelaunch_2022}. Ground control points are created, and a second order polynomial is fit to these points to get a smooth quantification of the smile effect. Ground control points further than 1 pixel from the fitted model are removed, and the model is fitted again. After quantifying the smile effect, a fourth Libya-4 capture is used for each satellite to validate the correction. Figure~\ref{fig:smile_corrected} shows a zoom of the O$_2$ A-band before and after smile correction, which visualizes the correction, while Table~\ref{tab:smile_validation} presents the quantitative results. The largest remaining smile is 2.0 pixels (0.8 nm) for HYPSO-1 and 1.2 pixels (0.5 nm) for HYPSO-2, down from 9.3 and 10.0 pixels, and is below one pixel for most features.

\begin{table}[htbp]
\centering
\caption{Maximum smile of each absorption feature along the slit in the validation capture of HYPSO-1 and HYPSO-2, before and after smile correction, in pixels unless otherwise stated.}
\label{tab:smile_validation}
\setlength{\tabcolsep}{4pt}
\begin{tabular}{@{}lcccc@{}}
\toprule
& \multicolumn{2}{c}{HYPSO-1} & \multicolumn{2}{c}{HYPSO-2}\\
Feature, $\lambda$ [nm] & Before & After & Before & After\\
\midrule
F (H$\beta$), 486 & 6.4 & 1.4 & 7.4 & 1.2 \\
b2 (Mg), 517 & 7.0 & 0.8 & 8.3 & 0.4 \\
E2 (Fe), 527 & 6.2 & 2.0 & 7.7 & 1.0 \\
D (Na), 589 & 7.2 & 1.0 & 8.6 & 0.1 \\
C (H$\alpha$), 656 & 7.8 & 0.4 & 8.7 & 0.0 \\
B (O$_2$), 686 & 8.7 & 0.6 & 8.7 & 0.0 \\
H$_2$O, 718 & 8.3 & 0.3 & 9.5 & 0.8 \\
A (O$_2$), 759 & 8.8 & 0.7 & 9.4 & 0.7 \\
Z (O$_2$), 822 & 9.3 & 1.1 & 10.0 & 1.2 \\
\midrule
\textbf{Max offset} & \textbf{9.3} & \textbf{2.0} & \textbf{10.0} & \textbf{1.2} \\
\textbf{Max offset [nm]} & \textbf{3.5} & \textbf{0.8} & \textbf{3.8} & \textbf{0.5} \\
\bottomrule
\end{tabular}
\end{table}


\begin{figure*}[t!]
    \centering
    \begin{subfigure}[b]{0.49\textwidth}
        \centering
        \includegraphics[width=\linewidth]{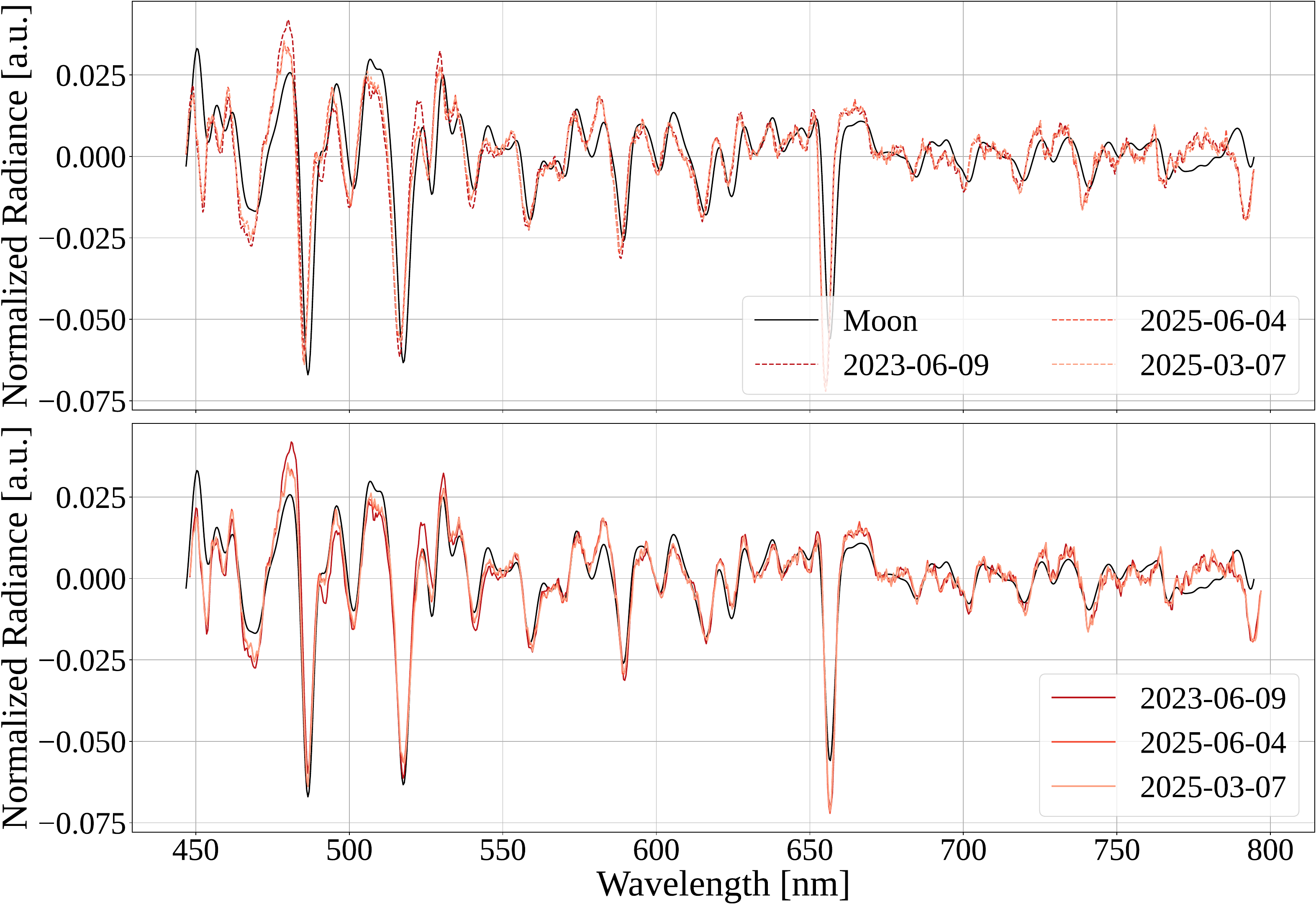}
        \caption{HYPSO-1 compared to LIME.}
        \label{fig:hypso_moon_spectra_h1}
    \end{subfigure}
    \hfill
    \begin{subfigure}[b]{0.49\textwidth}
        \centering
        \includegraphics[width=\linewidth]{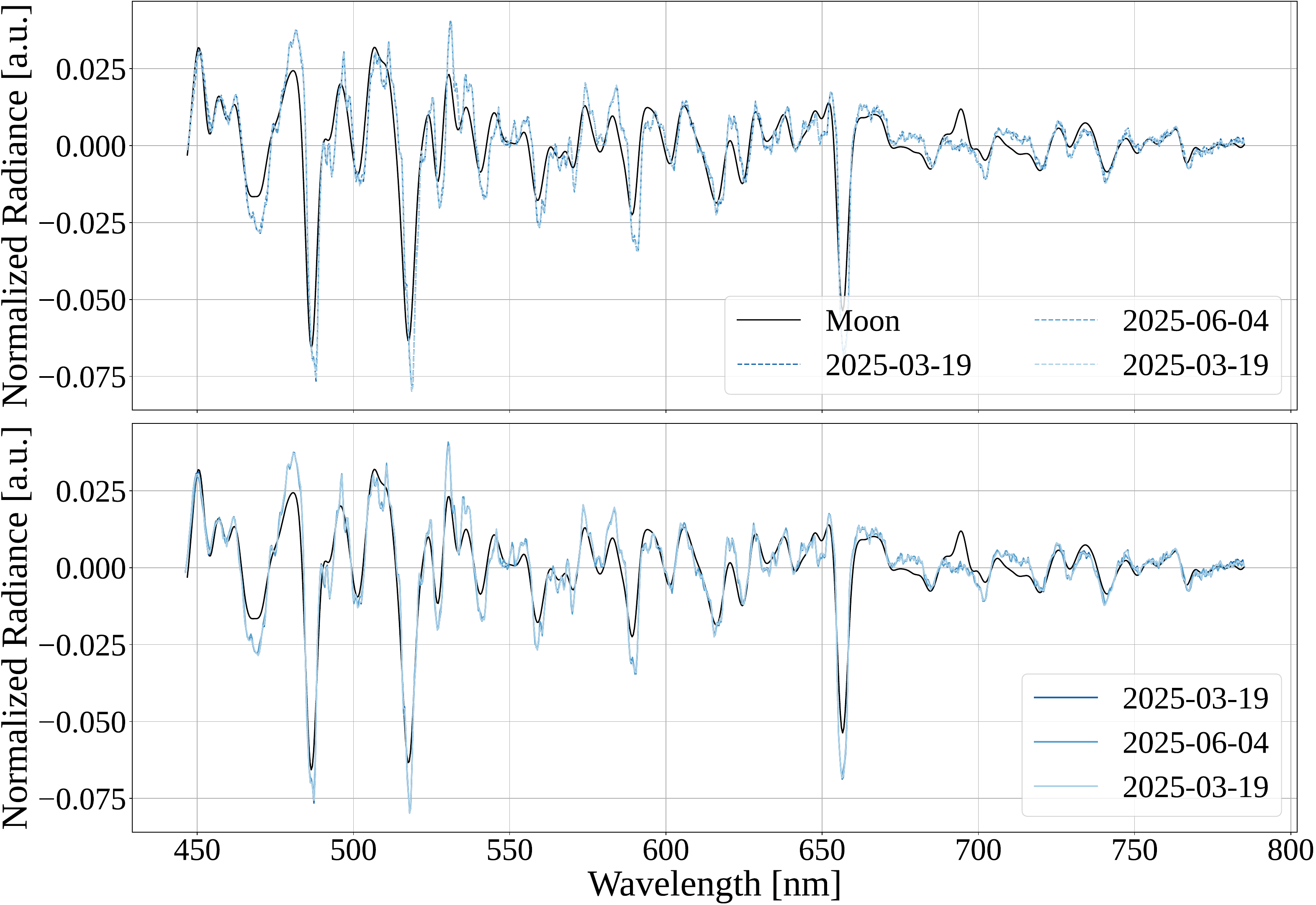}
        \caption{HYPSO-2 compared to LIME.}
        \label{fig:hypso_moon_spectra_h2}
    \end{subfigure}
    \caption{Band-pass filtered Moon spectra of HYPSO-1, HYPSO-2, and LIME. The dotted and solid lines show the spectra before and after spectral calibration, respectively.}
    \label{fig:hypso_moon_spectra}
\end{figure*}

\subsection{Spectral Calibration} \label{sec:spectral}
To start with, the measured positions of the most prominent Fraunhofer lines in the Lunar spectra are compared to their expected wavelengths. For the HYPSO satellites, the absorption lines $F$, $b2$, $D$, and $C$ were examined because they exhibited sufficiently distinct dips.\footnote{The $D_1$ and $D_2$ lines are analyzed as a single dip, $D$, due to their close proximity.} This analysis confirmed a shift in the spectral axis; however, a few individual lines are too few reference points to determine how the spectral axis is distorted. Consequently, a full-spectrum comparison approach was taken for calibration, while the individual lines are used for validation afterwards (Section~\ref{sec:spectral_validation}).

To exploit information across the full wavelength axis, the measured Lunar spectra, smile corrected using the coefficients from the previous section, were compared to a reference spectrum generated by the LIME model. Since the spectral resolution is estimated after the spectral calibration, the LIME spectrum is convolved to an assumed HYPSO resolution of 4 nm for this step, informed by the pre-launch lab characterization of HYPSO-1 \cite{henriksen_prelaunch_2022}. Prior to comparison, both the measured and reference spectra were normalized and band‑pass filtered to emphasize higher‑frequency spectral features while suppressing noise. We denote the reference spectrum as $y_\text{moon}$, its normalized form as $\hat{y}_\text{moon}$, and its band‑pass filtered form as $\hat{y}_\text{moon\_bp}$. Likewise, we denote the measured Lunar spectrum as $x_\text{moon}$, its normalized form as $\hat{x}_\text{moon}$, and its band‑pass filtered form as $\hat{x}_\text{moon\_bp}$. Following this, we calculate the polynomial transformation of the wavelength axis that gives the most consistent alignment between the measured spectra and reference spectra. To favor parsimonious corrections, we fit the polynomial hierarchically: starting with degree 0 and incrementally increasing the degree, each time warm-starting the solver with the parameters from the lower‑order fit, thereby biasing the solution toward lower‑order terms whenever possible.

\begin{figure}[htbp]
    \centering
    \includegraphics[width=\linewidth]{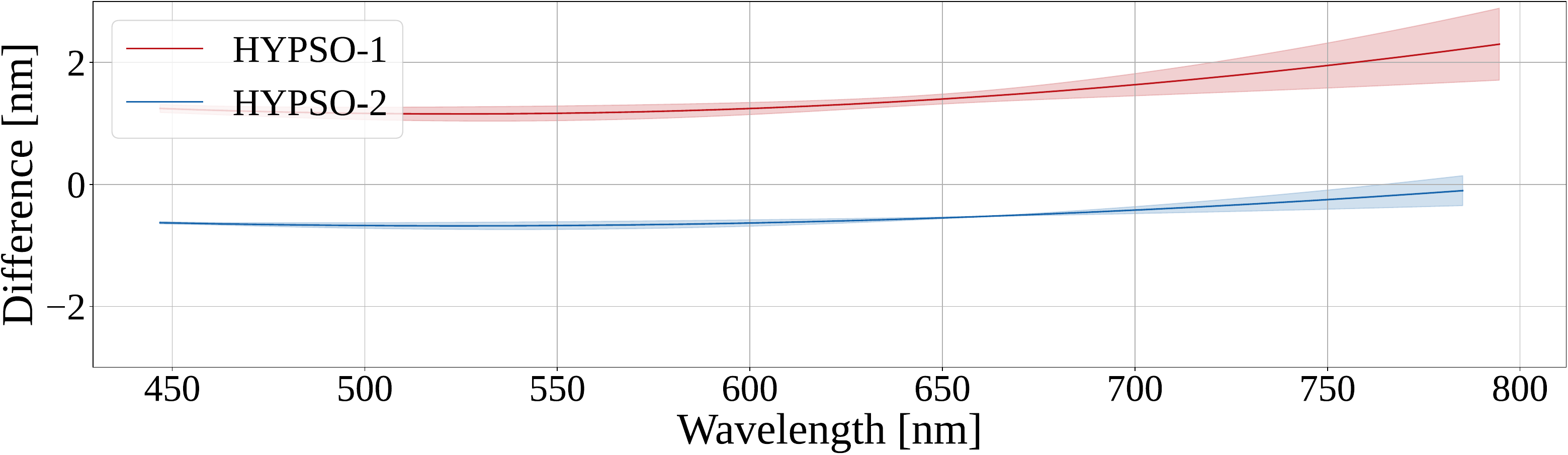}
    \caption{Difference between the spectral axis before and after the fit against the LIME Lunar spectrum, for HYPSO-1 and HYPSO-2.}
    \label{fig:wavelength_diff}
\end{figure}

For the HYPSO satellites a maximum polynomial order of 2 is used, while a fifth order zero-phase digital Butterworth filter is used for band-pass filtering, using cut-off values that pass spectra between 1 nm and 50 nm. Figure \ref{fig:wavelength_diff} shows the wavelength difference between the original and polynomial transformed spectral axis, for HYPSO-1 and HYPSO-2.

Here we see primarily a static offset across the wavelengths, with only a slight bend, which likely indicates a horizontal movement of the slit and a slight tilt, respectively. Including the bend improves the agreement with the Fraunhofer lines in the independent validation (Section~\ref{sec:spectral_validation}), and the second order polynomial, averaged over the three captures of each satellite, is therefore used to adjust the wavelength axis. The fit covers 450--790 nm, which is almost the entire range used for nominal captures, with the most uncertain bands between 400--450\,nm removed. Outside this range the polynomial is extrapolated.

Figures~\ref{fig:hypso_moon_spectra_h1} and~\ref{fig:hypso_moon_spectra_h2} show $\hat{x}_\text{moon\_bp}$ and $\hat{y}_\text{moon\_bp}$ before and after spectral calibration, for HYPSO-1 and HYPSO-2 respectively. Here the dotted lines show the spectra before calibration, while the solid lines show the spectra after calibration.\footnote{The analysis is done using a select region of the sensor to avoid the more noisy outer edges of the spectrum.}

\subsubsection{Assessment on Fraunhofer Lines in Libya-4 Captures} \label{sec:spectral_validation}

\begin{figure}[htbp]
    \centering
    \includegraphics[width=\linewidth]{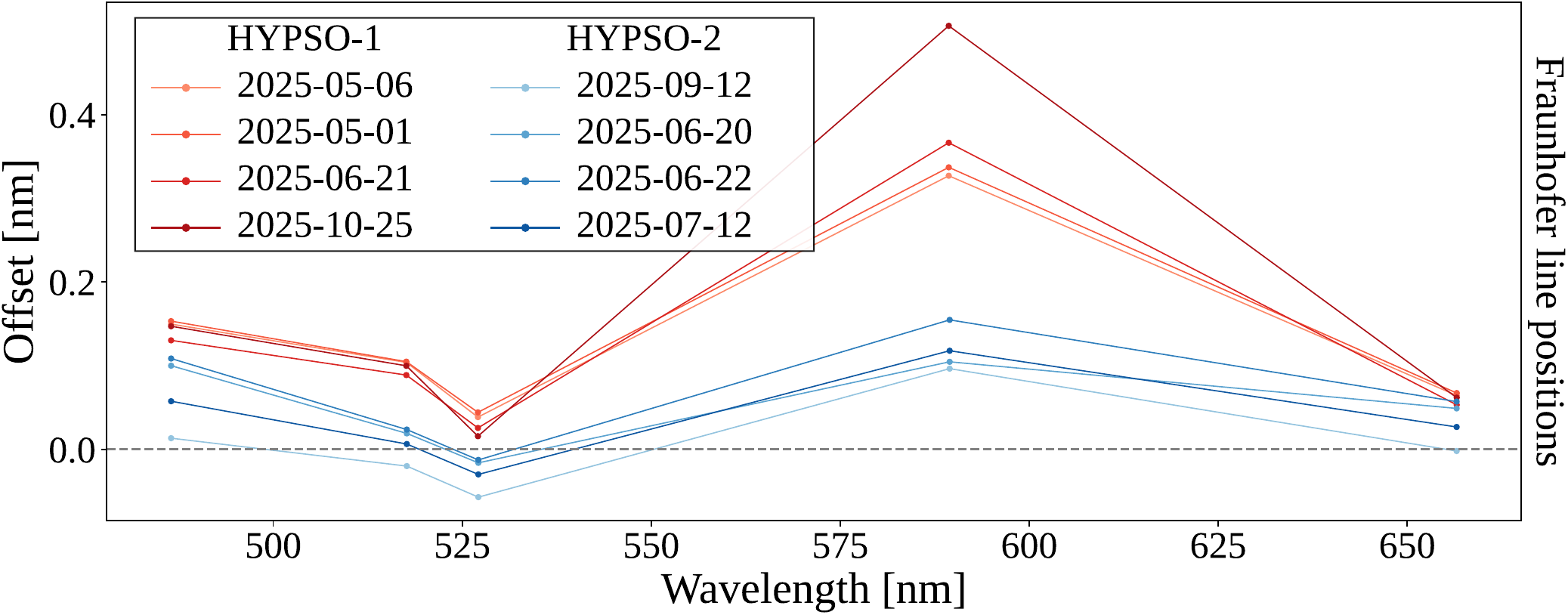}
    \caption{Spectral calibration error (measured $-$ expected wavelength) at the Fraunhofer lines in each smile corrected Libya-4 capture of HYPSO-1 and HYPSO-2, using the in-orbit spectral coefficients. The dashed line marks zero error.}
    \label{fig:spectral_validation}
\end{figure}

Spectral calibration was also verified at Fraunhofer lines $F$, $b2$, $E2$, $D$, and $C$ (486–656 nm). Table~\ref{tab:spectral_validation} shows the measured wavelength of these lines in the four smile corrected Libya-4 captures of HYPSO-1 and HYPSO-2 compared to their expected wavelengths, while Figure~\ref{fig:spectral_validation} visually shows the deviation for each line.

The results show that the mean deviation is +0.14 $\pm$ 0.13 nm for HYPSO-1 and +0.04 $\pm$ 0.06 nm for HYPSO-2, indicating sub-pixel accuracy for both satellites. It is also noteworthy that both satellites show the largest deviation for $D$ (Na), likely because $D_1$ and $D_2$ are blended at HYPSO resolution, making the measured center sensitive to the assumed spectral response.

\begin{table}[htbp]
\centering
\caption{Spectral calibration error (measured $-$ expected wavelength) of HYPSO-1 and HYPSO-2 at Fraunhofer lines in smile corrected full-frame Libya-4 captures, given as mean $\pm$ standard deviation over the four captures of each satellite.}
\label{tab:spectral_validation}
\begin{tabular}{@{}lll@{}}
\toprule
& \multicolumn{2}{c}{Error [nm]}\\
Line, $\lambda$ [nm] & HYPSO-1 & HYPSO-2\\
\midrule
F (H$\beta$), 486.1 & +0.15 $\pm$ 0.01 & +0.07 $\pm$ 0.04 \\
b2 (Mg), 517.3 & +0.10 $\pm$ 0.01 & +0.01 $\pm$ 0.02 \\
E2 (Fe), 527.0 & +0.03 $\pm$ 0.01 & $-$0.03 $\pm$ 0.02 \\
D (Na), 589.3 & +0.38 $\pm$ 0.07 & +0.12 $\pm$ 0.02 \\
C (H$\alpha$), 656.3 & +0.06 $\pm$ 0.01 & +0.03 $\pm$ 0.02 \\
\midrule
\textbf{All lines} & \textbf{+0.14 $\pm$ 0.13} & \textbf{+0.04 $\pm$ 0.06} \\
\bottomrule
\end{tabular}
\end{table}

\begin{figure*}[t!]
    \centering
    \begin{subfigure}[b]{0.49\textwidth}
        \centering
        \includegraphics[width=\linewidth]{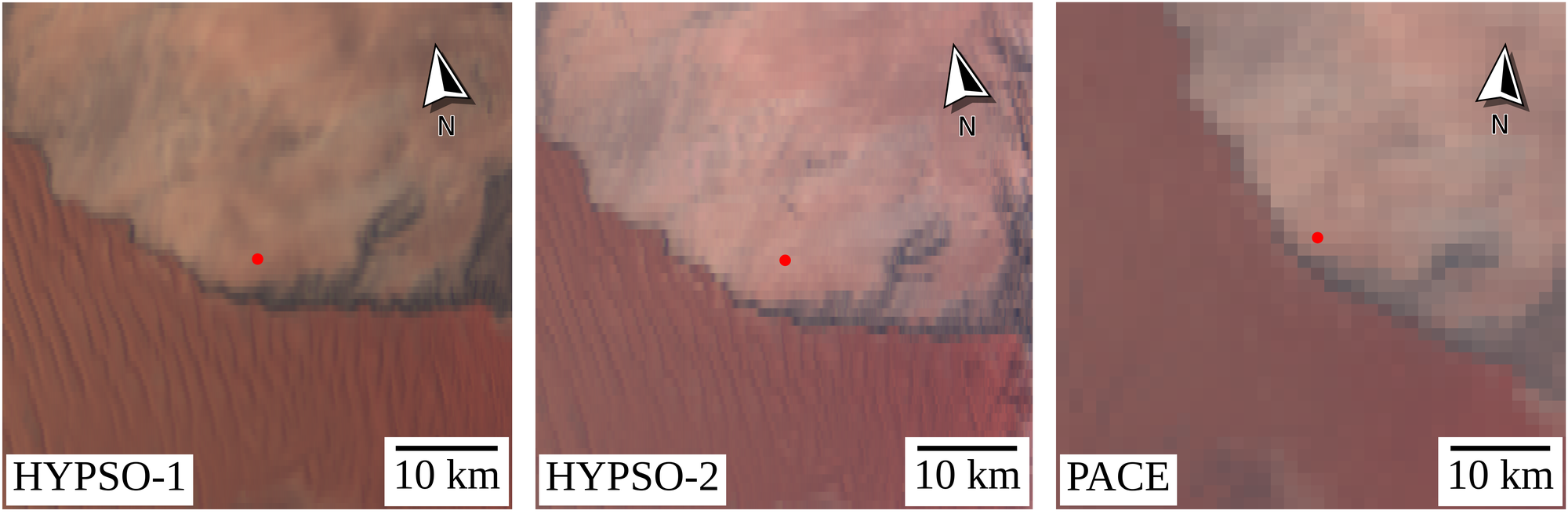}\\
        \includegraphics[width=\linewidth]{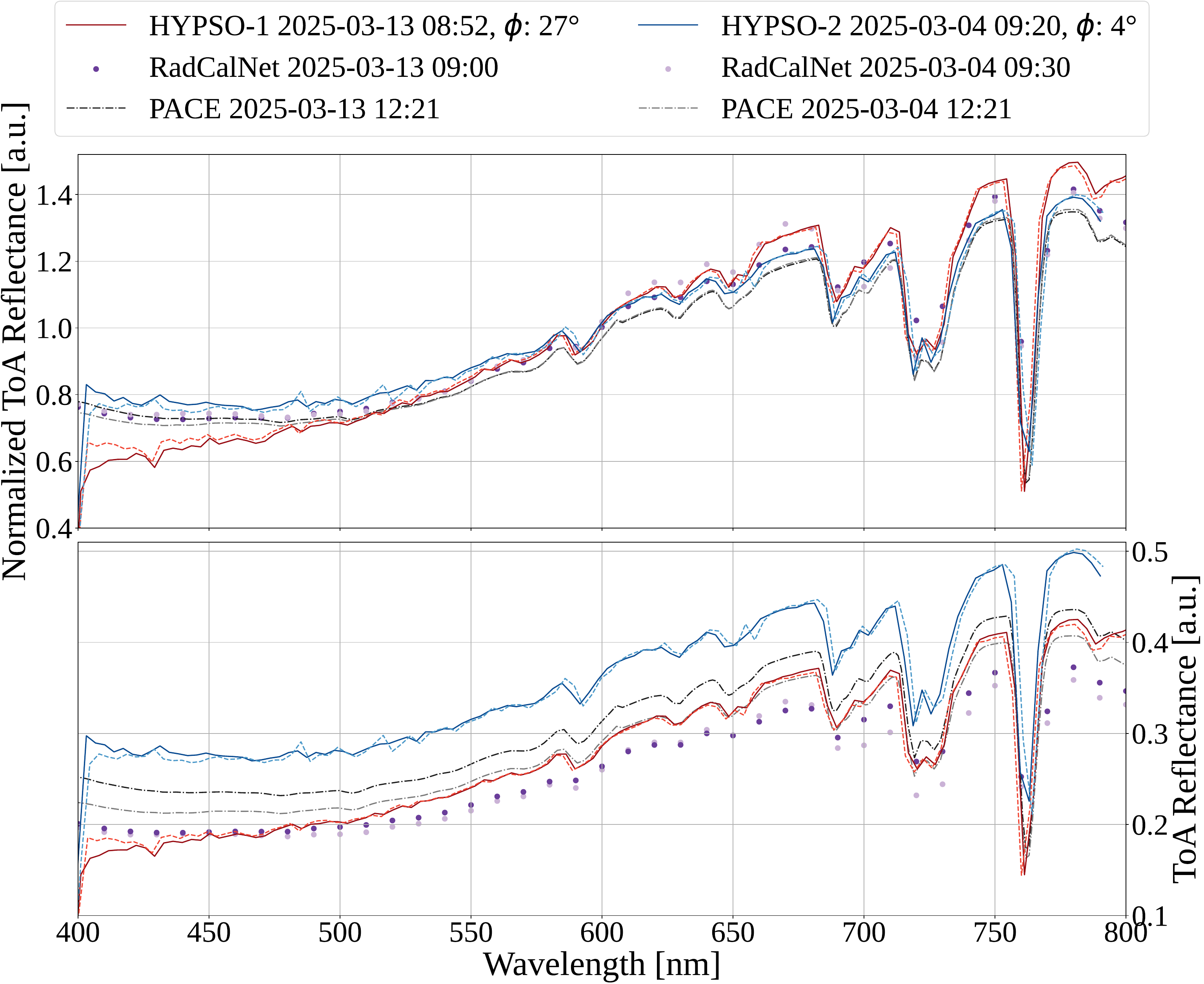}
        \caption{Gobabeb.}
        \label{fig:move_gobabeb}
    \end{subfigure}
    \hfill
    \begin{subfigure}[b]{0.49\textwidth}
        \centering
        \includegraphics[width=\linewidth]{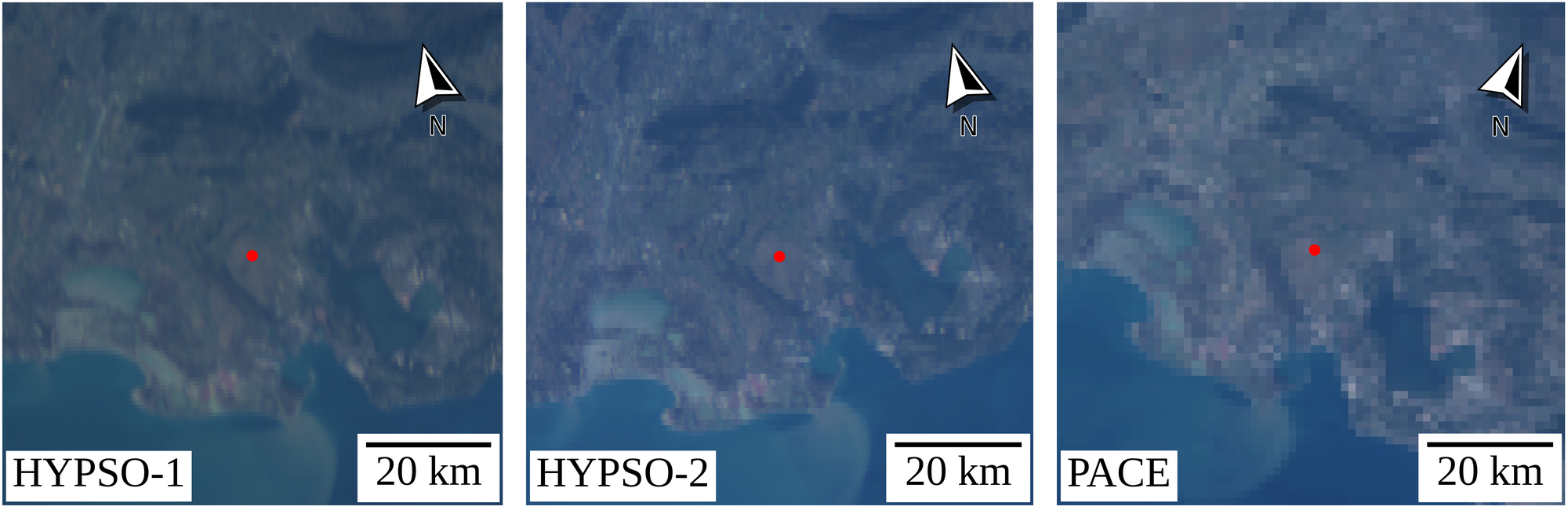}\\
        \includegraphics[width=\linewidth]{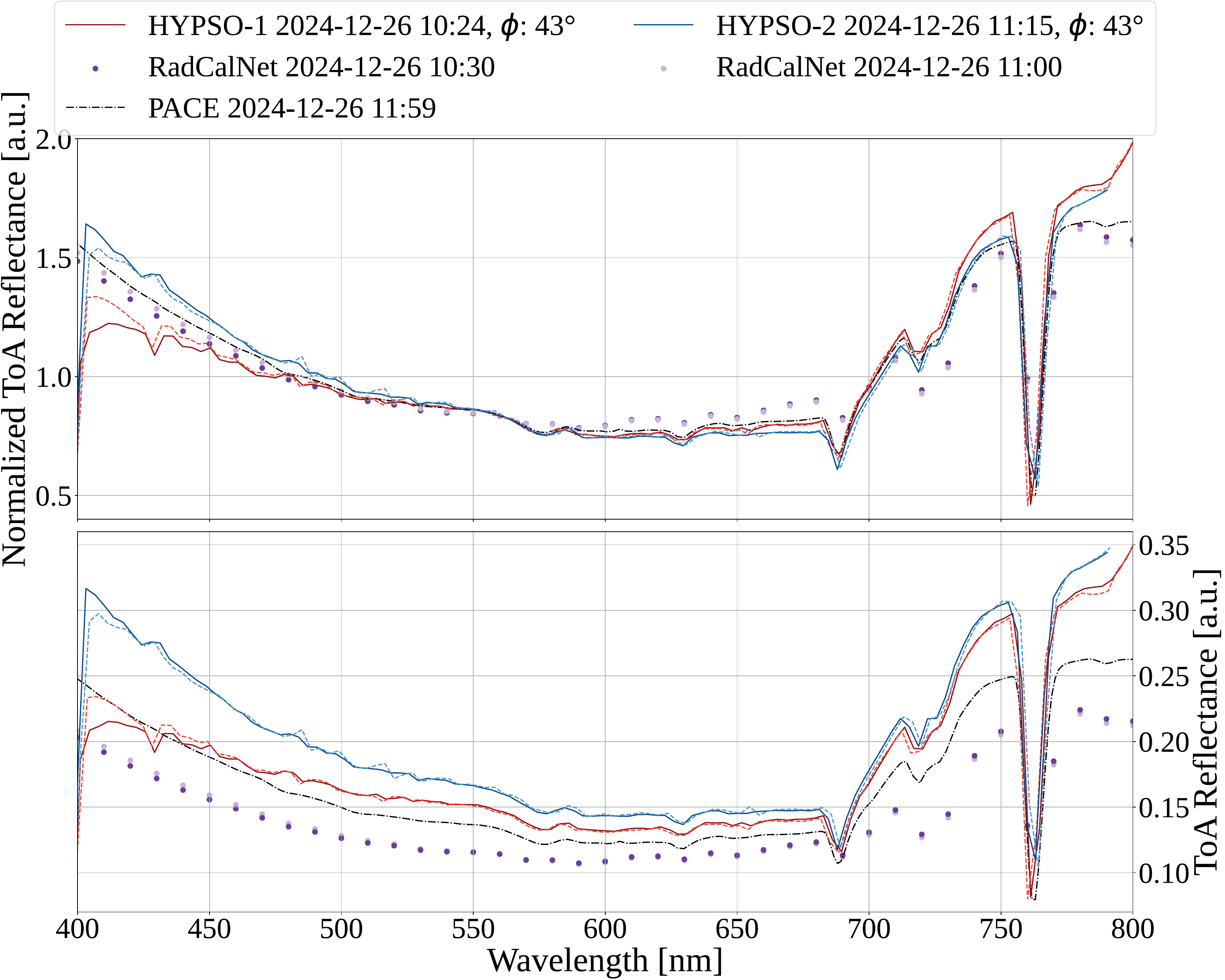}
        \caption{La Crau.}
        \label{fig:move_lacrau}
    \end{subfigure}
    \caption{ToA reflectance for HYPSO-1, HYPSO-2, PACE, and RadCalNet measurements over Gobabeb and La Crau, with RGB cut-outs of the sites where the red dot marks the location of the spectra. The dotted and solid lines show the spectra before and after re-alignment of radiometric coefficients and spectral calibration, respectively.}
    \label{fig:moved_radiometric}
\end{figure*}

\subsection{Correction of Misalignment} \label{sec:move}
In addition to serving as a method for adjusting the spectral coefficients, the spectral shift identified in Section~\ref{sec:spectral} also provides a way of quantifying potential horizontal movements resulting from launch. Such a shift may result from movement of either the slit or the camera sensor relative to the optical train. Since radiometric coefficients are assigned on a per-pixel basis, this movement might leave them misaligned with respect to their corresponding pixels. The measured shift in wavelength can therefore be translated to pixels and used to re-align the radiometric coefficients. For the HYPSO satellites, the spectral shift corresponds to a movement of 3 pixels toward longer wavelengths for HYPSO-1, and 2 pixels toward shorter wavelengths for HYPSO-2. The radiometric coefficients of both satellites are adjusted accordingly prior to further cal/val procedures.

\subsubsection{Assessment on RadCalNet Sites}
\begin{figure}[htbp]
  \centering
  \begin{subfigure}{0.4\textwidth}
    \includegraphics[width=\linewidth]{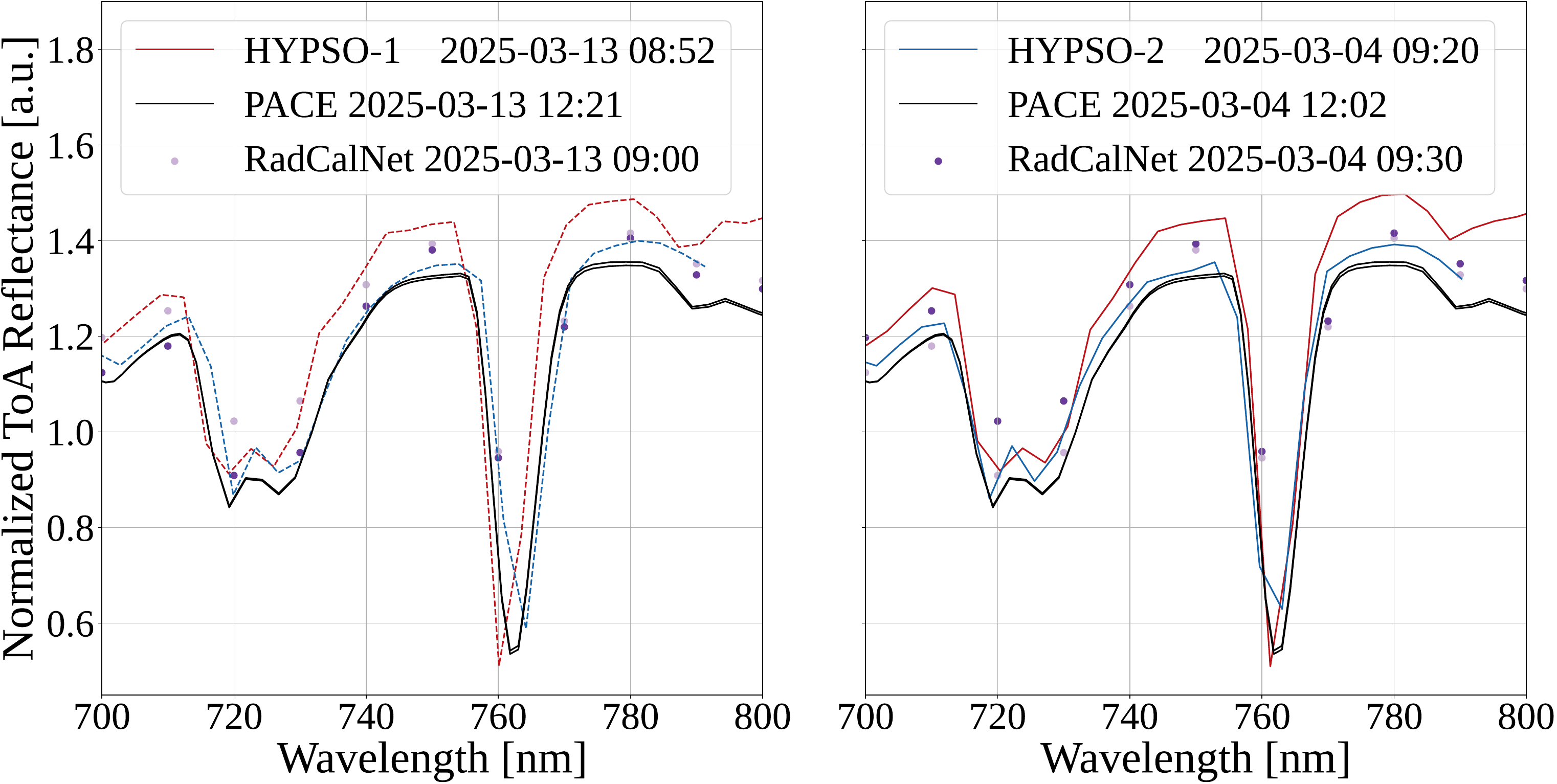}
    \caption{Gobabeb.}
  \end{subfigure}
  \hfill
  \begin{subfigure}{0.4\textwidth}
    \includegraphics[width=\linewidth]{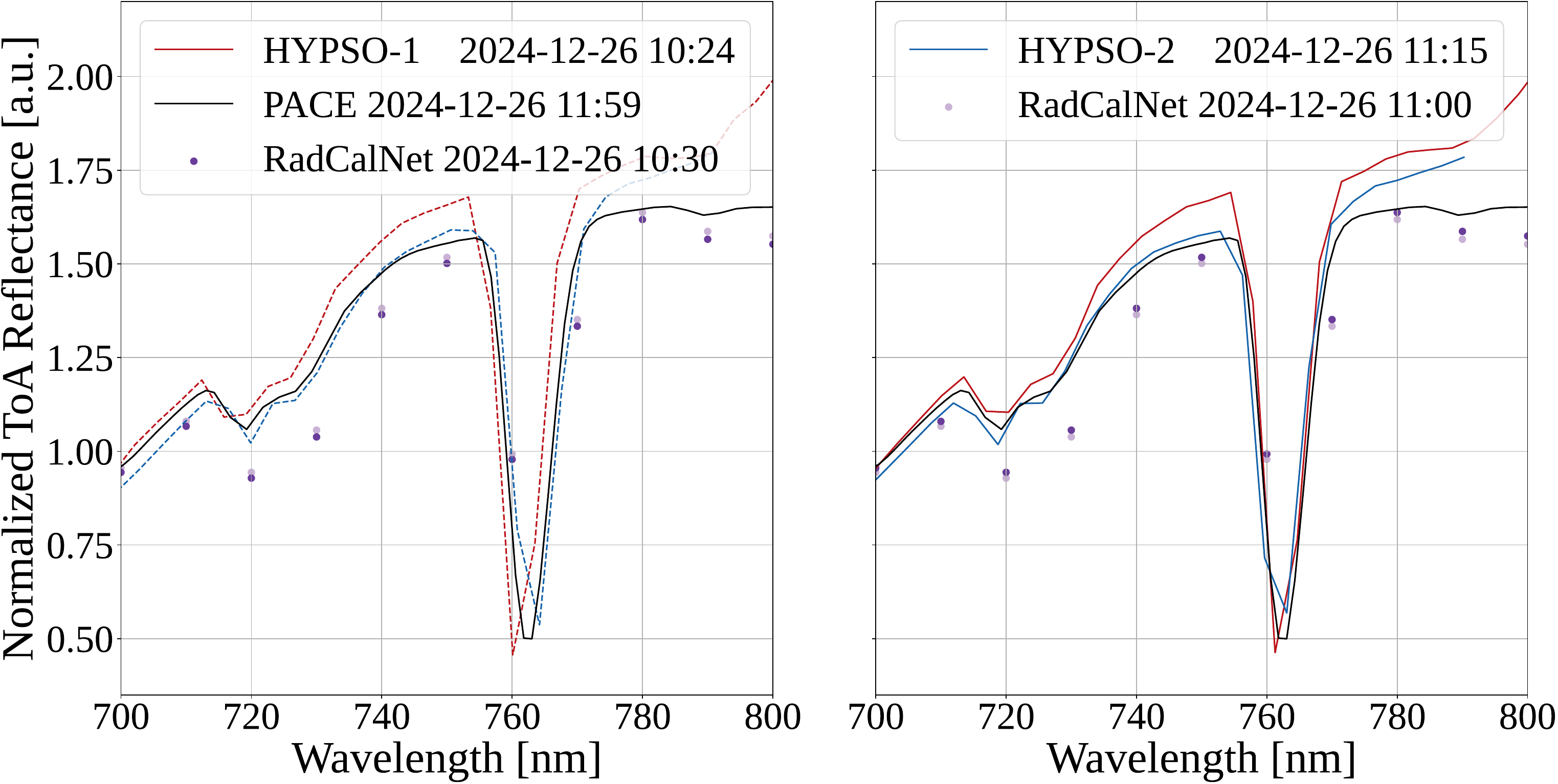}
    \caption{La Crau.}
  \end{subfigure}
  \caption{Comparison of spectral position of $O_2$ dip for HYPSO-1, HYPSO-2, RadCalNet, and PACE before and after spectral calibration.}
  \label{fig:O2_radcal}
\end{figure}

Figures \ref{fig:move_gobabeb} and \ref{fig:move_lacrau} show the ToA reflectance measured by HYPSO-1, HYPSO-2, PACE, and RadCalNet at Gobabeb and La Crau, before and after horizontal re-alignment of the radiometric coefficients and spectral calibration. All measurements are over the same coordinates, where the HYPSO spectra are derived as the average of a $2\times 2$ patch of pixels, while the PACE spectra are derived using only a single pixel (because of the larger pixel size). $\phi$ denotes the off-nadir viewing angle, the dotted lines show the spectra before calibration, and the solid lines show the spectra after calibration. Here we see that all of the HYPSO spectra become smoother, indicating that the original variations were the result of unaligned radiometric coefficients. We can also see that the oxygen A-band of HYPSO-1 and HYPSO-2 lines up better with PACE after calibration, as highlighted in Figure~\ref{fig:O2_radcal}, further validating the spectral calibration.

\subsection{Spectral Resolution} \label{sec:spec_res}
The LIME model can also be utilized when quantifying the spectral resolution. The LIME reference used here is reported to provide data with a spectral resolution of 3 nm with 1 nm sampling, with a Gaussian SRF already applied \cite{_downloads_}. Since we also model the SRF of the HYPSO satellites using Gaussian functions, and the convolution of two Gaussians is still Gaussian, we can convolve Gaussians onto the spectrum provided by LIME to model SRFs of different FWHM. The total FWHM is given by

\begin{equation}
    FWHM_\text{add}^{2} = FWHM_\text{L}^{2} + FWHM_\text{H}^{2},
\end{equation}
where $FWHM_\text{L}$ is the FWHM of the Gaussian that is already applied by LIME, $FWHM_\text{H}$ is the FWHM of the extra Gaussian that we convolve to simulate different resolutions, and $FWHM_\text{add}$ is the resultant FWHM corresponding to the resolution being simulated.\footnote{This is because the Fourier transform of a Gaussian function is still Gaussian, and a convolution in Fourier space is a multiplication. Given two Gaussian functions with standard deviation $\sigma_1$ and $\sigma_2$ the Fourier transform is $\mathcal{F}(e^{-\frac{x^2}{2\sigma^2}})=\sigma e^{-\frac{1}{2}\sigma^2 k^2}$, and $\sigma_1e^{-\frac{1}{2}\sigma_1^2k^2}\cdot \sigma_2e^{-\frac{1}{2}\sigma_2^2k^2} = \sigma_1\sigma_2e^{-\frac{1}{2}(\sigma_1^2+\sigma_2^2)k^2}$; which is a Gaussian function with standard deviation $\sigma^2 = \sigma_1^2 + \sigma_2^2$ when taken the inverse Fourier transform on. Lastly, since $FWHM = 2\sqrt{2\ln{2}}\cdot \sigma$ the same relation also holds for the FWHM.}

Initial experiments indicated that the effective spectral resolution of the LIME reference is coarser than the reported 3 nm. This was revealed when the analysis returned HYPSO resolutions below the instrument's theoretical limit of 3.33 nm \cite{prentice_design_2021}. Measuring the FWHM of the Fraunhofer lines $F$, $D$, and $C$ in the LIME spectrum gives an effective resolution closer to 3.4\,nm (see Figure \ref{fig:lime_fwhm} and Table \ref{tab:lime_fwhm} in the Supplementary Material). Results also show that the spectral resolution of LIME is consistently finer than that of both HYPSO-1 and HYPSO-2. The analysis was therefore repeated with $FWHM_\text{L} = 3.4$ nm.

For the HYPSO satellites spectral resolutions between 3.4 nm and 10 nm are tested using 0.01 nm steps over 8 spectral windows across the spectral range.\footnote{Here there is a trade-off where a smaller window size gives better granularity at the expense of lower accuracy.} Table \ref{tab:spec_res} shows the results of this analysis for the HYPSO satellites, where smile correction, the in-orbit spectral coefficients, and the re-aligned radiometric coefficients are applied to the signals prior to analysis. Results estimate that HYPSO-1 and HYPSO-2 have average spectral resolutions of 3.89 nm and 3.98 nm respectively, for the full spectral range. And for the central wavelengths both instruments have spectral resolutions close to their theoretical limit of 3.33 nm.\footnote{Values of 3.40\,nm in Table~\ref{tab:spec_res} lie at the lower bound of the tested range, and should be read as upper limits.}\footnote{The Moon spectra these results are derived from are not binned.}

\begin{table}[htbp]
\centering
\caption{Estimated spectral resolution of HYPSO-1 and HYPSO-2, based on un-binned Moon spectra, using 3 captures for each.}
\label{tab:spec_res}
\begin{tabular}{@{}lll@{}}
\toprule
& \multicolumn{2}{c}{Resolution [nm]}\\
 Range [nm]& HYPSO-1&HYPSO-2\\
\midrule
406--454 & 4.20 $\pm$ 0.05 & 6.64 $\pm$ 0.44 \\
454--502 & 4.07 $\pm$ 0.17 & 3.68 $\pm$ 0.09 \\
502--550 & 4.11 $\pm$ 0.12 & 3.40 $\pm$ 0.00 \\
550--598 & 4.54 $\pm$ 0.15 & 3.55 $\pm$ 0.05 \\
598--645 & 3.48 $\pm$ 0.05 & 3.40 $\pm$ 0.00 \\
646--693 & 3.59 $\pm$ 0.06 & 3.42 $\pm$ 0.01 \\
693--740 & 3.53 $\pm$ 0.06 & 3.47 $\pm$ 0.07 \\
741--787 & 3.58 $\pm$ 0.13 & 4.28 $\pm$ 0.35 \\
\midrule
\textbf{Full range: }& \textbf{3.89} $\pm$ \textbf{0.38} & \textbf{3.98} $\pm$ \textbf{1.06} \\
\bottomrule
\end{tabular}
\end{table}

A visual comparison between the HYPSO spectra and the LIME spectra simulated at seven different spectral resolutions between 3.4\,nm and 10\,nm further confirms that the estimates in Table \ref{tab:spec_res} are reasonable (Figure~\ref{fig:spec_res_comparison} in the Supplementary Material). All results show that the instruments have comparable spectral resolution, which is close to that of LIME.

\subsection{Radiometric Calibration} \label{sec:radiometric}
For the HYPSO satellites, reliable absolute radiometric calibration requires that the extracted Lunar signal $x_\text{moon}$ contain all of the Lunar radiance, and only Lunar radiance. However, HYPSO exhibits spectrally dependent smearing and a spectrally varying \gls{psf} (see Sections~\ref{sec:artifacts} and~\ref{sec:grd}), which makes simple Moon/background masking unreliable and favors the background-removal approach described in Section~\ref{ax:Extraction of Moon Signal} in the Supplementary Material. The resulting magnitude of $x_\text{moon}$ then depends on the effectiveness of that removal: any residual background or inadvertent loss of Lunar signal will bias the estimate. Further uncertainty arises from the Lunar sweep rate, which is only known from the attitude telemetry and affects both the oversampling of the Moon and the solid-angle calculations needed to convert between irradiance and radiance \cite{lee_pace_2024}. In several of the HYPSO captures the Moon is additionally partly out of frame, see Figure~\ref{fig:moon}, which introduces a further source of bias.

Because of these factors we only use the Moon for relative radiometric calibration and not absolute radiometric calibration, since the relative radiometric response is more reliable and less affected by the uncertainties described above.

\begin{figure}[htbp]
    \centering
    \includegraphics[width=\linewidth]{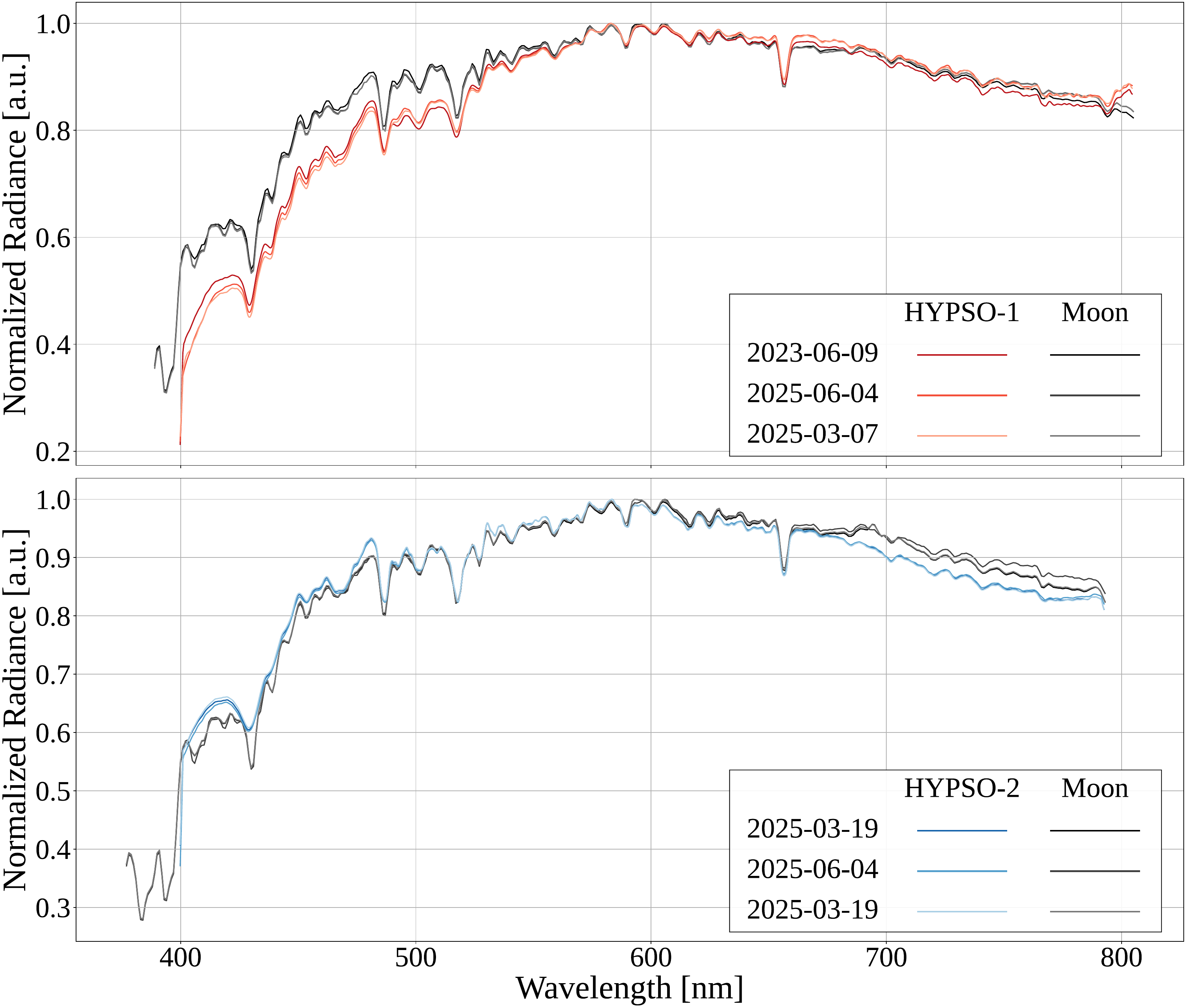}
    \caption{Normalized Moon radiance of HYPSO-1, HYPSO-2, and LIME, used for radiometric calibration.}
    \label{fig:lime_hypso_norm}
\end{figure}

Figure \ref{fig:lime_hypso_norm} shows the normalized radiance response of the Moon, comparing the HYPSO satellites to the LIME model, after smile correction, spectral calibration, and re-alignment of the radiometric coefficients, as described in Sections~\ref{sec:spectral} and~\ref{sec:move}, with LIME convolved to the estimated spectral resolution. Here we see that HYPSO-2 shows very good agreement with LIME in its spectral shape. The comparison also reveals a dip at short wavelengths for HYPSO-1 ($\approx$400--500\,nm) compared to the LIME model, that is not present for HYPSO-2. Such a discrepancy is important to correct if one aims to use the data across the constellation interchangeably.

We can use the LIME spectra as a reference, and utilize the residual between $\hat{x}_\text{moon}$ and $\hat{y}_\text{moon}$ to construct a new set of radiometric gain coefficients, that we call \textit{intra‑calibration gain factors}. These are an additional set of radiometric coefficients intended to harmonize the shape of the radiometric response of the two satellites. Since the underlying mechanism is not yet characterized, we refrain from directly adjusting the primary radiometric coefficients.

An initial hypothesis was that the dip was caused by increased spatial smearing for the short-wave bands of HYPSO-1. However, this is not evident in the FWHM of the Moon edge (Section \ref{sec:artifacts}), suggesting another cause.
\subsubsection{Temporal Stability}
In addition to inter-satellite consistency, the temporal stability of HYPSO-1 can also be assessed, by comparing the measurement taken in June 2023 with June 2025, shown in Figure \ref{fig:lime_hypso_norm}. This indicates that the spectrum has not undergone any major degradation in the two years. We can only see some slight differences for the short-wave and long-wave bands, which are not deemed significant enough to take into account when deriving the intra-calibration gain factors.

\subsubsection{Harmonization of Radiometric Magnitude} \label{sec:magnitude}
Since the Moon is only used to calibrate the shape of the radiometric response, a difference in magnitude between the satellites may remain. To quantify and harmonize this, the Libya-4 captures of HYPSO-1 and HYPSO-2 are compared to same-day captures by PACE, which here serves as a common reference for both satellites. For each of the 12 matchups (7 for HYPSO-1 and 5 for HYPSO-2), the ToA reflectance is averaged over a $0.2^\circ \times 0.2^\circ$ region at the center of Libya-4, after degrading the HYPSO spectra to the 5\,nm spectral response of PACE \cite{gorman_nasa_2019}; bands within atmospheric absorption features are left out, and saturated pixels are removed.

Because HYPSO and PACE pass over Libya-4 at different times of day (about 10:30 and 13:00 local time), they observe the site at different sun-target-sensor geometries, and the BRDF of the desert causes a difference in reflectance that is not related to calibration. Each HYPSO/PACE ratio is therefore corrected to equal viewing geometry using an empirical correction, fitted across the matchups of both satellites and applied equally to all bands, as described in the Supplementary Material (Section~\ref{sec:supp_brdf}).

\begin{figure}[htbp]
    \centering
    \includegraphics[width=\linewidth]{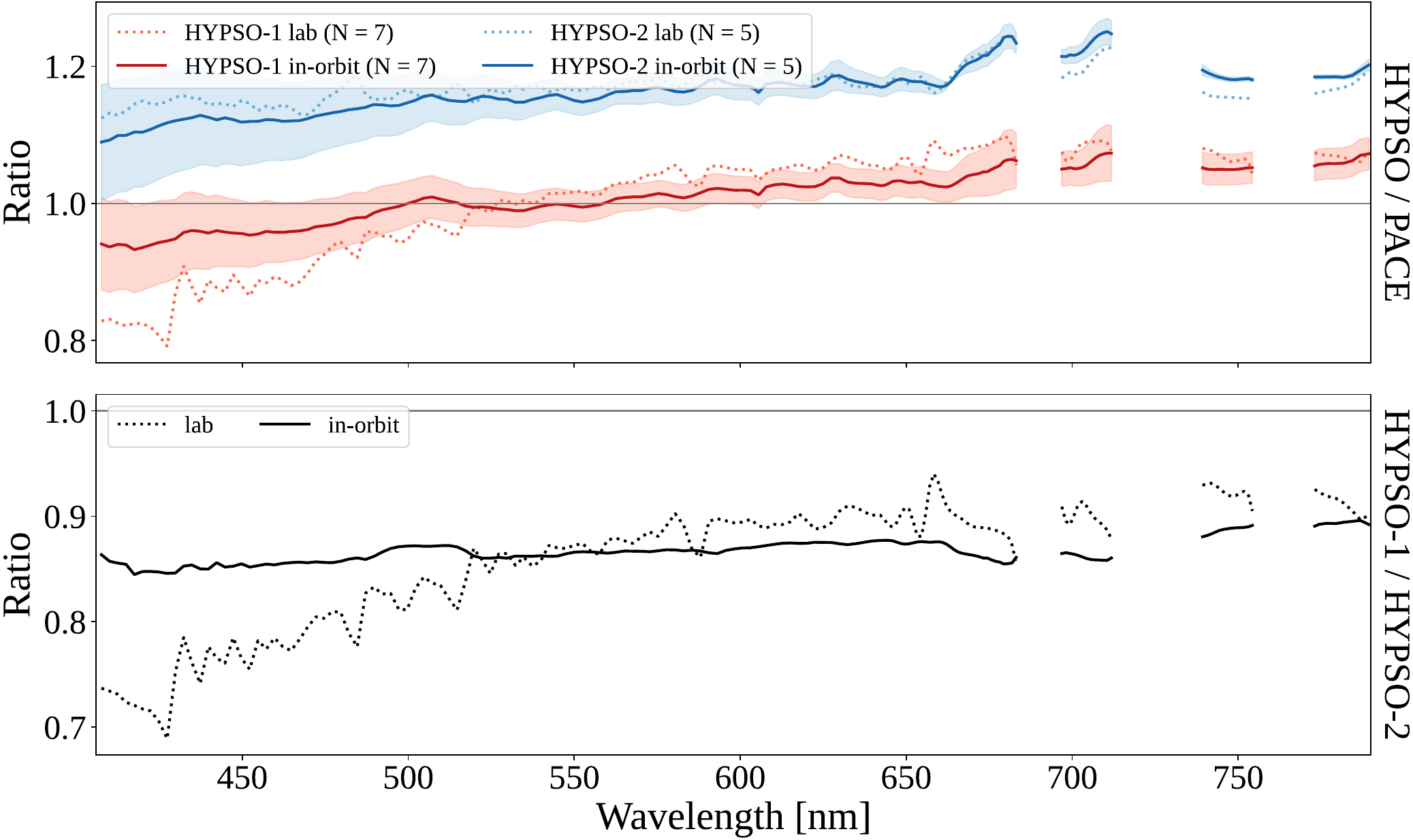}
    \caption{Ratio between the ToA reflectance of HYPSO and PACE over Libya-4 at equal viewing geometry (top), and between HYPSO-1 and HYPSO-2 (bottom), using lab (dotted) and in-orbit (solid) coefficients. The shaded areas show the standard deviation over matchups, and gaps are atmospheric absorption bands.}
    \label{fig:libya_pace_ratio}
\end{figure}

Figure~\ref{fig:libya_pace_ratio} shows the ratio between HYPSO and PACE at equal viewing geometry, using lab and in-orbit coefficients, and Table~\ref{tab:magnitude} quantifies the shape and magnitude of these ratios. Here we see that the in-orbit coefficients make the spectral shape of the two satellites more similar, as the standard deviation over wavelength of the ratio between HYPSO-1 and HYPSO-2 is reduced from 6.7\% to 1.3\%, and that the shape of HYPSO-1 becomes closer to PACE (from 7.8\% to 3.6\%). This confirms the relative radiometric calibration against a reference that is independent of the Moon. For HYPSO-2, which already agreed well with LIME, the agreement with PACE is slightly reduced (from 2.1\% to 3.0\%), which is expected since the in-orbit coefficients harmonize both satellites to LIME rather than to PACE.

The magnitude, however, is not corrected by the relative calibration, and HYPSO-2 measures 16.9\% above PACE while HYPSO-1 measures 1.4\% above. HYPSO-2 is therefore scaled to the magnitude of HYPSO-1, by applying a factor of 0.868 to its radiometric coefficients. HYPSO-1 is used as the reference since it is closest to PACE, and so that its data products remain unchanged. Note that this harmonizes the magnitude of the two satellites, but is not an absolute calibration. To validate the factor, each matchup is in turn left out of the fit, and its corrected magnitude is compared to the HYPSO-1 magnitude found from the remaining matchups. This gives an rms difference of 2.6\% and 2.7\% when leaving out HYPSO-1 and HYPSO-2 captures respectively. This is larger than the remaining difference in shape between them (1.3\%), and shows that the magnitude is the more uncertain part of the radiometric calibration.

Table~\ref{tab:libya_pace} shows how close HYPSO is to PACE after calibration, given as the mean absolute difference over the bands of each matchup. As measured, the in-orbit coefficients and the magnitude correction reduce the difference from 8.5\% to 5.9\% for HYPSO-1 and from 21.6\% to 7.7\% for HYPSO-2. With BRDF correction, the difference is reduced to 4.0\% for both satellites. The spectra of all 12 matchups are shown in Figure~\ref{fig:libya_pace_toa_reflectance} in the Supplementary Material.

\begin{table}[htbp]
\centering
\caption{ToA reflectance of HYPSO relative to PACE over Libya-4 with the lab and the in-orbit coefficients, with BRDF correction (7 and 5 matchups). Magnitude: band-averaged ratio minus one. Shape: standard deviation of the mean ratio over wavelength.}
\label{tab:magnitude}
\begin{tabular}{@{}lccc@{}}
\toprule
& HYPSO-1 & HYPSO-2 & HYPSO-1/HYPSO-2 \\
\midrule
\multicolumn{4}{@{}l}{\textit{Magnitude [\%]}} \\
\quad Lab & +1.3 $\pm$ 1.0 & +17.3 $\pm$ 1.4 & $-$13.6 $\pm$ 1.3 \\
\quad In-orbit & +1.4 $\pm$ 1.0 & +16.9 $\pm$ 1.4 & $-$13.2 $\pm$ 1.3 \\
\midrule
\multicolumn{4}{@{}l}{\textit{Shape [\%]}} \\
\quad Lab & 7.8 & 2.1 & 6.7 \\
\quad In-orbit & 3.6 & 3.0 & 1.3 \\
\midrule
\multicolumn{4}{@{}l}{Correction factor for HYPSO-2: 0.868 $\pm$ 0.013} \\
\bottomrule
\end{tabular}
\end{table}

\begin{table}[htbp]
\centering
\caption{Mean absolute difference between the ToA reflectance of HYPSO and PACE over Libya-4, $|$HYPSO/PACE $-$ 1$|$ averaged over the bands of each matchup, given as mean $\pm$ standard deviation over matchups (7 for HYPSO-1 and 5 for HYPSO-2), as measured and with BRDF correction.}
\label{tab:libya_pace}
\begin{tabular}{@{}lcc@{}}
\toprule
Coefficients & HYPSO-1 [\%] & HYPSO-2 [\%] \\
\midrule
\multicolumn{3}{@{}l}{\textit{As measured}} \\
Lab & 8.5 $\pm$ 1.2 & 21.6 $\pm$ 6.5 \\
In-orbit + correction & 5.9 $\pm$ 1.1 & 7.7 $\pm$ 3.7 \\
\midrule
\multicolumn{3}{@{}l}{\textit{With BRDF correction}} \\
Lab & 7.1 $\pm$ 1.9 & 17.3 $\pm$ 1.9 \\
In-orbit + correction & 4.0 $\pm$ 1.7 & 4.0 $\pm$ 1.5 \\
\bottomrule
\end{tabular}
\end{table}

\subsubsection{Assessment on RadCalNet Sites}
\begin{figure}[htbp]
    \centering
    \includegraphics[width=1\linewidth]{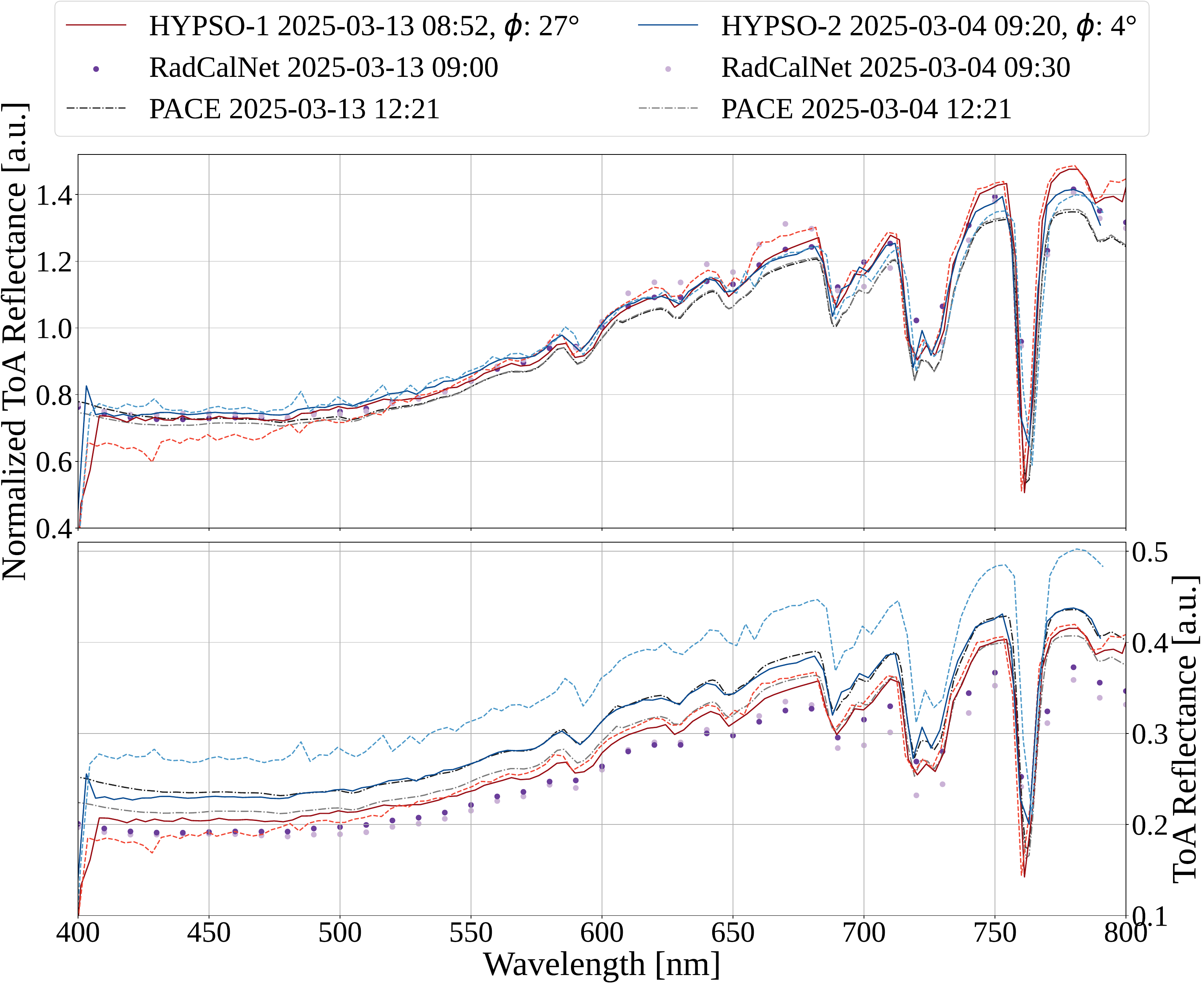}
    \caption{ToA reflectance for HYPSO-1, HYPSO-2, PACE, and RadCalNet measurements over the Gobabeb site, normalized (top) and not normalized (bottom). The dotted and solid lines show the spectra before and after radiometric calibration and harmonization of magnitude, respectively.}
    \label{fig:radiometric_gobabeb}
\end{figure}

\begin{figure}[htbp]
    \centering
    \includegraphics[width=1\linewidth]{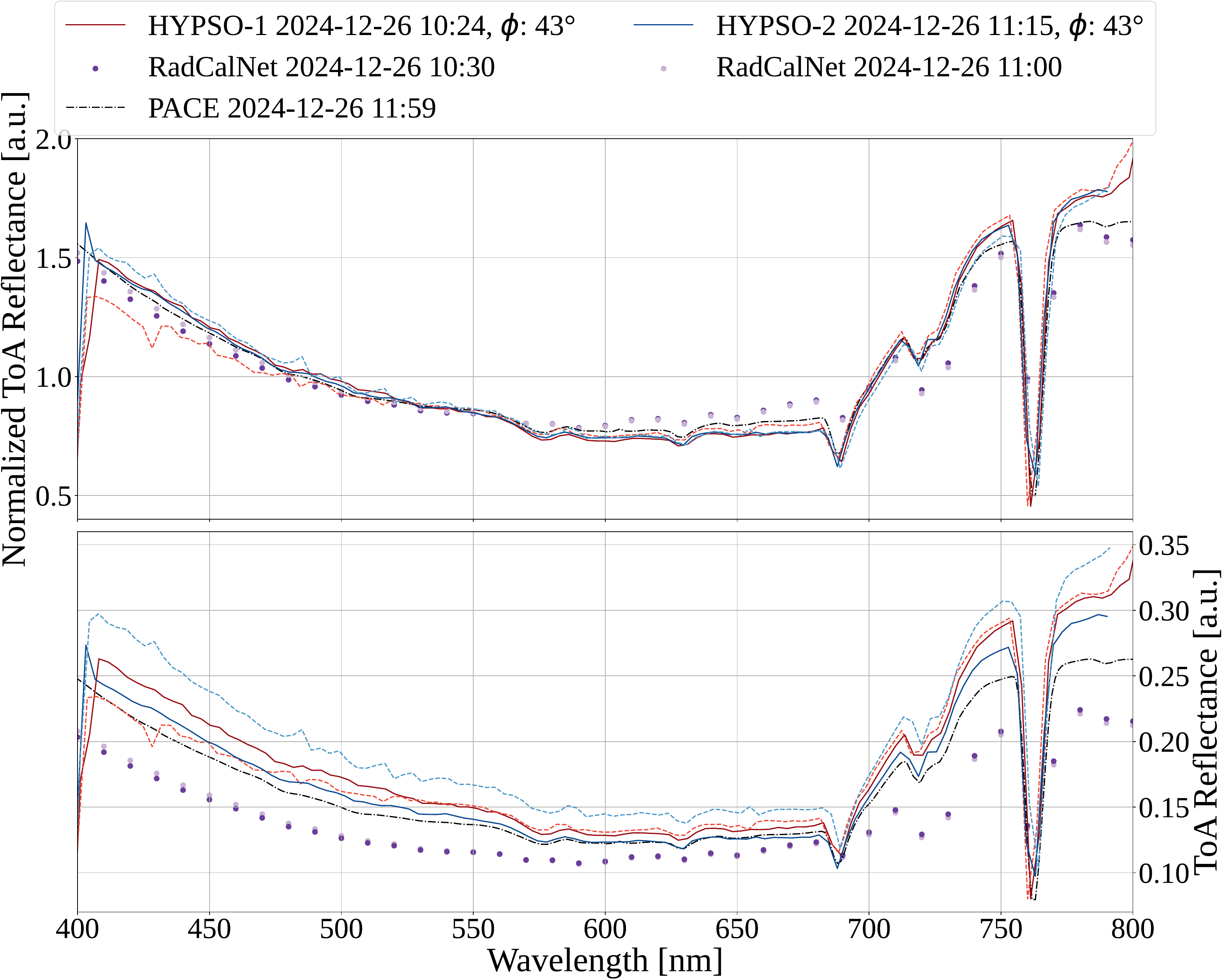}
    \caption{ToA reflectance for HYPSO-1, HYPSO-2, PACE, and RadCalNet measurements over the La Crau site, normalized (top) and not normalized (bottom). The dotted and solid lines show the spectra before and after radiometric calibration and harmonization of magnitude, respectively.}
    \label{fig:radimetric_lacrau}
\end{figure}

The radiometric coefficients are also assessed by cross-validation over the RadCalNet sites, as in Section~\ref{sec:move}, see Figures~\ref{fig:radiometric_gobabeb} and~\ref{fig:radimetric_lacrau}. As these captures are taken in the nominal (binned) capture mode and over other targets than Libya-4, they give an independent, but qualitative, comparison. Unlike for Libya-4, no BRDF correction is applied, and the captures differ in time and viewing angle: RadCalNet provides ToA reflectance for a nadir view, while the HYPSO captures are taken up to $43^\circ$ off-nadir and up to 3.5 hours before PACE. Still, we see that the spectra of HYPSO-1 and HYPSO-2 become more similar to each other in both shape and magnitude at both sites, including at La Crau where both satellites captured the site on the same day and at the same viewing angle, and that they generally come closer to PACE.

\begin{figure}[htbp]
    \centering
    \includegraphics[width=1\linewidth]{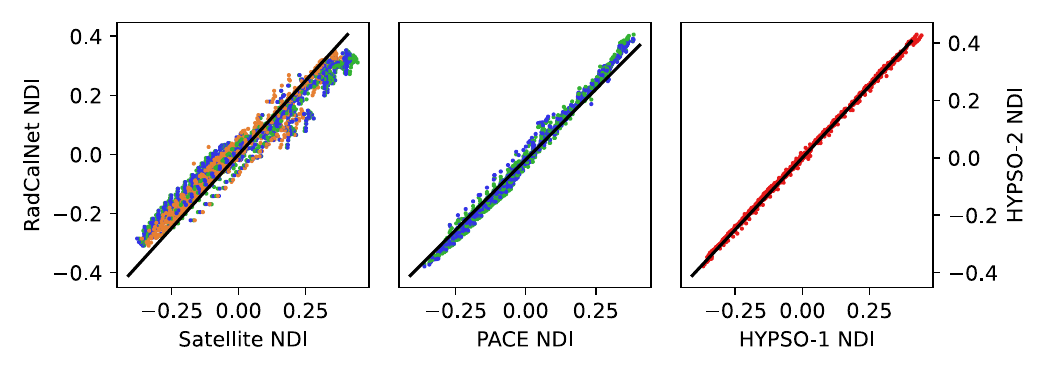}
    \caption{NDIs over the La Crau site of HYPSO-1, HYPSO-2, and PACE against RadCalNet (left), HYPSO-1 and HYPSO-2 against PACE (center), and HYPSO-2 against HYPSO-1 (right). The black line shows equality.}
    \label{fig:NDI_lacrau}
\end{figure}

To assess what the low shape uncertainty implies for the data products discussed in Section~\ref{sec:data_products}, NDIs are compared over the RadCalNet sites. The satellite spectra are first smoothed and resampled to the RadCalNet wavelengths, and NDIs are then computed for all pairs of bands separated by less than 200\,nm, see Figure~\ref{fig:NDI_lacrau}. At both sites, the NDIs of HYPSO agree slightly better with PACE than with RadCalNet. This indicates that part of the discrepancy with RadCalNet is shared by the satellites, and may arise from how the RadCalNet product is derived, such as its modeled atmosphere and nadir viewing geometry. At La Crau, where both HYPSO satellites captured the site on the same day, HYPSO-1 and HYPSO-2 spectra are most consistent. The corresponding results for Gobabeb, and a spatial analysis over the La Crau scene, are given in the Supplementary Material (Section~\ref{sec:Spatialvalidation}). This shows that the spectral shape, and thereby the shape-based data products derived from it, is consistent within the constellation and with PACE, despite the remaining uncertainty in magnitude.

\subsection{Ground Resolvable Distance}\label{sec:grd}
Captures of the Lake Pontchartrain bridge at different wavelengths show that the across-track GRD varies by wavelength. The bridge becomes wider towards both ends of the wavelength range, indicating increased spatial smearing for these bands (Figure~\ref{fig:bridge_combined} in the Supplementary Material). The FWHM of the bridge is estimated by fitting a curve to its intensity profile along the red line in Figure~\ref{fig:grd_bridge}. These estimates become unreliable below around 500\,nm and above around 700\,nm for HYPSO-1, and below around 450\,nm and above 700\,nm for HYPSO-2. These wavelength ranges are therefore omitted from Table \ref{tab:grd}, which shows the estimated across-track GRD values for HYPSO-1 and HYPSO-2.

\begin{table}[htbp]
    \centering
    \caption{Across-track FWHM and GRD of HYPSO-1 and HYPSO-2 for different wavelength ranges.}
    \label{tab:grd}
    \begin{tabular}{@{}lllll@{}}
    \toprule
    & \multicolumn{2}{c}{FWHM [pixels] }& \multicolumn{2}{c}{GRD [m] }\\
 Range [nm]& HYPSO-1& HYPSO-2& HYPSO-1&HYPSO-2\\
    \midrule
        447--485 &  --   &  10.2 & --  & 713 \\
        489--527 &  12.5 &   6.7 & 669 & 470 \\
        531--569 &   9.4 &   2.7 & 507 & 186 \\
        572--610 &   5.0 &   1.7 & 267 & 121 \\
        614--652 &   3.1 &   1.2 & 164 &  85 \\
        655--693 &   3.3 &   2.0 & 177 & 144 \\
    \midrule
    \textbf{Full range: }    & \textbf{6.7} &  \textbf{4.1} & \textbf{357}  &   \textbf{287} \\
    \bottomrule
    \end{tabular}
\end{table}

Comparing the across-track GRD between 489 nm and 693 nm, where we have reliable estimates for both satellites, we find that HYPSO-1 has a mean GRD of 357 m while HYPSO-2 has a mean GRD of 201 m. This difference is also visible in Figure~\ref{fig:bridge_combined}, where the bridge signal of HYPSO-1 degrades more rapidly towards the two ends of the wavelength range than that of HYPSO-2. It should also be noted that this advantage is in spite of HYPSO-1's lower altitude, giving it a lower GSD. The difference in resolution is assumed to mainly be the result of more careful assembly and more rigorous calibration efforts for HYPSO-2, although uncontrolled factors like impact of launch have probably also played a significant role.

\subsection{SNR}
\begin{figure}[htbp]
    \centering
    \includegraphics[width=1\linewidth]{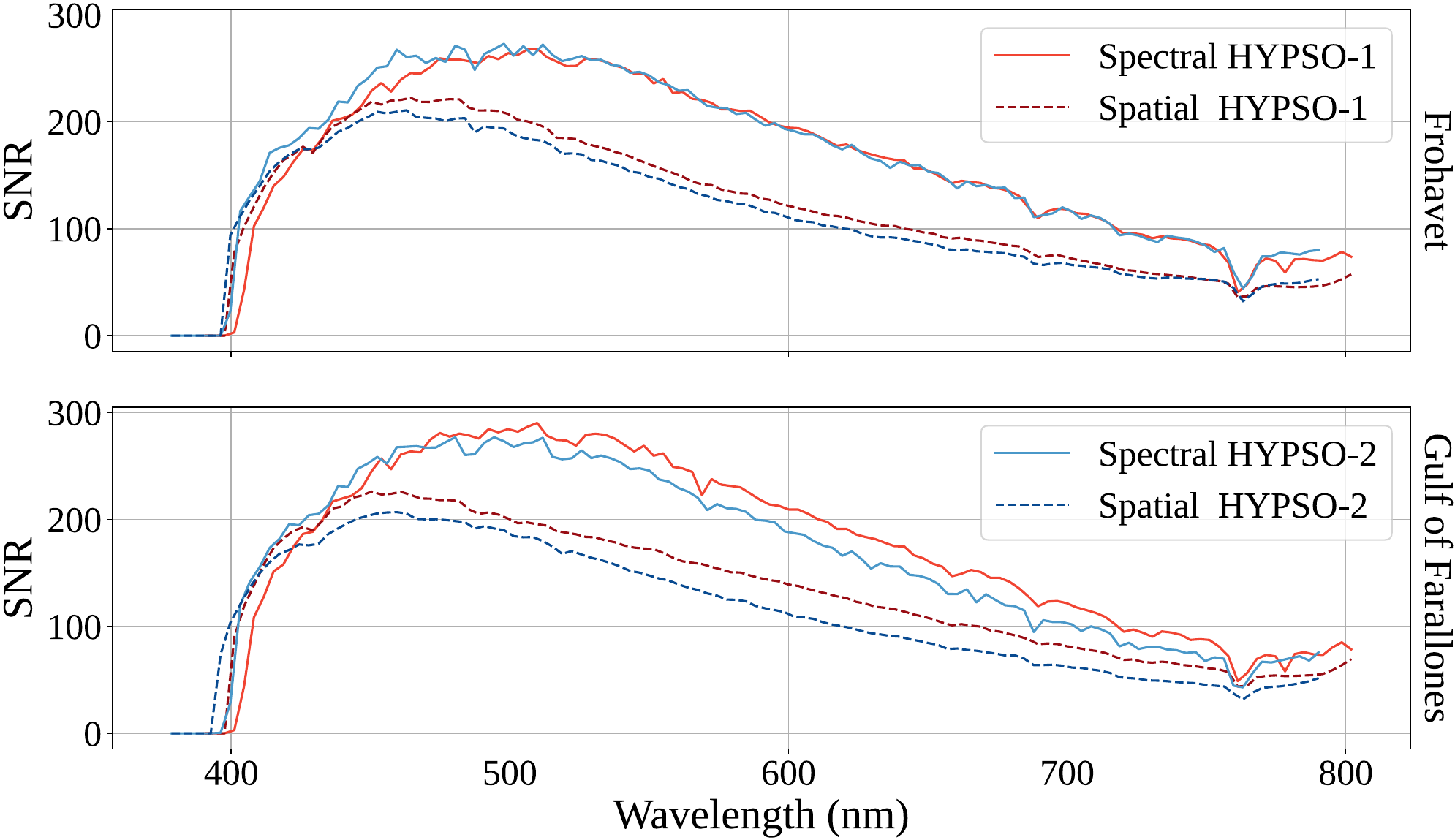}
    \caption{Estimated spatial and spectral SNR for HYPSO-1 and HYPSO-2 over uniform dark ocean patches, in Frohavet and Gulf of Farallones. }
    \label{fig:snr_analysis}
\end{figure}

Figure \ref{fig:snr_analysis} presents the spectral and spatial SNR derived for HYPSO‑1 and HYPSO‑2 from the uniform ocean patches shown in Figure \ref{fig:snr_targets}. The results indicate that both satellites exhibit comparable SNR, with peak spectral values near 300 and peak spatial values around 200. HYPSO‑1 shows a slight SNR advantage, which can potentially be attributed to its lower altitude. However, these differences may also stem from non‑uniformities in the water or atmosphere, as the selected scenes are located relatively close to the coast and thus do not fully represent open‑ocean conditions. A clear wavelength dependence is also evident, with spectral SNR peaking near 500 nm and spatial SNR near 450 nm. We can also notice that above 700 nm, the SNR generally drops below 100. This decline in the longer‑wavelength region is consistent with the results shown in Section \ref{sec:artifacts}.

\subsection{Artifacts} \label{sec:artifacts}
\begin{figure*}[t!]
    \centering
    \begin{subfigure}[b]{0.49\textwidth}
        \centering
        \includegraphics[width=\linewidth]{figures/moon_edge.png}
        \vspace*{0.1mm}
        \caption{Moon captures across the spectral range by HYPSO-1 (left) and HYPSO-2 (right).}
        \label{fig:moon_edge}
    \end{subfigure}
    \hfill
    \begin{subfigure}[b]{0.49\textwidth}
        \centering
        \includegraphics[width=\linewidth]{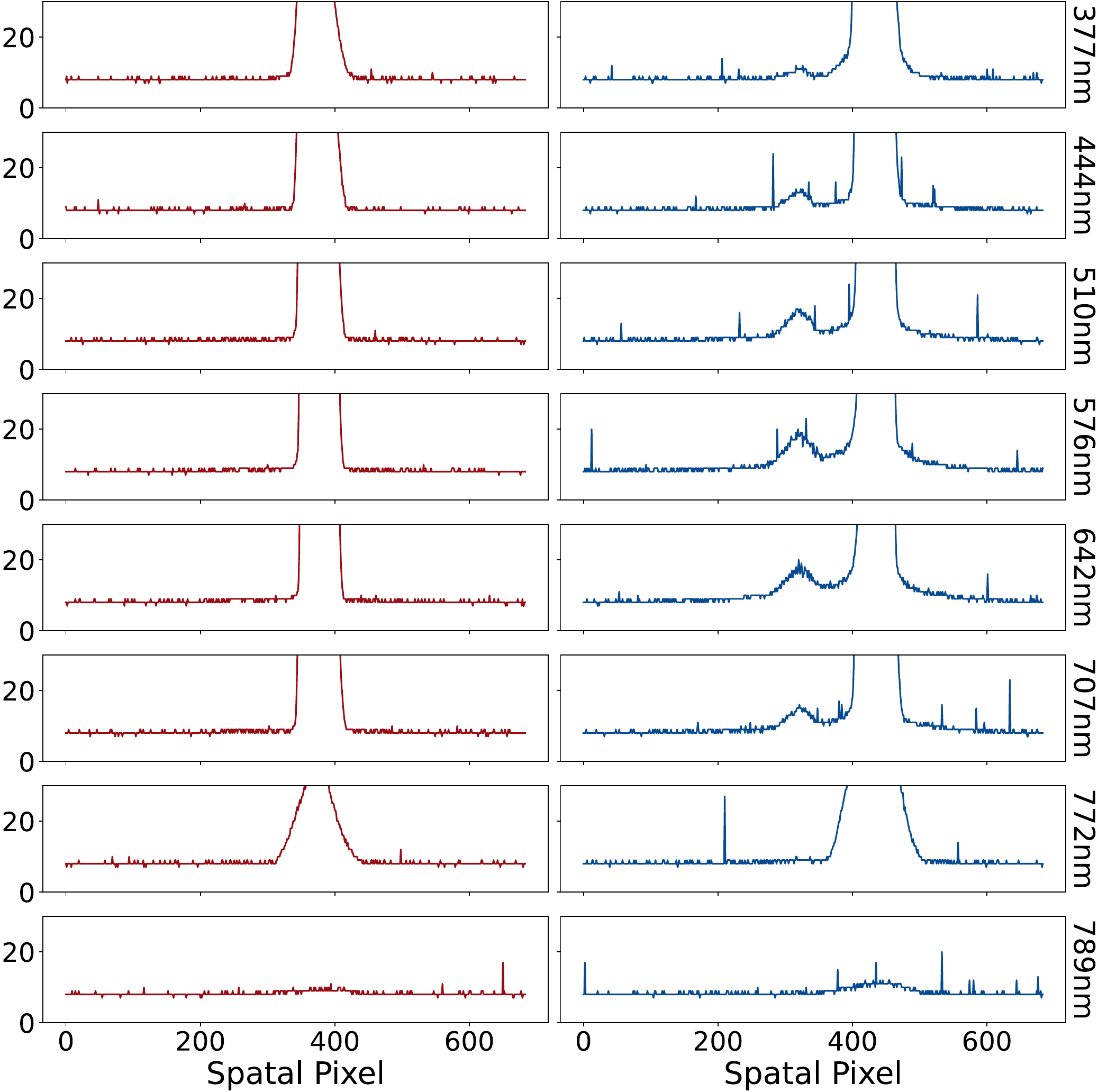}
        \caption{Moon intensity along the central line for HYPSO-1 (left) and HYPSO-2 (right).}
        \label{fig:moon_edge_spectra}
    \end{subfigure}
    \caption{Moon captures across the spectral range for HYPSO-1 and HYPSO-2, with corresponding intensity profiles, used to investigate artifacts. The along-track direction is vertical in these images.}
    \label{fig:moon_across_combined}
\end{figure*}

For the HYPSO satellites the spatial extent of targets varies with wavelength, as can be seen in Figure \ref{fig:moon_across_combined}. Figure \ref{fig:moon_edge_spectra} shows the intensity across the center frame of Figure \ref{fig:moon_edge}, highlighting the change in intensity across the swath width for different wavelengths. The artifact seen in the captures is known as cross-talk or smearing. To quantify the smearing effect we can look into the FWHM of the LSF of the Moon edge across-track, as shown in Table \ref{tab:moon_fwhm}. Based on the captures shown in Figure~\ref{fig:moon_across_combined} it can be assumed that the LSF estimates become unreliable on the last wavelength range. This wavelength range is therefore omitted from the table.

The results show that the FWHM of the LSF increases towards the two ends of the wavelength range for both HYPSO-1 and HYPSO-2, and that both satellites are similarly affected by the artifact.

\begin{table}[htbp]
\centering
\caption{Estimated FWHM of Moon captures for HYPSO-1 and HYPSO-2, using three captures for each satellite.}
\label{tab:moon_fwhm}
\begin{tabular}{@{}lll@{}}
\toprule
& \multicolumn{2}{c}{FWHM [pixels]}\\
 Range [nm]& HYPSO-1&HYPSO-2\\
\midrule
    383--425 & 8.1 $\pm$ 2.4 & 7.9 $\pm$ 0.6 \\
    425--467 & 6.6 $\pm$ 2.5 & 6.5 $\pm$ 0.2 \\
    467--509 & 5.4 $\pm$ 2.2 & 5.4 $\pm$ 0.2 \\
    509--551 & 4.8 $\pm$ 1.7 & 4.8 $\pm$ 0.4 \\
    551--592 & 4.3 $\pm$ 0.9 & 4.1 $\pm$ 1.0 \\
    593--634 & 3.9 $\pm$ 0.9 & 3.7 $\pm$ 1.3 \\
    634--675 & 4.1 $\pm$ 1.1 & 3.8 $\pm$ 1.2 \\
    675--716 & 3.8 $\pm$ 1.2 & 4.5 $\pm$ 0.3 \\
    716--757 & 6.1 $\pm$ 2.0 & 8.2 $\pm$ 0.3 \\
\midrule
\textbf{Full range: }& \textbf{5.8} $\pm$ \textbf{1.5} & \textbf{5.1} $\pm$ \textbf{0.5} \\
\bottomrule
\end{tabular}
\end{table}

In addition to the spectrally dependent smearing, an artifact is visible around pixel 300 for HYPSO-2 in Figure~\ref{fig:moon_edge_spectra}. Moon captures with saturation in the central bands show that this artifact is a mirroring of the Moon about the optical axis (Figure~\ref{fig:moon_artifact} in the Supplementary Material), which might point to internal reflections in the optical system. However, this effect is only significant during saturation and with a dark surrounding, so it is not likely to have a large effect when capturing ground targets as long as saturation is avoided. Notably, the bright bridges against the dark waters used to assess the GRD were not found to have any ghost artifact.

\section{Conclusion}
In this paper we have demonstrated a cal/val procedure on the HYPSO hyperspectral small satellite constellation, requiring no onboard calibration hardware. The procedure relies on Lunar captures analyzed against LIME, complemented by a small selection of terrestrial targets, such as Libya-4 for smile characterization. The Moon proved to be a versatile calibration reference, enabling spectral calibration, radiometric calibration, estimation of spectral resolution, assessment of post-launch misalignment of laboratory coefficients, and artifact analysis, where several of the methods are only applicable because the instruments are hyperspectral.

Two sets of radiometric calibration coefficients were derived: one correcting for post-launch misalignment of laboratory coefficients, and one harmonizing the radiometric response across the constellation to enable joint algorithms that can utilize data from both satellites interchangeably. These coefficients are application-agnostic and allow the instruments to be used across a wide range of downstream tasks. For applications such as ocean color retrieval, ocean color system vicarious calibration (OC-SVC) should subsequently be applied, as it provides absolute radiometric adjustment that accounts for both the instrument and the chosen atmospheric correction model \cite{_ocean_}. It should also be noted that ocean color algorithms such as band-ratio methods, NDIs, and spectral unmixing depend on the shape of the measured spectrum rather than its absolute magnitude (Section~\ref{sec:data_products}), and normalizing water spectra to emphasize shape over magnitude has become common practice for optically complex waters \cite{bi_holistic_2024}.

The results show a clear difference in how reliably the different parts of the instrument response can be calibrated without onboard calibration hardware, and of the procedures demonstrated, spectral calibration was the most successful. The spectral axes of both satellites were found to be mainly offset by a constant shift, corresponding to a post-launch misalignment of 3 pixels for HYPSO-1 and 2 pixels for HYPSO-2. After calibration, the Fraunhofer lines in Libya-4 captures deviate by $+0.14 \pm 0.13$\,nm for HYPSO-1 and $+0.04 \pm 0.06$\,nm for HYPSO-2 from their expected positions, and the spectral alignment with PACE is improved. The relative radiometric response also showed improved agreement with PACE and RadCalNet after calibration, as well as improved internal consistency across the constellation: HYPSO-2 was found to agree well with the LIME spectral shape, whereas HYPSO-1 exhibited a dip in the short-wave bands that was corrected using the LIME reference, although the underlying mechanism remains uncharacterized and should be investigated further. Compared to PACE over Libya-4, the difference in spectral shape between the satellites was reduced from 6.7\% to 1.3\%, while a difference in magnitude of 13.2\% remained, which was harmonized to within 2.7\% (rms of held-out captures). The radiometric magnitude is therefore the least certain part of the calibration, whereas the spectral axis is verified to within a fraction of a pixel using Fraunhofer lines, and confirmed by the oxygen A-band position of PACE. While the Moon proved reliable for the shape of the radiometric response, the magnitude of the extracted Lunar signal is uncertain due to captures where the Moon is partly out of frame, uncertainty in the oversampling estimated from the Lunar sweep rate, and the spectrally dependent smearing of the instruments. A reference satellite was therefore needed to harmonize the magnitude. After calibration and harmonization, the HYPSO spectra over Libya-4 differ from PACE by 4.0\% on average after correction for viewing geometry, compared to 7.1\% and 17.3\% for HYPSO-1 and HYPSO-2 with the laboratory coefficients, showing that a CubeSat with COTS instrumentation and no onboard calibration hardware can come close to a well-calibrated reference mission. Detection and correction of the smile artifact was also demonstrated in-orbit using a Sobel edge detection filter on Libya-4 captures, reducing the largest smile from 9.3 to 2.0 pixels (0.8 nm) for HYPSO-1 and from 10.0 to 1.2 pixels (0.5 nm) for HYPSO-2, with the remaining smile below one pixel for most absorption features. The results further show that the spectral resolution of both satellites is close to the theoretical limit of 3.33\,nm in the central wavelengths, with averages of 3.89 and 3.98\,nm over the full range, and that their SNR is comparable, but that HYPSO-2 has a notably higher spatial resolution, with a mean across-track GRD of 201\,m compared to 357\,m for HYPSO-1. These differences are important to account for when developing algorithms intended to operate across both satellites.

We expect that other dispersive hyperspectral imagers can achieve a similar spectral accuracy, as long as their spectral resolution is sufficient to resolve the Fraunhofer lines and atmospheric absorption features. The radiometric magnitude, on the other hand, remains less certain than the spectral accuracy, as it relies on cross-calibration against a reference satellite. Absolute calibration against the Moon could reduce this dependence, but was not feasible here: in several captures the Moon is partly out of frame, the oversampling depends on an uncertain estimate of the Lunar sweep rate, and the spectrally dependent smearing complicates the extraction of the Lunar signal. Until the magnitude is better constrained, algorithms that depend on the spectral shape rather than the magnitude, such as NDIs, are the most robust choice for joint use of data across the constellation and with larger missions, and initial tests show that NDIs varying by less than one percent of the signal are consistent across the satellites (Supplementary Material, Section~\ref{sec:Spatialvalidation}).

For other missions applying a similar procedure, we recommend acquiring full-frame captures without binning for the spectral characterization, and scheduling Lunar captures such that the whole Moon passes through the frame, within the phase range of LIME, and with measures taken to minimize uncertainty in sweep rate. With such captures, the Moon may also be used to constrain the absolute radiometric magnitude. More matchups would further improve the calibration; here only three Lunar captures and fewer than ten Libya-4 matchups per satellite were usable, and good RadCalNet matchups were only found at two sites. Libya-4 captures should therefore be planned with viewing geometries close to that of the reference sensor. Future missions can also utilize HYPERNETS to obtain additional matchups at higher spectral resolution, in addition to RadCalNet. Finally, the smearing, SNR, and GRD of the HYPSO instruments all degrade towards the ends of the wavelength range, which may motivate taking measures to mitigate these effects during instrument design if possible.

In future work the radiometric coefficients should be validated further using the DCC and Rayleigh scattering methods, and absolute radiometric calibration using the Moon should also be investigated further. Regular Lunar measurements, taken at least monthly, are recommended for monitoring sensor changes over time \cite{stone_moon_2009}, and regular DCC captures may serve a similar purpose \cite{fougnie_monitoring_2009}. Artifacts not addressed here, including keystone, striping, dead pixels, and stray light, should also be characterized and corrected where possible. Finally, the results show that spectrally dependent smearing is a prominent artifact degrading the data quality, and its correction should therefore be investigated.

The HYPSO constellation contributes openly to the remote sensing community: data from both satellites is freely available through the HYPSO data portal at \url{https://hypso.space/dataportal/en/}, and the NTNU Small Satellite Lab has developed an open-source Python package for processing HYPSO data, available at \url{https://github.com/NTNU-SmallSat-Lab/hypso-package}. We hope that the cal/val approach demonstrated here provides a useful reference for the growing number of hyperspectral small satellite missions operating without onboard calibration hardware.

\section*{Acknowledgment}

The authors would like to thank the Research Council of Norway and industry partners for funding through grant number 337495 (SoS), 325961 (HYPSCI), 333229 (ARIEL), as well as the European Space Agency and Norwegian Space Agency through Prodex Grant 4000134666. The research is also funded by the European Union under the Horizon Europe Programme, Grant Agreement No. 101082004 (DiverSea). Views and opinions expressed are however those of the author(s) only and do not necessarily reflect those of the European Union or European Research Executive Agency (REA). Neither the European Union nor the granting authority can be held responsible for them.

\ifCLASSOPTIONcaptionsoff
  \newpage
\fi

\printbibliography

\clearpage
\setcounter{figure}{0}
\setcounter{table}{0}
\setcounter{section}{0}
\setcounter{equation}{0}
\fancyhead{} 

\renewcommand{\thetable}{S-\Roman{table}}
\renewcommand{\thefigure}{S\arabic{figure}}
\renewcommand{\thesection}{S\arabic{section}}
\renewcommand{\theequation}{S\arabic{equation}}

\section*{Supplementary Material}

\section{Extraction of Moon Signal} \label{ax:Extraction of Moon Signal}

Algorithm \ref{alg:bg_removal} shows a method for background signal removal that inputs the hyperspectral cube $X \in \mathbb{R}^{w\times h \times b}$, and outputs the summed radiance spectrum of the Moon across the capture $x_\text{moon}\in \mathbb{R}^{b}$. Here $w$ is the number of frames, $h$ is the number of pixels across-track, and $b$ is the number of bands. For the HYPSO satellites $\tilde{X}_\text{bg}$ is taken as the initial $50\times 50$ pixels that are in the corner of the capture (because these are far away from the Moon), $k=3$, and $m=5$. After these computations $x_\text{moon}$ can subsequently be compared against the irradiance spectra given by models like LIME for different types of analysis. Unmixing is not used for background removal because it results in a trade-off between spectral granularity and accuracy.

\begin{algorithm}[htbp]
\SetAlgoLined
\KwData{Hyperspectral cube $X \in \mathbb{R}^{w\times h \times b}$}
\KwResult{Total Moon spectrum $x_\text{moon}\in \mathbb{R}^{b}$}
Use a median filter of size $k\times k$ across $h$ and $b$ of $X$, and store as $\tilde{X} \in \mathbb{R}^{w\times h \times b}$ \;
Extract $l\times l$ square of deep space pixels from $\tilde{X}$, and store as $\tilde{X}_\text{bg} \in \mathbb{R}^{l\times l \times b}$\;
Take mean over $w$ and $h$ of $\tilde{X}_\text{bg}$, and store as $\overline{\tilde{X}}_\text{bg} \in \mathbb{R}^{1\times 1\times b}$ \;
Use a median filter of size $m$ across $b$ of $\overline{\tilde{X}}_\text{bg}$, and store as $\tilde{\overline{\tilde{X}}}_\text{bg} \in \mathbb{R}^{1\times 1\times b}$ \;
Compute $\tilde{X} - \tilde{\overline{\tilde{X}}}_\text{bg}$, and store as $X_\text{moon} \in \mathbb{R}^{w\times h \times b}$\;
Calibrate $X_\text{moon}$, and store as $X_\text{moon\_cal}\in \mathbb{R}^{w\times h \times b}$\;
Sum over $w$ and $h$ of $X_\text{moon\_cal}$, and store as $x_\text{moon}\in \mathbb{R}^{b}$\;
\caption{Background Removal Algorithm}
\label{alg:bg_removal}
\end{algorithm}

\section{Spectral Resolution}

\begin{figure}[htbp]
    \centering
    \includegraphics[width=\linewidth]{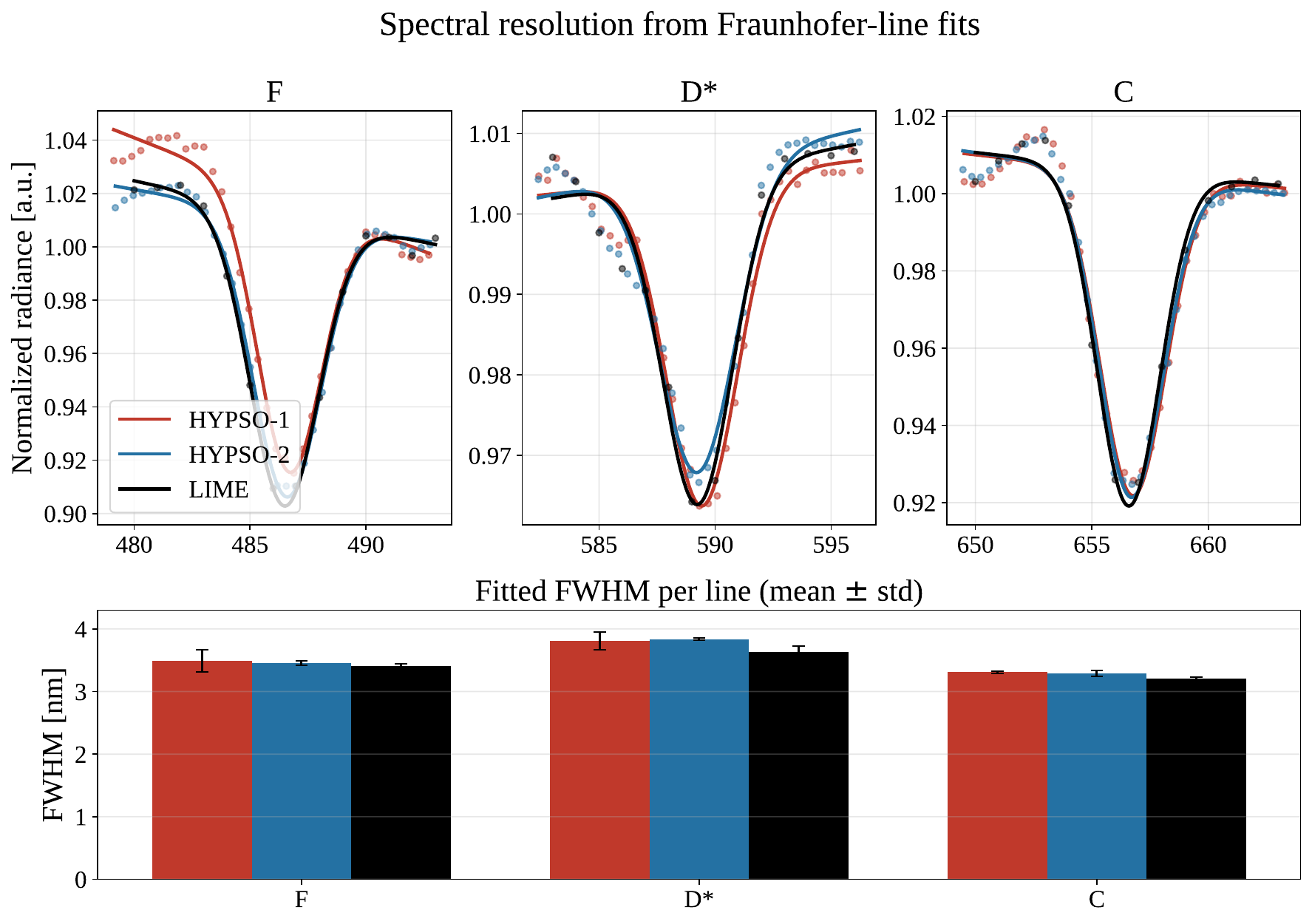}
    \caption{Analysis of Fraunhofer-line FWHM for HYPSO-1, HYPSO-2, and LIME.}
    \label{fig:lime_fwhm}
\end{figure}

Figure~\ref{fig:lime_fwhm} and Table~\ref{tab:lime_fwhm} show the FWHM of the Fraunhofer lines $F$, $D$, and $C$, used to determine the effective spectral resolution of the LIME reference in Section~\ref{sec:spec_res}.

\begin{table}[htbp]
    \centering
    \caption{FWHM of Fraunhofer lines $F$, $D$, and $C$ for HYPSO-1, HYPSO-2, and LIME.}
    \label{tab:lime_fwhm}
    \begin{tabular}{@{}llll@{}}\toprule
    & \multicolumn{3}{c}{Resolution [nm]}\\
         Line&  HYPSO-1&  HYPSO-2& LIME\\\midrule
         $F$&  3.49&  3.45& 3.40\\
         $D$&  3.81&  3.84& 3.63\\
         $C$&  3.31&  3.29& 3.21\\
         \midrule
         \textbf{Mean}&  \textbf{3.54}&  \textbf{3.53}& \textbf{3.41}\\ \bottomrule
    \end{tabular}
\end{table}

\begin{figure}[htbp]
    \centering
    \includegraphics[width=\linewidth]{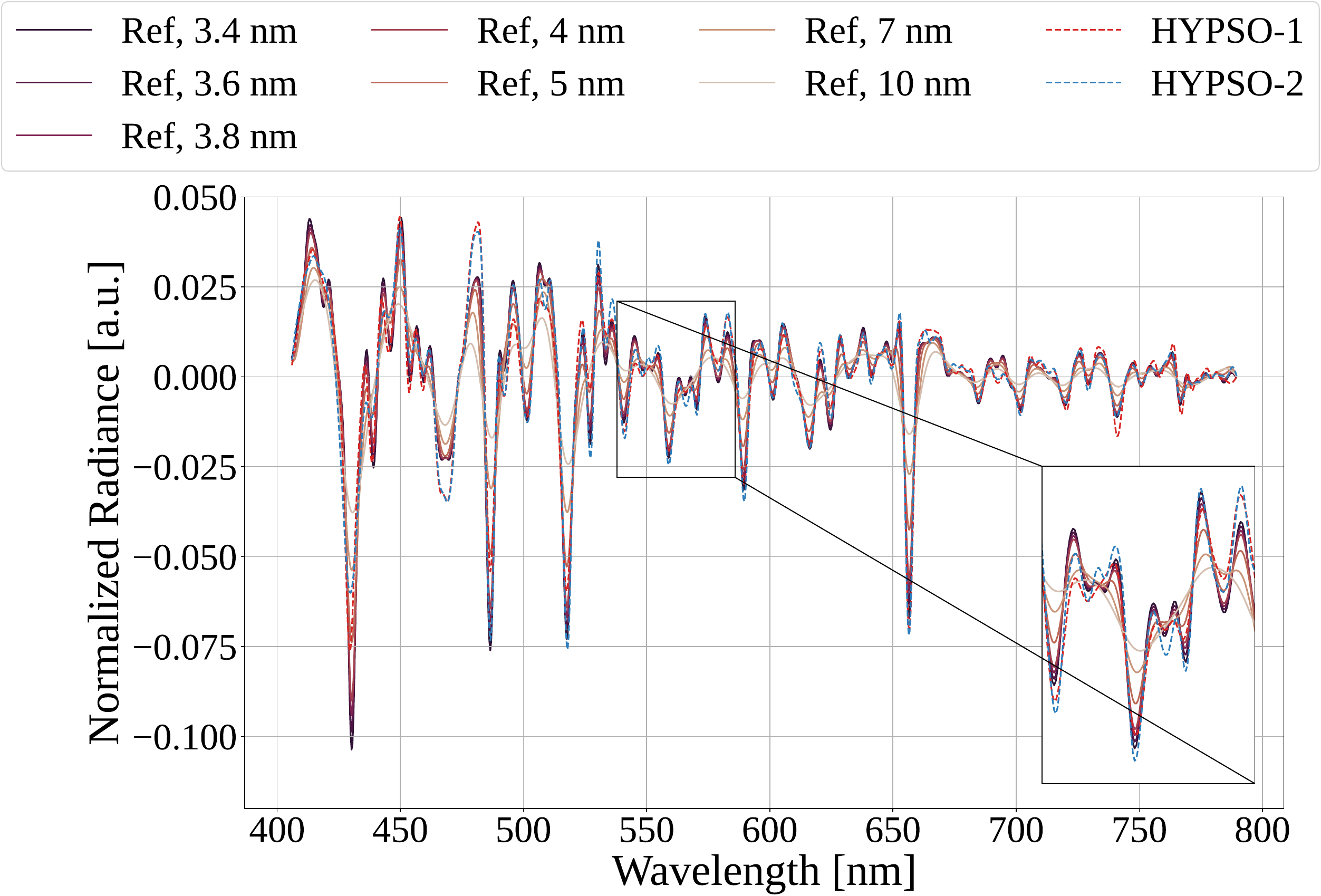}
    \caption{Comparison between band-pass filtered Moon spectra of HYPSO-1, HYPSO-2, and LIME simulated at seven different spectral resolutions, used to investigate the spectral resolution.}
    \label{fig:spec_res_comparison}
\end{figure}

Figure~\ref{fig:spec_res_comparison} shows the band-pass filtered Moon spectra of HYPSO-1 and HYPSO-2 together with LIME spectra simulated at seven different spectral resolutions between 3.4\,nm and 10\,nm, supporting the estimates in Table~\ref{tab:spec_res}.

\section{BRDF Correction over Libya-4} \label{sec:supp_brdf}

\begin{figure*}[htbp]
    \centering
    \includegraphics[width=\linewidth]{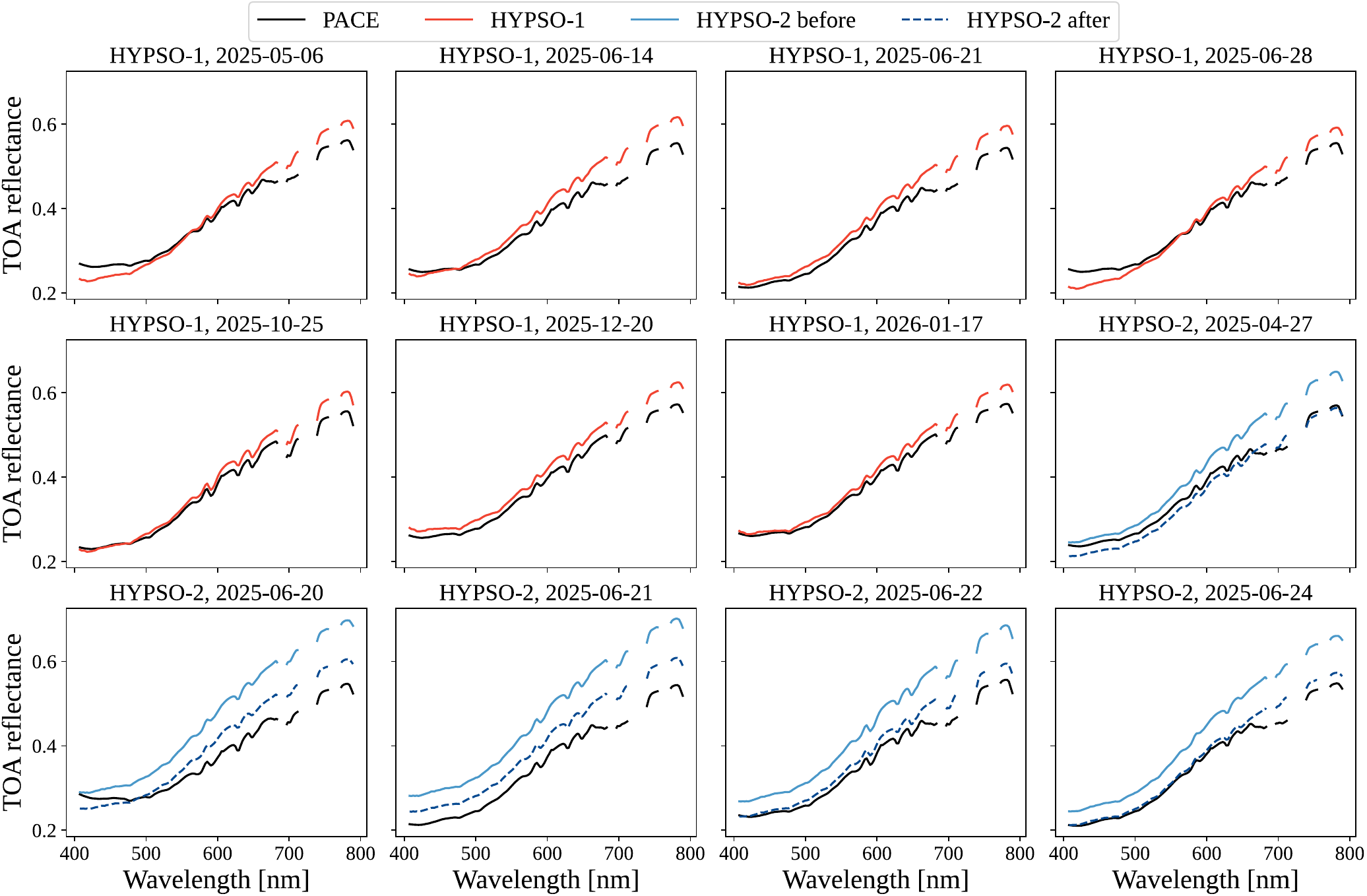}
    \caption{HYPSO-1 and HYPSO-2 ToA reflectance over Libya-4, compared to PACE, after BRDF correction. For HYPSO-2, the solid line shows the spectra before magnitude harmonization, and the dashed line the spectra after.}
    \label{fig:libya_pace_toa_reflectance}
\end{figure*}

\begin{figure}[htbp]
    \centering
    \includegraphics[width=\linewidth]{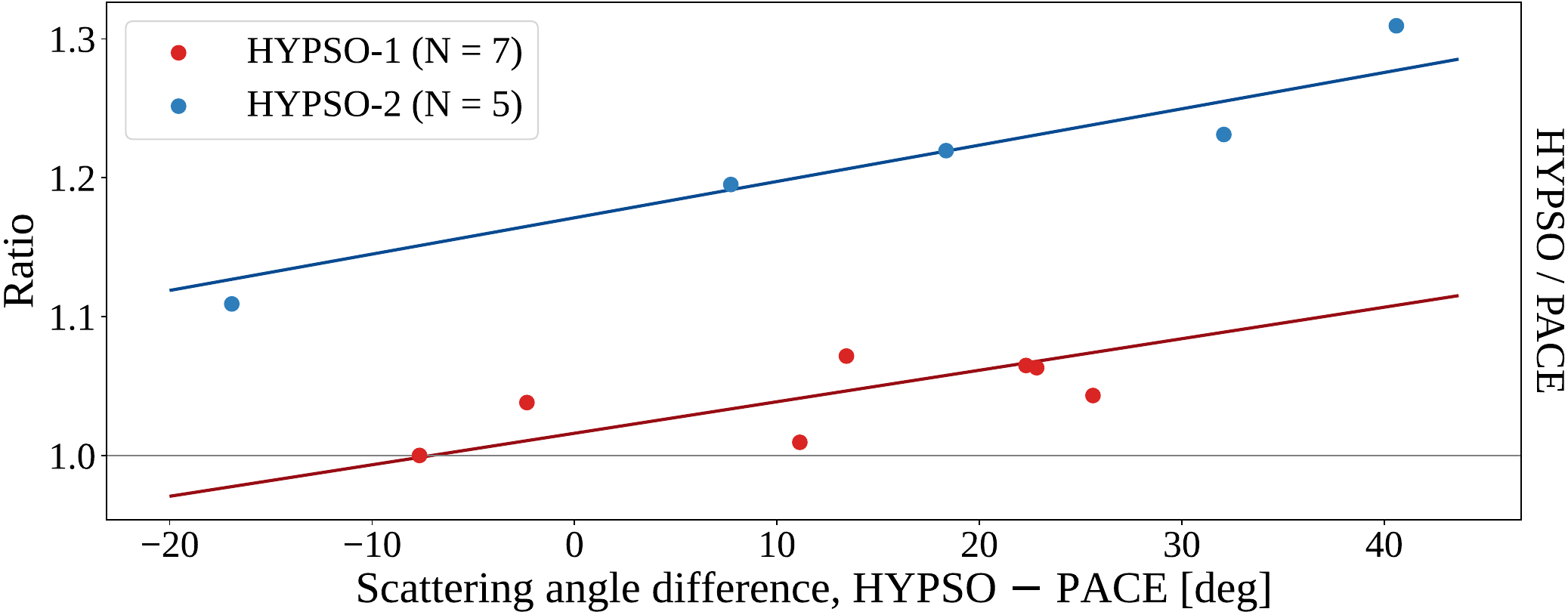}
    \caption{Band-averaged ratio between the ToA reflectance of HYPSO and PACE over Libya-4, using in-orbit coefficients, against the difference in viewing geometry between HYPSO and PACE, before geometry correction. The lines show the fit with a slope shared by both satellites and one intercept per satellite.}
    \label{fig:libya_pace_geometry}
\end{figure}

As described in Section~\ref{sec:magnitude}, HYPSO and PACE observe Libya-4 at different viewing geometries. Instead of a full BRDF model, we use an empirical correction based on how the difference in viewing geometry $\Delta\Theta$ between HYPSO and PACE changes the ratio $r$ between the two, where a larger viewing geometry is closer to the backscatter direction, in which the desert appears brighter. HYPSO mostly observes the site at larger viewing geometries than PACE (Figure~\ref{fig:libya_pace_geometry}). Since both the calibration offset and the BRDF act as gains on the signal, and therefore have a multiplicative effect, we fit $r$ by linear regression in log space,
\begin{equation}
    \ln r = a_i + k\,\Delta\Theta,
\end{equation}
with one intercept $a_i$ per satellite and a slope $k$ shared by both satellites, since the BRDF is a property of the site. The intercept gives the log of the ratio at equal geometry ($\Delta\Theta = 0$), and thereby the calibration offset that we want to find,
\begin{equation}
    a_i = \ln(1+b_i),
\end{equation}
where $b_i$ is the calibration offset of satellite $i$ relative to PACE, and the $+1$ turns the offset into a ratio. The fit, shown in Figure~\ref{fig:libya_pace_geometry}, gives $k = 0.22 \pm 0.04$\,\%/deg, with a residual scatter of 1.9\%. Each HYPSO/PACE ratio is subsequently corrected for the difference in geometry by the factor $e^{-k\Delta\Theta}$, applied equally to all bands, as a simple BRDF correction.

Figure~\ref{fig:libya_pace_toa_reflectance} shows the 12 Libya-4 matchups used to validate the radiometric response of the HYPSO satellites against PACE, with the BRDF correction applied. For HYPSO-2, the solid line shows the spectra before magnitude harmonization, and the dashed line the spectra after.

\section{Ground Resolvable Distance}

\begin{figure*}[t!]
    \centering
    \begin{subfigure}[b]{0.49\textwidth}
        \centering
        \includegraphics[width=\linewidth]{figures/bridge_per_band.png}
        \vspace*{1mm}
        \caption{Captures of the bridge at 10 wavelengths by HYPSO-1 (left) and HYPSO-2 (right). The along-track direction is horizontal in these images.}
        \label{fig:bridge_per_band_combined}
    \end{subfigure}
    \hfill
    \begin{subfigure}[b]{0.49\textwidth}
        \centering
        \includegraphics[width=\linewidth]{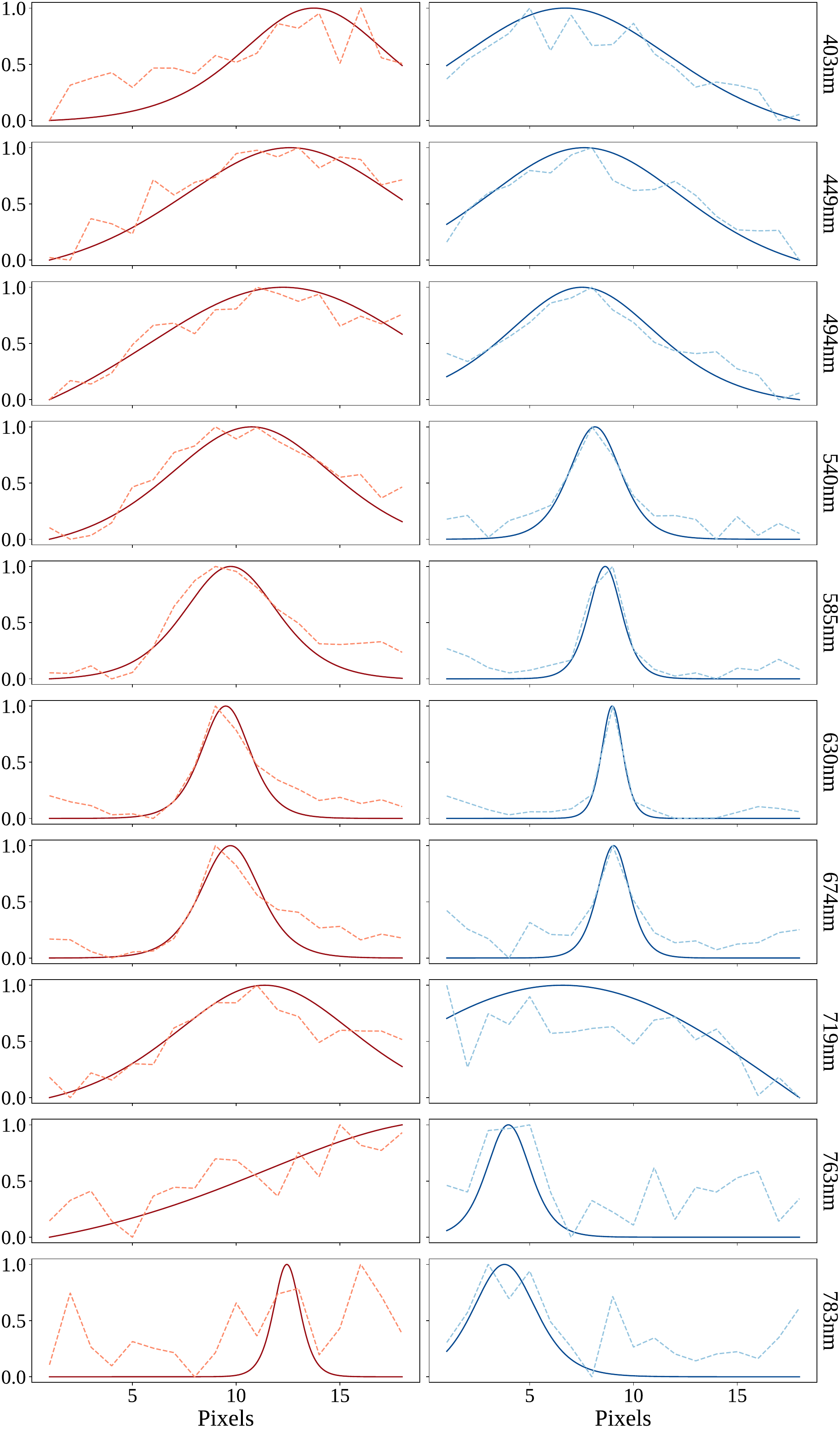}
        \caption{Normalized bridge intensity at 10 wavelengths for HYPSO-1 (left) and HYPSO-2 (right).}
        \label{fig:bridge_spectrum_per_band_combined}
    \end{subfigure}
    \caption{Lake Pontchartrain bridge across the wavelength range for HYPSO-1 and HYPSO-2.}
    \label{fig:bridge_combined}
\end{figure*}

Figure~\ref{fig:bridge_per_band_combined} shows the Lake Pontchartrain bridge captures at 10 wavelengths, where the bridge becomes visibly wider towards the two ends of the wavelength range. Figure~\ref{fig:bridge_spectrum_per_band_combined} shows the pixel intensity of the bridge along the red line in Figure~\ref{fig:grd_bridge}. The dotted line shows the raw measurements, and the solid line shows the fitted curve used to calculate the FWHM. These profiles are used to determine the wavelength ranges where the FWHM estimates are reliable (Section~\ref{sec:grd}).

\section{Artifacts}

\begin{figure}[htbp]
    \centering
    \includegraphics[width=\linewidth]{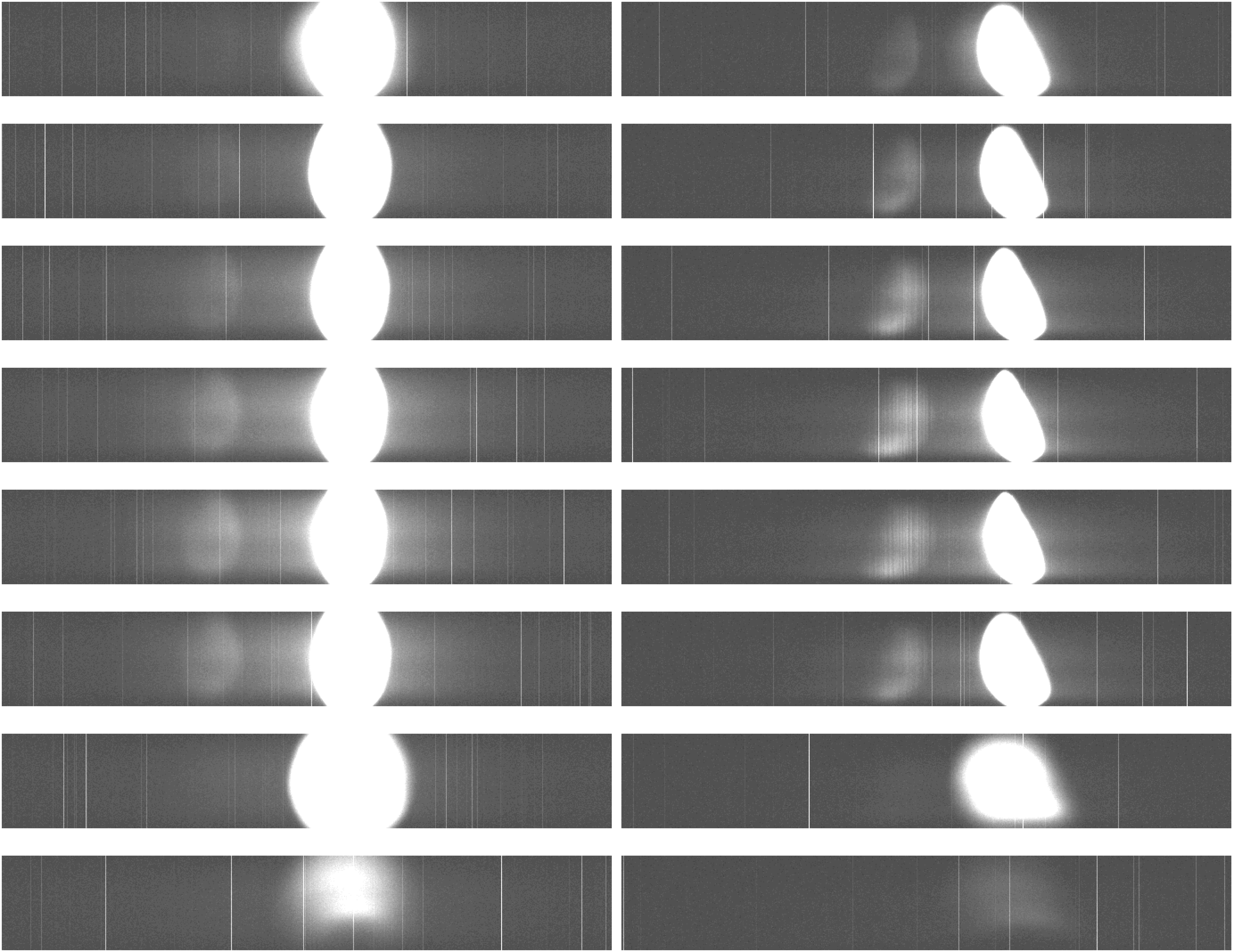}
    \caption{Moon captures across the spectral range with magnified intensity, by HYPSO-1 (left) and HYPSO-2 (right), used to investigate the ghosting artifact.}
    \label{fig:moon_artifact}
\end{figure}

Figure~\ref{fig:moon_artifact} shows two Moon captures by HYPSO-1 and HYPSO-2 with saturation in the central bands, where the intensity is magnified ten-fold to make the ghosting artifact discussed in Section~\ref{sec:artifacts} visible. In the HYPSO-1 capture the Moon is also stretched in the along-track direction, which is a clear indication of oversampling.

\section{Spatial Validation}
\label{sec:Spatialvalidation}

\begin{figure}[htbp]
    \centering
    \includegraphics[width=1\linewidth]{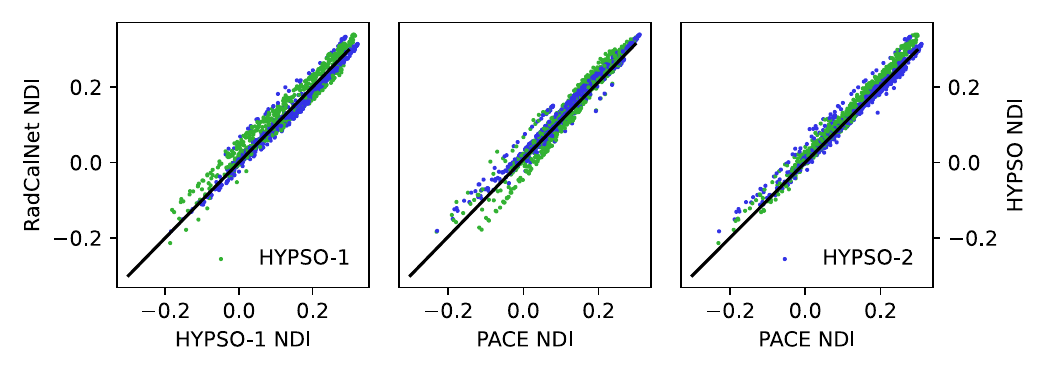}
    \caption{NDIs over the Gobabeb site of HYPSO-1 against RadCalNet (left), PACE against RadCalNet (center), and HYPSO against PACE (right). The black line shows equality.}
    \label{fig:NDI_gobabeb}
\end{figure}

The NDIs over the Gobabeb site, corresponding to Figure~\ref{fig:NDI_lacrau} for La Crau, are shown in Figure~\ref{fig:NDI_gobabeb}.

\begin{figure}[htbp]
    \centering
    \includegraphics[width=1\linewidth]{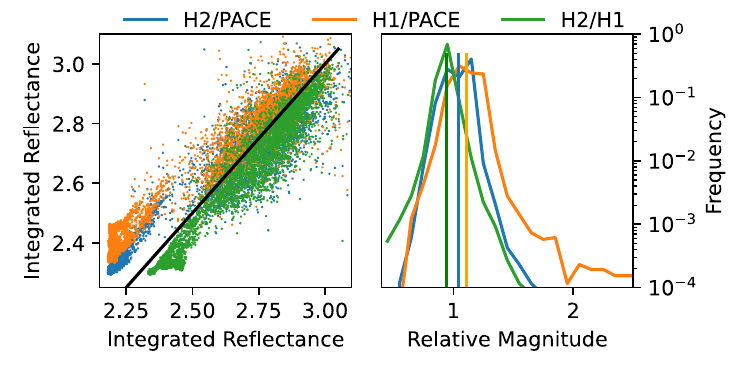}
    \caption{For each pixel in the aligned La Crau image from each satellite, the \gls{toa} reflectance is integrated over all wavelengths. The integrated reflectances between the satellites are parallel to equality (black line), although there is an offset between the water pixels at lower reflectance and the land pixels (left). The integrated reflectance ratios between the satellites show standard deviations of 10\% for HYPSO-2/PACE, 13\% for HYPSO-1/PACE, and 6.4\% for HYPSO-1/HYPSO-2, while the mean ratios on the scene are 1.04, 1.11, and 0.94, respectively (right).}
    \label{fig:lacrau_magnitude_comparison}
\end{figure}

In the main text, it was found that normalized spectra showed much less variation between the satellites compared to spectra without normalization (Table~\ref{tab:magnitude}). The implications of the relative amounts of \textit{magnitude} and \textit{shape} uncertainty for data processing are explored using the La Crau capture (Figure~\ref{fig:radimetric_lacrau}). First, data from the three satellites is mapped to a common spatial/spectral grid, spatially defined by the PACE image and spectrally defined by the HYPSO-2 wavelengths. For each pixel in the region where the three satellites overlap, the 25 nearest HYPSO pixels are selected and averaged. The PACE and HYPSO-1 spectra are resampled to the HYPSO-1 wavelengths, with the former blurred slightly to prevent aliasing. These three aligned images are used for all the validation tests described in this section.

\begin{figure}[htbp]
    \centering
    \includegraphics[width=1\linewidth]{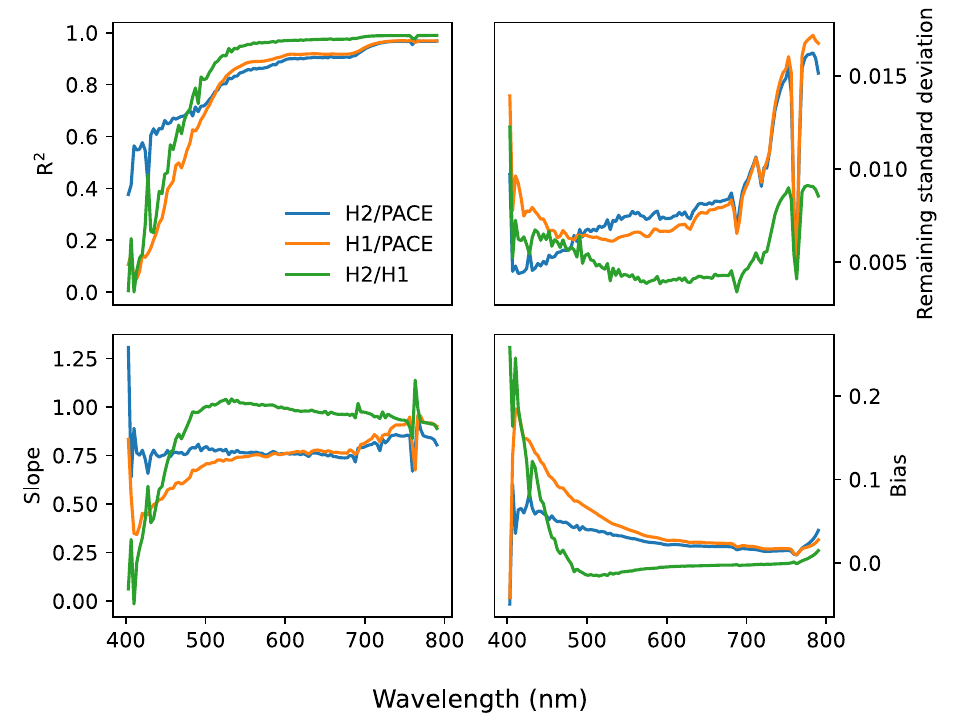}
    \caption{Linear regression of the variation of the \gls{toa} reflectance between the HYPSO-1, HYPSO-2, and PACE satellites over La Crau. The $R^{2}$ values show consistency between all three satellites above 500\,nm, although the HYPSO satellites are slightly more consistent with each other than with PACE. The remaining standard deviation is similarly the lowest between 500--700\,nm and also shows consistency between the HYPSO satellites. Taking into account the signal level at the RadCalNet site, this would correspond to $\approx$5\% uncertainty in the radiometric accuracy for these bands. The slope of the linear regression shows that the HYPSO satellites underestimate the variance observed in PACE, but are very consistent with each other, having a slope near one at wavelengths longer than 500\,nm (except for a few gas absorption lines). The bias is largest near the ultraviolet regime.}
    \label{fig:lacrau_toaR_fits}
\end{figure}

First, changes in the magnitude, or integrated reflectance, across the scene are analyzed. The \gls{toa} reflectance for each pixel is summed across all wavelengths and compared between the satellites (Figure~\ref{fig:lacrau_magnitude_comparison}). Although a correlation between the integrated reflectances is clear, there is also considerable variation between pixels. Notably, HYPSO-1 and HYPSO-2 are more consistent with each other, with a standard deviation of only 6.4\% of the signal, versus 10\% or more when compared to PACE. This level of variation is consistent with the uncertainty levels found in the main text.

Second, each band is regressed independently between the satellites in order to better understand the differing contributions of shape and magnitude uncertainty (Figure~\ref{fig:lacrau_toaR_fits}).
Over much of the range, the PACE data explain most of the variance of the HYPSO signal. However, at wavelengths shorter than 500\,nm, the variance is no longer well explained by PACE signal variation. Again, more variation is explained in the intra-HYPSO comparison than in the comparison with PACE. In particular, the slope of the HYPSO regression is nearly 1 over much of the range, which is what would be expected if both satellites measure the same underlying quantity. The reduced slope of the regression between the HYPSO satellites and PACE perhaps reflects the effect of different imaging angles.

\begin{figure}[htbp]
    \centering
    \includegraphics[width=1\linewidth]{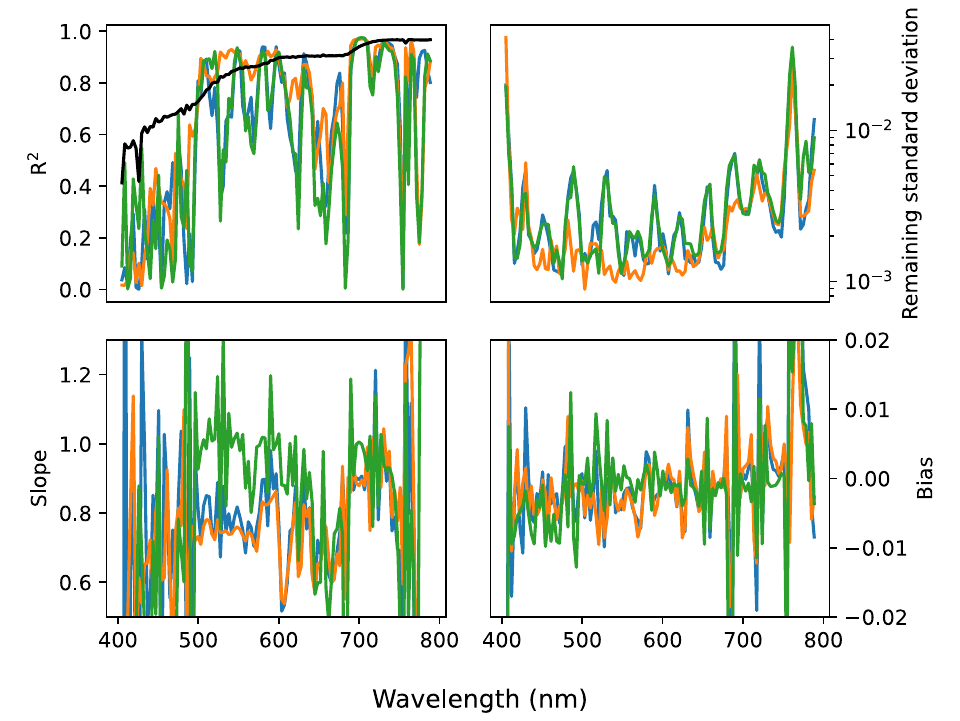}
    \caption{Linear regression of the variation of the 1st spectral normalized-difference derivative of \gls{toa} reflectance between the HYPSO-1, HYPSO-2, and PACE satellites over La Crau. The $R^{2}$ shows consistency between all three satellites above 500\,nm, although some regions in 600--700\,nm are notably worse. The HYPSO-2/PACE $R^{2}$ for single bands is overlaid in black for comparison. The remaining standard deviation is less than 0.002 over 500--700\,nm. This would correspond to uncertainty levels of $\approx$0.1\% in the original data, lower than suggested by the radiometric calibration. This validates that the spectral shape shows less uncertainty than the overall radiometric calibration. The slope of the linear regression shows that the HYPSO satellites underestimate the variance observed in PACE, but are relatively consistent with each other (and with the analysis in Figure~\ref{fig:lacrau_toaR_fits}).}
    \label{fig:lacrau_D1_fits}
\end{figure}

\begin{figure}[htbp]
    \centering
    \includegraphics[width=1\linewidth]{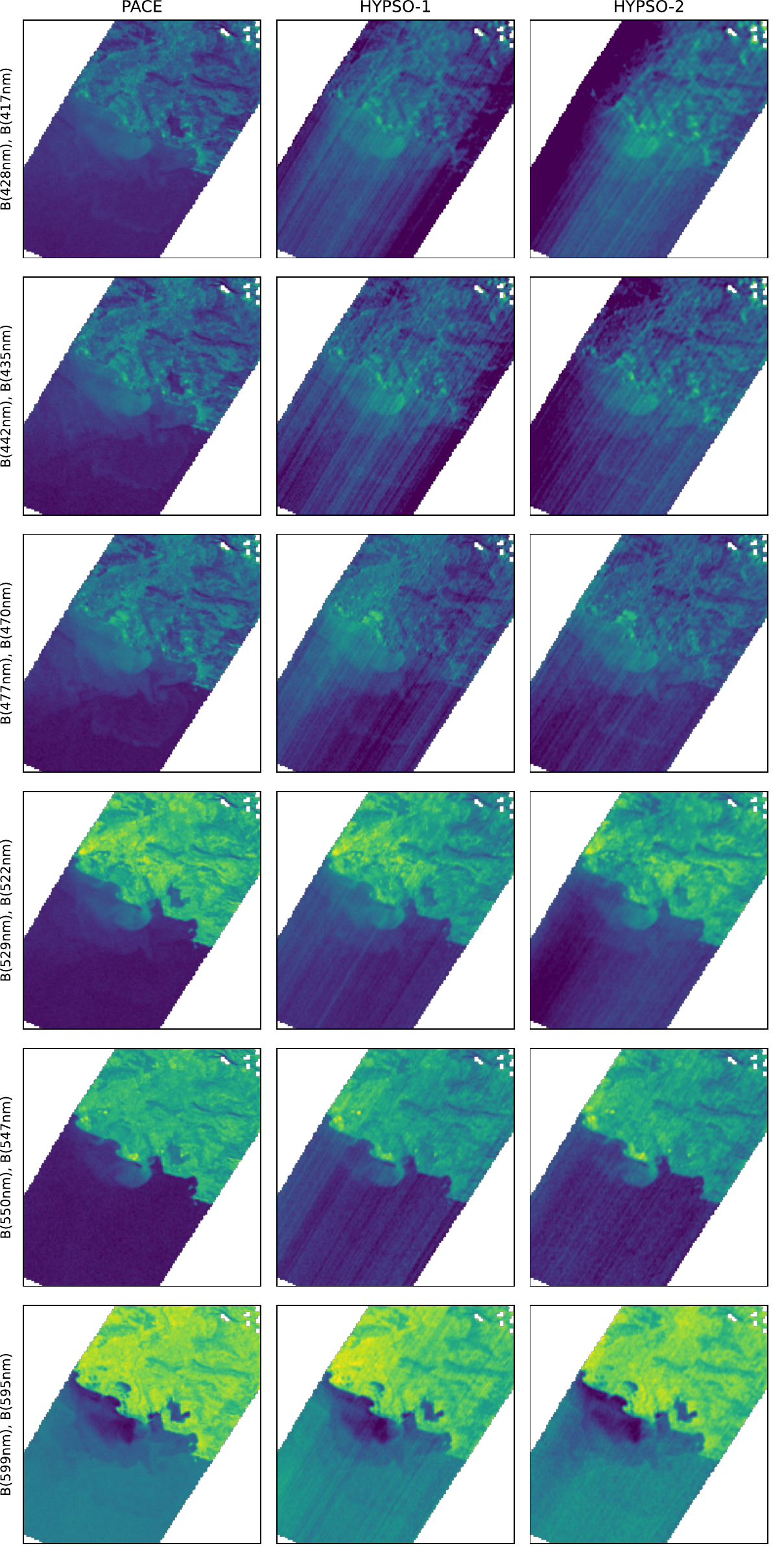}
    \caption{The NDI is calculated for neighboring wavelengths for all three satellites. The offset between the images is removed, and the images are then plotted on the same scale, with the minimum and maximum values defined by the PACE image. Even though striping artifacts are visible at the shortest wavelengths, similar spatial patterns are observed between the satellites. This illustrates the high $R^{2}$ and sensitivities found in Figure~\ref{fig:lacrau_D1_fits}. The spatial images suggest that the low values of the slope found in that figure may originate from the difference in the appearance of water and land. NDIs for wavelengths longer than 600\,nm are shown in Figure~\ref{fig:Dcomparisonred}.}
    \label{fig:Dcomparisonblue}
\end{figure}

Third, the 1st-derivative NDI is calculated across the spectrum of each pixel. This quantity separates the overall spectral magnitude from the geometry of the separate bands. Moreover, it is critical for many types of hyperspectral data analysis, for example any 4th-derivative search of absorption extrema. In addition, the form of the NDI allows variation to be interpreted in terms of percent variation of the original signal. More specifically, variation in the NDI corresponds to $2\times$ the percent variation of the original signal.

It is not surprising that the regression between the different signals shows considerably more variation than the regression between the individual bands, as the spectra tend to be quite smooth except at atmospheric absorption lines. Even so, PACE explains most of the variance in the signals of the HYPSO satellites in the ranges 500--600\,nm and 700--750\,nm. The computed slopes are similar to those calculated above, although they show a dip between 600--700\,nm. The remaining variance suggests that signals well below 1\% of the relative radiometric intensity can be resolved. Thus, the shape uncertainty assessment in the main text underestimates the consistency between satellites for many bands, because some regions of the spectrum, namely the blue, contribute a disproportionate amount of shape uncertainty.

\begin{figure}[htbp]
    \centering
    \includegraphics[width=1\linewidth]{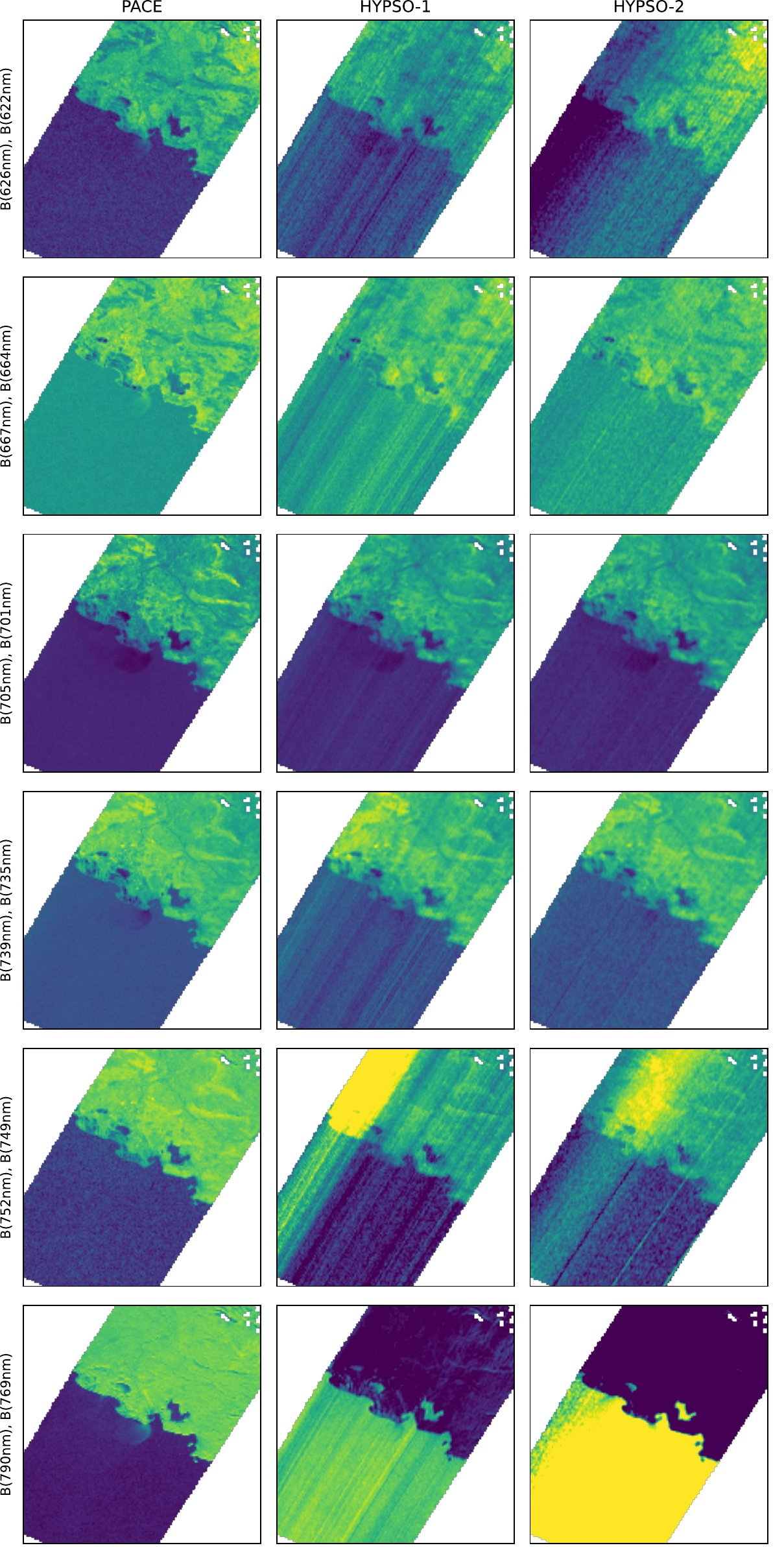}
    \caption{The NDI is calculated for neighboring bands in the red and near-infrared part of the spectrum. Similar spatial patterns are observed up to about 750\,nm, but above that, the spatial patterns of the NDI signal between HYPSO and PACE show little relationship, and become inverted at wavelengths longer than 780\,nm.}
    \label{fig:Dcomparisonred}
\end{figure}

The 1st-derivative NDI signals show considerable spatial consistency between the satellites, except at wavelengths longer than 750\,nm (Figures~\ref{fig:Dcomparisonblue} and \ref{fig:Dcomparisonred}). Considerable amounts of striping are visible in the HYPSO images, but otherwise the spatial patterns are generally consistent with PACE. This supports the suggestion from Figure~\ref{fig:lacrau_D1_fits} that the uncertainty in some of the bands is much smaller than the average. Moreover, this analysis also suggests that further improvements in signal clarity could be achieved by improving the de-striping presented in \cite{bakken_hypso1_2023}.

\begin{figure}[htbp]
    \centering
    \includegraphics[width=1\linewidth]{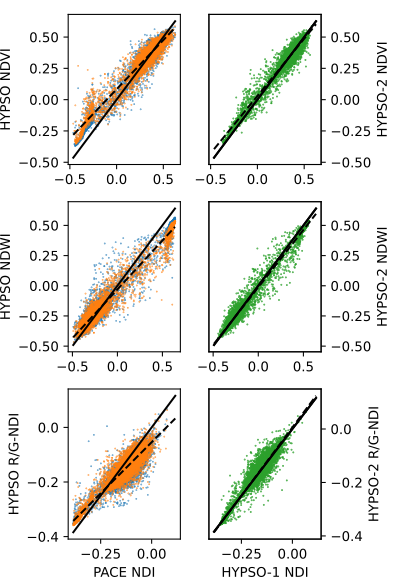}
    \caption{Classic NDIs, designed for multispectral imagers, focus on the comparison of bands further separated in wavelength. Here, the normalized difference vegetation index, the normalized difference water index, and the red/green normalized difference index are compared between the sensors. The solid black lines show the ideal correlation, while the dashed black lines show the relationship found by regression.}
    \label{fig:NDI_metrics}
\end{figure}

\begin{figure}[htbp]
    \centering
    \includegraphics[width=1\linewidth]{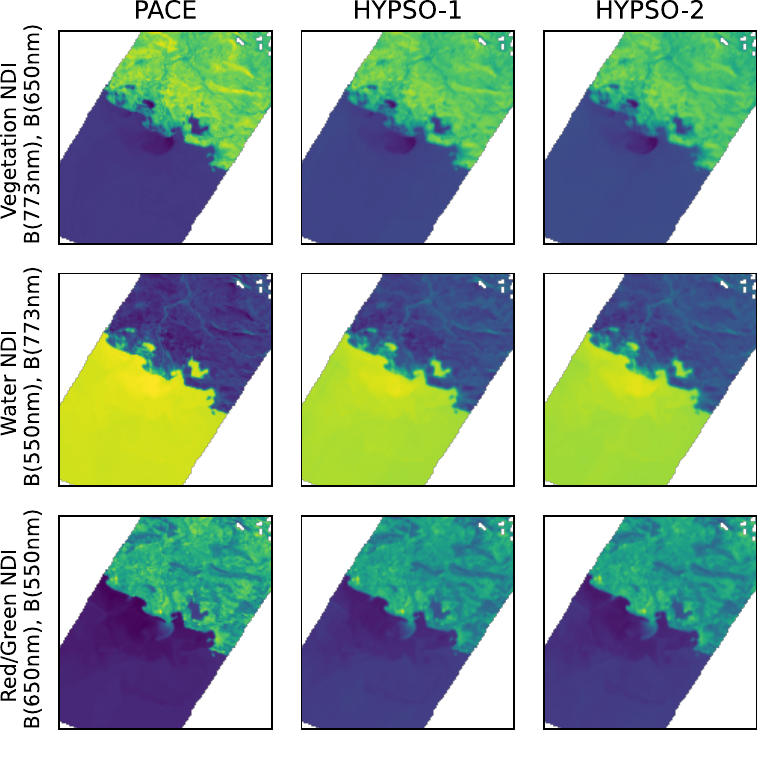}
    \caption{The classic NDIs show similar spatial patterns over the La Crau scene for the different satellites.}
    \label{fig:ndix_lacrau}
\end{figure}

Many classical NDIs are designed for multispectral satellites and thus use much larger intervals between bands. The HYPSO constellation is also tested with three of these indices (Figures~\ref{fig:NDI_metrics} and \ref{fig:ndix_lacrau}). Again, HYPSO shows considerable consistency with PACE, even though the consistency within the HYPSO constellation is somewhat larger.

\end{document}